%% file: main.tex
\documentclass[11pt,a4paper]{article}
\usepackage[T1]{fontenc}
\usepackage[utf8]{inputenc}
\usepackage[margin=1in]{geometry}
\usepackage{amsmath,amssymb,amsthm}
\usepackage{graphicx,booktabs,rotating}
\usepackage{microtype}
\usepackage[authoryear,round]{natbib}
\usepackage[hidelinks]{hyperref}
\newsavebox{\maintablebox}
\newcommand{\expit}{\operatorname{expit}}


\theoremstyle{plain}
\newtheorem{proposition}{Proposition}
\newtheorem{theorem}{Theorem}
\theoremstyle{definition}

\newcommand{\E}{\mathbb E}

\title{Source projection for calibrated transfer learning in survival prediction}
\author{Fangzhou Cao \and Ting Tian }
\date{School of Mathematics, Sun Yat-sen University, Guangzhou, China}
\begin{document}
\maketitle
\noindent\textbf{Corresponding author:} Ting Tian, \href{mailto:tiant55@mail.sysu.edu.cn}{tiant55@mail.sysu.edu.cn}.

\begin{abstract}
Transfer learning can improve survival prediction in cohorts with few events. An external model may capture useful covariate associations yet systematically overestimate or underestimate risk in the target population, creating a need for selective borrowing. We propose Source Projection Transfer (SPT) for fixed-horizon prediction with a target generalized additive model. SPT requires only source predictions at target covariates, supporting generalized additive, Cox and neural-network source models without individual-level source data. It projects source log-odds scores onto the target basis and borrows through two separately adapted channels: a projection-centered penalty transfers effect shapes, while intercept-recalibrated soft labels inform probabilities at the target risk level. Under local asymptotics, projection avoids the strictly positive risk cost of estimating an uninformative additional source direction. Each channel asymptotically switches off under its corresponding persistent-mismatch condition. In log-logistic and Weibull simulations with covariate shift, SPT improves mean Brier score and discrimination over target-only modeling in all 24 setting--distribution--horizon combinations. Matched comparisons show improvements over borrowing through source probabilities alone across settings, and over borrowing the complete source score under structural mismatch. In UK Biobank-to-SPRINT mortality prediction among standard-treatment participants with reduced kidney function, SPT improves mean discrimination at 24, 30 and 36 months, with Brier improvements at 24 and 36 months. SPT enables flexible source learners to inform a parsimonious target model while adapting to differences between populations.
\end{abstract}

\section{Introduction}
Predicting death or disease progression within a specified period is a common goal in clinical risk assessment. Even a cohort with more than a thousand participants may contain few events from which to estimate risk relationships. In our application, the target cohort comprises 1126 SPRINT standard-treatment participants with reduced kidney function, among whom 67 deaths were observed by 36 months. We use a source cohort of 7059 UK Biobank participants to inform mortality prediction in this target population \citep{Sudlow2015,SPRINT2015}. Modeling event probabilities at individual horizons allows predictor associations to change over time \citep{Zheng2006,Uno2007}. A generalized additive model (GAM) represents nonlinear associations through smooth covariate effects \citep{Wood2017}; the challenge is to estimate these curves reliably when target events are sparse and follow-up is censored.

Existing transfer methods borrow either regression parameters or predictions. Parameter shrinkage requires corresponding source and target coefficients \citep{Li2022,Tian2023,Li2023}; methods that combine source and target likelihood contributions, such as CoxTL, additionally use individual-level source data \citep{Lu2026}. Prediction-based methods accept external risks from different model classes. Cox and discrete-time KL regularization encourage agreement with external predictions \citep{CoxKL2026,Wang2025}, and prediction-oriented transfer learning (POTL) uses an overall tuning parameter to regularize a semiparametric survival model toward a flexible external predictor \citep{Gu2026}. When a single weight controls agreement with external risks, miscalibration can lead to down-weighting the source predictions as a whole, even when their covariate-effect shapes remain useful. This distinction matters when risk levels and effect shapes differ between populations. UK Biobank exhibits a healthy-volunteer effect, with lower mortality than the general population \citep{Fry2017}. A clinical target population may therefore require a different risk level even when some source effect shapes remain useful. Intercept and slope recalibration shift and rescale the source score as a whole \citep{Steyerberg2004}.

We propose Source Projection Transfer (SPT), which uses predictions from a fitted source model to guide a target GAM. The source model may be a GAM, a Cox model or a neural network; SPT requires its predictions at target covariates, rather than individual-level source records. Projection expresses the source log-odds score in the target spline basis. SPT then borrows through two separately adapted channels. The projected coefficients center the smoothness penalty, encouraging target effects to follow useful source shapes while allowing smooth target corrections. Intercept-recalibrated source probabilities provide soft labels that inform fitted risks after adjustment to the target risk level. Target data determine the strength of each contribution. This construction lets a flexible external predictor inform a parsimonious target model, and deployment requires only the fitted target GAM.

Our contributions are threefold. First, projection-centered smoothing transfers source information into the target model's own coefficient space and estimates how strongly that information should guide its smooth effects. Second, under local asymptotics, borrowing the projection avoids the strictly positive risk cost of estimating an uninformative additional source direction, and each borrowing channel asymptotically switches off under its corresponding persistent-mismatch condition. Third, matched simulations identify the contributions of source-centered smoothing and calibrated probability borrowing. Across 24 combinations of setting, distribution and prediction horizon, SPT improves mean Brier score and discrimination over target-only modeling. Structural-mismatch experiments demonstrate the value of projection over direct-score borrowing, and the UK Biobank-to-SPRINT application shows improved mean discrimination at all three clinical horizons.

Section~\ref{sec:methods} presents SPT and its theoretical properties. Sections~\ref{sec:simulation} and~\ref{sec:realdata} report the simulation study and clinical application. Section~\ref{sec:discussion} discusses the findings and their implications. Algorithms, proofs and additional results appear in the Supplementary Material.


\section{Methods}\label{sec:methods}
SPT borrows through two channels: a shape channel via the penalty center (Section~\ref{sec:projection_center}) and a probability channel via calibrated soft labels (Section~\ref{sec:soft_labels}). The first guides the target smooth effects toward the projected source shape; the second encourages agreement with source probabilities after adjustment to the target risk level.

\subsection{Notation and censoring-adjusted target prediction}
Consider independent source ($S$) and target ($T$) populations with sample sizes $n_S$ and $n_T$. In the target training sample, observe $(X_i,Y_i,\delta_i)$, $i=1,\ldots,n$, with $n=n_T$, $Y_i=\min(T_i^*,C_i)$ and $\delta_i=I(T_i^*\leq C_i)$, where $T_i^*$ and $C_i$ are event and censoring times. At a specified horizon $h$, the target is $p_{T,h}(x)=P_T(T^*\leq h\mid X=x)$. Modeling this probability directly allows covariate effects and borrowing strength to vary across clinically relevant horizons without imposing proportional hazards \citep{Zheng2006}. Each fit supplies a horizon-specific risk; separate fits do not impose monotonicity across horizons.

Within the target training data, estimate the censoring survival $G(t)=P_T(C>t)$. Censoring adjustment for the source model is part of source fitting. With $y_i=I(Y_i\leq h,\delta_i=1)$, define the target inverse-probability-of-censoring weight
\begin{equation}
 w_i=\frac{I(Y_i\leq h,\delta_i=1)}{\widehat G(Y_i-)}+
     \frac{I(Y_i>h)}{\widehat G(h)}.\label{eq:ipcw}
\end{equation}
Here $G(t-)=P_T(C\geq t)$ uses the left limit at an event time. Marginal weights recover the full-data risk under $C\perp(T^*,X)$ and positive censoring survival; conditional independence requires a corresponding conditional censoring estimator \citep{Zheng2006,Uno2007}. Supplementary Section~S1 specifies the empirical time convention and numerical stabilization used in the reported analyses.

Write $\ell(y,p)=-y\log p-(1-y)\log(1-p)$ for logistic cross-entropy. Let $B(x)\in\mathbb R^d$ contain the target intercept, unpenalized parametric terms (such as binary covariates), and centered spline basis functions. The target-only GAM (T-GAM) has predictor $\eta_{\mathrm{TG}}(x)=B(x)^\mathsf T\widehat\theta_{\mathrm{TG}}$, estimated by minimizing the weighted logistic loss with penalty $\theta^\mathsf T\Omega\theta/2$. The matrix $\Omega=\sum_j\widehat\lambda_j\Omega_j$ combines basis-specific penalties with smoothing parameters estimated by restricted maximum likelihood (REML) in T-GAM \citep{Wood2003,Wood2011,Wood2017}. The intercept and parametric terms are unpenalized. Basis construction and smoothing estimation are detailed in Supplementary Section~S1.

The source model supplies an estimated event probability $\widehat p_{S,h}(x)$ at the same horizon. It may be a survival model, such as Cox regression, supplying $1-\widehat S_S(h\mid x)$, or a probabilistic classifier trained with a censoring-adjusted loss, such as an IPCW logistic GAM or neural network. In each case the output estimates horizon-specific event risk, rather than an uncalibrated ranking score. Define
\[
 \eta_S(x)=\operatorname{logit}\{\widehat p_{S,h}(x)\}.
\]
SPT uses these source scores at target training covariates. It therefore requires a fitted source model that can be evaluated on harmonized target covariates, but not the source individual-level data, and the source model need not share the form of the target GAM.

\subsection{Source projection and an adaptive penalty center}\label{sec:projection_center}
To express source information in the target model's coordinates, let $B$ stack the rows $B(X_i)^\mathsf T$ and let $\boldsymbol\eta_S=\{\eta_S(X_i)\}_{i=1}^{n_T}$. A penalized Gaussian projection gives
\begin{equation}
 c=\arg\min_v\left\{\tfrac12\|\boldsymbol\eta_S-Bv\|_2^2+\tfrac12v^\mathsf TPv\right\},
 \qquad \widehat a_S(x)=B(x)^\mathsf Tc.\label{eq:projection}
\end{equation}
Using target training covariates directs this approximation toward the empirical target covariate distribution, without estimating a density ratio. Conditional on the selected target basis, the projection uses source scores as responses, with no target outcome loss or IPCW weights. Its penalty $P=\sum_j\widehat\lambda_j^{\mathrm{proj}}\Omega_j$ controls the complexity and stability of the approximation; the projection smoothing parameters are estimated by a separate Gaussian REML fit \citep{Wood2011,Wood2017}. The two penalties serve distinct roles: $P$ smooths the source-score approximation within the target basis, while $\Omega$ regularizes estimation of the target outcome model.

SPT combines the projected source function with a smooth target correction:
\begin{equation}
 \eta_{\mathrm{SPT}}(x)=B(x)^\mathsf T\beta+\rho\widehat a_S(x),
 \qquad \operatorname{pen}(\beta)=\tfrac12\beta^\mathsf T\Omega\beta,
 \qquad 0\leq\rho\leq1.
 \label{eq:projected_score_form}
\end{equation}
The correction $B^\mathsf T\beta$ adapts the borrowed shape to target outcomes. The interval $[0,1]$ places the penalty center between zero and the projected source coefficients, excluding a reversed or amplified source shape. The final target effects remain free to exceed the source effects through the correction.

Because $\widehat a_S=B^\mathsf Tc$, the predictor is fully described by $\theta=\beta+\rho c$. We therefore compute the fit in these target coordinates, where the correction penalty becomes
\begin{equation}
 \tfrac12\beta^\mathsf T\Omega\beta
 =\tfrac12(\theta-\rho c)^\mathsf T\Omega(\theta-\rho c).
 \label{eq:profile}
\end{equation}
For each $\theta$, minimization over the penalty center has the analytic solution
\begin{equation}
 \widehat\rho(\theta)=\Pi_{[0,1]}\left\{\frac{c^\mathsf T\Omega\theta}{c^\mathsf T\Omega c}\right\},
 \label{eq:rho}
\end{equation}
where $\Pi_{[0,1]}$ truncates a scalar to $[0,1]$. If $c^\mathsf T\Omega c=0$, take $\widehat\rho=0$. Within an interior profile region, profiling leaves local movements along $c$ unpenalized and adds at most one effective degree of freedom relative to zero-centered smoothing at the same fixed loss curvature (Supplementary Section~S2.2). Only $\theta$ requires numerical optimization. For fixed training quantities and fixed $\tau$, the profiled fitting criterion is convex; its profiled penalty has a Lipschitz-continuous gradient (Supplementary Section~S2). Throughout, $\Omega$ is held at its T-GAM value within each training split, so SPT reduces exactly to T-GAM when $\rho=\tau=0$.

A natural comparison is direct-score transfer (DST):
\begin{equation}
 \eta_{\mathrm{DST}}(x)=B(x)^\mathsf T\beta+v\eta_S(x),
 \qquad \operatorname{pen}(\beta)=\tfrac12\beta^\mathsf T\Omega\beta,
 \qquad 0\leq v\leq1.
 \label{eq:dst}
\end{equation}
DST borrows the complete source score, whereas SPT borrows its target-space projection. With the same source fit, target basis, penalty and soft-label criterion, this comparison isolates the effect of the borrowed representation. When the source score contains a direction outside the target representation that carries no target signal, estimating its coefficient can increase prediction risk (Section~\ref{sec:theory}).

\subsection{Calibrated soft labels and estimation}\label{sec:soft_labels}
The second channel acts on fitted probabilities, carrying information about risk level and unpenalized effects, such as sex and smoking in Section~\ref{sec:realdata}, that the penalty center cannot transmit directly. We first align the source risk level with the target by estimating an intercept shift:
\begin{equation}
 \sum_iw_i\left[\expit\{\eta_S(X_i)+\alpha\}-y_i\right]=0,
 \qquad q_i=\expit\{\eta_S(X_i)+\widehat\alpha\}.\label{eq:calibration}
\end{equation}
This censoring-adjusted intercept recalibration preserves the scale and ordering of the source score while adjusting its overall risk level \citep{Steyerberg2004}. The resulting $q_i$ are the calibrated soft labels. Their contribution is controlled by $\tau$, separately from the penalty-center coefficient $\rho$.

Denote $B_i=B(X_i)$ and $p_i=\expit(B_i^\mathsf T\theta)$. For fixed $\tau\geq0$, minimize
\begin{equation}
 \mathcal J_\tau(\theta,\rho)=
 \underbrace{\vphantom{\sum_i}\sum_iw_i\ell(y_i,p_i)}_{\text{\strut Target outcome loss}}
 +\underbrace{\vphantom{\sum_i}\tau\sum_i\ell(q_i,p_i)}_{\text{\strut Calibrated soft-label loss}}
 +\underbrace{\vphantom{\sum_i}\tfrac12(\theta-\rho c)^\mathsf T\Omega(\theta-\rho c)}_{\text{\strut Adaptive source-centered penalty}}.
 \label{eq:objective}
\end{equation}
Soft-label cross-entropy differs from the Bernoulli KL divergence $\operatorname{KL}\{\operatorname{Bern}(q_i)\Vert\operatorname{Bern}(p_i)\}$ only by the entropy of the fixed soft label. Related prediction-based borrowing appears in discrete-time KL integration \citep{Wang2025} and prediction-oriented survival transfer \citep{Gu2026}; CoxKL instead defines its KL discrepancy through risk-set probabilities in the Cox partial likelihood \citep{CoxKL2026}. It constrains predicted probabilities, complementing the center's action on smooth coefficients. The soft-label term is unweighted and includes all target training participants, including those censored before $h$, because its responses are source predictions rather than observed outcomes.

We select $\tau$ by three-fold target cross-validation over $\{0,0.1,0.5,1\}$. A paired one-standard-error rule retains the reference $\tau=0.5$ unless another candidate improves held-out log loss by more than one estimated standard error. This intermediate reference avoids changing the borrowing weight for small validation differences. Supplementary Section~S1 provides the detailed estimation and cross-validation procedure.

\begin{center}
\fbox{\begin{minipage}{0.94\linewidth}\small
\textbf{SPT at a fixed prediction horizon}
\begin{enumerate}
\item Obtain source probabilities and their logit scores $\eta_S$.
\item Within each target training fold, fit T-GAM to obtain $B$ and $\Omega$, project $\eta_S$ to obtain $c$, and estimate the calibration intercept.
\item Fit each $\tau$ candidate, profiling $\rho\in[0,1]$ analytically, and evaluate held-out IPCW log loss.
\item Select $\tau$ using the paired one-standard-error rule with reference 0.5.
\item Re-estimate the target training quantities and fit the selected candidate on the full target training sample.
\end{enumerate}
\end{minipage}}
\end{center}

The final estimator and its penalty-center coefficient are
\begin{equation}
 \widehat\theta_{\mathrm{SPT}}
 =\arg\min_\theta\mathcal J_{\widehat\tau}\{\theta,\widehat\rho(\theta)\},
 \qquad \widehat\rho_{\mathrm{SPT}}=\widehat\rho(\widehat\theta_{\mathrm{SPT}}).
 \label{eq:final_estimator}
\end{equation}
Restoring the horizon index $h$, the event risk and corresponding survival probability are
\begin{equation}
 \widehat p_{T,h}(x)=\expit\{B_h(x)^\mathsf T\widehat\theta_{\mathrm{SPT},h}\},
 \qquad \widehat S_T(h\mid x)=1-\widehat p_{T,h}(x).
 \label{eq:final_prediction}
\end{equation}
Prediction requires only the fitted target GAM, so the source model is unnecessary at deployment.


\subsection{Theoretical properties}\label{sec:theory}
We study how the projected center changes target prediction risk and how the two borrowing channels adapt to source mismatch. Throughout this section, matched estimators use the same target basis $B$, penalty matrix $\Omega$ and IPCW weights. The first result identifies exactly when the center is switched off. Theorem~\ref{thm:risk} compares borrowing the projected source score with borrowing the complete source score at $\tau=0$, without soft-label regularization. Theorem~\ref{thm:safety} then considers cross-validation and the full-sample refit. Complete conditions and proofs are given in Supplementary Section~S2.

For fixed $\tau$, define the loss without the smoothness penalty by
\[
 F_\tau(\theta)=\sum_iw_i\ell(y_i,p_i)+\tau\sum_i\ell(q_i,p_i),
 \qquad p_i=\expit(B_i^\mathsf T\theta).
\]
These are the target outcome and calibrated soft-label terms in \eqref{eq:objective}.

\begin{proposition}[Exact rejection of the penalty center]\label{prop:direction}
Suppose $F_\tau(\theta)$ is differentiable and strictly convex, and let $\widehat\theta_{\mathrm{PR},\tau}$, the prediction-regularized fit with $\rho=0$, be its finite minimizer after adding $\theta^\mathsf T\Omega \theta/2$. For $\Omega\succeq0$ and $c^\mathsf T\Omega c>0$, the joint SPT minimizer is unique and
\[
 \widehat\rho=0
 \quad\Longleftrightarrow\quad
 c^\mathsf T\Omega \widehat\theta_{\mathrm{PR},\tau}\leq0.
\]
On this event, $\widehat\theta=\widehat\theta_{\mathrm{PR},\tau}$. If $c^\mathsf T\Omega c=0$, take $\widehat\rho=0$.
\end{proposition}
Thus profiling retains the projected source shape when it aligns with the zero-centered fit in the penalty metric. At $\tau=0$, rejecting the center recovers the matched T-GAM estimator.

For the risk comparison, condition on an independently fitted source model, and fix a horizon and an identifiable target basis $B(x)\in\mathbb R^d$. Suppose the true event probability is $p^\star(x)=\expit\{B(x)^\mathsf T\theta^\star\}$. For a fitted logit $\eta$, define its excess log risk by
\begin{equation}
 \mathcal K(\eta)=\E_X\!\left[
 \operatorname{KL}\!\left\{\operatorname{Bern}(p^\star(X))\,\Vert\,
 \operatorname{Bern}(\expit\{\eta(X)\})\right\}\right].
 \label{eq:main_logrisk}
\end{equation}
This is the expected logistic loss above the Bayes risk; an outer expectation averages over target training samples conditional on the source model and fixed basis. Write $\eta_S=B^\mathsf Tc_n+r_n$ to distinguish the source projection from the remaining source direction. DST in \eqref{eq:dst} retains this direction, while SPT remains in the target representation.

Let $\Omega_n$ be the common penalty matrix in the sum-loss criterion. Under the local scaling $A_n=\Omega_n/\sqrt n\to_p A\succeq0$, denote the limiting centering gain by $G_K$ and the extra-direction cost by $D(k)>0$. Here $k$ captures the local penalty-induced shift along the additional direction. Their full expressions, including the curvature and IPCW score variance, are given in Supplementary Section~S2.

\begin{theorem}[Expected log risk of projected and direct-score borrowing]\label{thm:risk}
Suppose the target model is correctly specified, the bounded augmented design $(B^\mathsf T,r)^\mathsf T$ is identifiable, and the weighted-score, smoothing and moment conditions C1--C4 in Supplementary Section~S2 hold. For matched estimators with $\tau=0$, convergent source projections and residuals, and $A_n\to_p A$ as above,
\begin{align}
 n\{\E\mathcal K(\widehat\eta_{\mathrm{TG}})-\E\mathcal K(\widehat\eta_{\mathrm{SPT},0})\}&\longrightarrow G_K,\nonumber\\
 n\{\E\mathcal K(\widehat\eta_{\mathrm{DST}})-\E\mathcal K(\widehat\eta_{\mathrm{TG}})\}&\longrightarrow D(k)>0.
 \label{eq:main_risk_limits}
\end{align}
Consequently, the scaled DST-minus-SPT risk difference converges to $G_K+D(k)$. If $G_K\geq0$, SPT has strictly smaller expected log risk than DST for sufficiently large $n$. If $A=0$, then $G_K=0$ and the DST-minus-SPT risk difference is $\nu/(4sn)+o(n^{-1})$, where $s>0$ and $\nu>0$ are the residual curvature and limiting efficient-score variance defined in Supplementary Section~S2.
\end{theorem}
The two terms distinguish source-centered smoothing from estimation of an additional source direction. To interpret $G_K$, let $H=\E[p^\star(1-p^\star)BB^\mathsf T]$ be the target log-risk curvature, $M=AH^{-1}A$, and $\widetilde\rho$ the limiting profiled center weight defined in Supplementary Section~S2. When $c^\mathsf TAc>0$, the limiting centering gain is positive exactly when the center moves toward $\theta^\star$ without overshooting in the $M$ metric: $G_K>0$ if and only if $0<\widetilde\rho<2c^\mathsf TM\theta^\star/(c^\mathsf TMc)$ (Supplementary equation~S18). The term $D(k)$ is the expected first-order risk cost of estimating an additional coefficient constrained to be nonnegative; even when $k=0$, $D(0)=\nu/(4s)>0$. In the $A=0$ regime, SPT and T-GAM have the same first-order expected log risk, while DST retains the extra-direction cost.

To study automatic borrowing, use the finite-grid, paired one-standard-error rule and full-sample refit described above. Assume a correctly specified fixed target model, consistent censoring weights, bounded validation losses, a fixed number of folds, $\|\Omega_m\|_{\mathrm{op}}/m\to_p0$ and stable projected centers on each training sample of size $m$. The detailed conditions N1--N5 are given in Supplementary Section~S2.

\begin{theorem}[Asymptotic rejection of mismatched borrowing channels]\label{thm:safety}
Under the preceding regularity conditions:
\begin{enumerate}\renewcommand{\labelenumi}{(\alph{enumi})}
\item Calibrated soft labels retain a systematic discrepancy in directions represented by the target basis. If the calibrated soft-label function converges uniformly to $q_\infty$ with $\E[B(X)\{q_\infty(X)-p^\star(X)\}]\ne0$, then $\Pr(\widehat\tau=0)\to1$.
\item The projected center is either inactive or persistently opposed to the target coefficient vector in the penalty metric. Specifically, if, for some $\gamma>0$,
\[
 \Pr\!\left\{\Omega_nc_n=0\ \text{or}\ 
 \frac{(\Omega_nc_n)^\mathsf T\theta^\star}{\|\Omega_nc_n\|}\leq-\gamma\right\}\to1,
\]
then the $\tau=0$ candidate satisfies $\Pr(\widehat\rho_0=0,\widehat\theta_{n,0}=\widehat\theta_{\mathrm{TG}})\to1$.
\item When both source conditions hold, the selected full-sample estimator satisfies
\[
 \Pr\{\widehat\tau=0,\widehat\rho_{\widehat\tau}=0,
       \widehat\theta_{n,\widehat\tau}=\widehat\theta_{\mathrm{TG}}\}\to1.
\]
\end{enumerate}
\end{theorem}
The theorem assigns a separate adaptation mechanism to each borrowing channel: validation detects persistent source-probability discrepancy, while profiling rejects an adverse source center. Together they recover the matched target-only estimator when both channels are unsuitable.


\section{Simulation studies}\label{sec:simulation}
\subsection{Simulation settings}
We evaluate SPT under covariate shift and four relationships between the source and target outcome models. The design varies the transferability of main effects and an interaction while retaining the same measured predictors. We consider log-logistic and Weibull event times to distinguish a generating signal that directly determines fixed-horizon log odds from one that determines log cumulative hazards.

Let $X=(X_1,\ldots,X_7)^\mathsf T$, with independent components within each population. In the source cohort,
\[
 X_1\sim\operatorname{Bernoulli}(0.4),\qquad
 X_2\sim\operatorname{Bernoulli}(0.5),\qquad
 X_j\sim\operatorname{Unif}(-\sqrt3,\sqrt3),\quad j=3,\ldots,7.
\]
In the target cohort, $X_1\sim\operatorname{Bernoulli}(0.5)$, while $X_2$ and $X_4,\ldots,X_7$ retain their source distributions. The target density of $X_3$ is
\begin{equation}
 f_{T,3}(x)=\frac{1+0.5x/\sqrt3}{2\sqrt3},
 \qquad -\sqrt3\leq x\leq\sqrt3.
 \label{eq:sim_covariates}
\end{equation}
Thus the cohorts differ in one binary and one continuous marginal distribution but have common covariate support. The target test sample follows the same covariate distribution as the target training sample.

Define $z_1=(x_1-0.4)/\sqrt{0.24}$ and $z_2=(x_2-0.5)/0.5$. The additive signal and interaction are
\begin{equation}
\begin{aligned}
 a(x)={}&0.6z_1-0.45z_2+0.75x_3-0.5x_5+0.35x_7\\
        &+0.55\frac{x_4^2-1}{\sqrt{0.8}}+0.45\sin(\pi x_6/\sqrt3),
 \qquad u(x)=z_1x_3.
\end{aligned}\label{eq:sim_signal}
\end{equation}
The source predictor is $g_S(x)=a(x)+0.8u(x)$ in all four settings. The target predictors are
\begin{equation}
\begin{aligned}
\text{A: }&g_T(x)=a(x)+0.8u(x),\\
\text{B: }&g_T(x)=a(x)-0.55x_3+0.85x_5+0.8u(x),\\
\text{C: }&g_T(x)=a(x)+0.48u(x),\\
\text{D: }&g_T(x)=a(x).
\end{aligned}\label{eq:settings}
\end{equation}
Setting A preserves the conditional outcome relationship despite covariate shift. Setting B changes two main-effect coefficients while retaining the interaction. Setting C retains 60\% of the source interaction strength, and setting D removes the interaction. These settings examine borrowing under agreement, main-effect change, partial interaction retention and loss of a source effect, respectively.

For population $D\in\{S,T\}$, generate $U\sim\operatorname{Unif}(0,1)$ independently of $X$ and set
\begin{equation}
\begin{aligned}
 T_D^{*,\mathrm{LL}}&=36\exp\{-g_D(X)\}
             \left(\frac{U}{1-U}\right)^{\sigma_{\mathrm{LL}}},\\
 T_D^{*,\mathrm{W}}&=36\exp\{-g_D(X)+\sigma_{\mathrm{W}}\gamma_E\}
             \{-\log(1-U)\}^{\sigma_{\mathrm{W}}},
\end{aligned}\label{eq:sim_times}
\end{equation}
where $\gamma_E$ is Euler's constant and $\sigma_{\mathrm{LL}}$ and $\sigma_{\mathrm{W}}$ are distribution-specific scales, fixed across populations and settings. Their numerical values are given in the Supplementary Material. The corresponding conditional survival functions are
\begin{equation}
\begin{aligned}
 S_D^{\mathrm{LL}}(t\mid x)
 &=\left[1+(t/36)^{1/\sigma_{\mathrm{LL}}}
          \exp\{g_D(x)/\sigma_{\mathrm{LL}}\}\right]^{-1},\\
 S_D^{\mathrm{W}}(t\mid x)
 &=\exp\left[-(t/36)^{1/\sigma_{\mathrm{W}}}
          \exp\{g_D(x)/\sigma_{\mathrm{W}}-\gamma_E\}\right].
\end{aligned}\label{eq:sim_survival}
\end{equation}
Each replication contains $n_S=3000$ source observations, $n_T=500$ target training observations and 5000 independent target test observations. Independent censoring times follow $C\sim\operatorname{Exp}(\lambda)$, with rates calibrated separately for each setting and event-time distribution to expected censoring proportions of 40\% in the source sample, 50\% in the target training sample and 30\% in the target test sample. We observe $Y=\min(T^*,C)$ and $\delta=I(T^*\leq C)$. Prediction horizons Q25, Q50 and Q75 are the 25th, 50th and 75th percentiles of target-training observed $Y$ in each replication.

In each source and target training sample, generalized cross-validation (GCV) compares an additive GAM with a GAM that also includes the $X_1$--$X_3$ interaction. The selected formula is then fitted by REML, using shrinkage thin-plate bases with $k_S=10$ and $k_T=5$. By design, the source and target parameterize the interaction differently: the source uses a varying-coefficient smooth of $X_3$ by numeric $X_1$, while the target uses an ordered-factor-by smooth with the binary main effect retained. Target structure selection is repeated within every inner training fold and on the full training sample. T-GAM, SPT and their matched transfer variants use the same selected target basis and outcome penalty. Pooled GAMs use shared smooth effects including the interaction, with their smoothing parameters estimated by REML. This design evaluates borrowing when the target representation is learned from its training outcomes.

We conduct 100 independent replications per configuration, using the same training and test samples across methods within a replication. The primary tables compare SPT with target-only GAM (T-GAM), source-only GAM (Source GAM), Stratified GAM, CoxTL \citep{Lu2026}, POTL-Cox \citep{Gu2026}, random survival forests (RSF) \citep{Ishwaran2008} and Oracle. Stratified GAM pools both cohorts with shared smooth effects and cohort-specific intercepts. Oracle uses the known target event risk from \eqref{eq:sim_survival}. Further comparisons and implementation details are given in the Supplementary Material, together with experiments using $n_T=200$ or 1000 and 13 additional irrelevant covariates in B and D.

Evaluation uses IPCW Brier score \citep{Graf1999}, time-dependent AUC \citep{Uno2007,Blanche2013} and truncated Uno concordance $C_U(h)$ \citep{Uno2011} on censored test outcomes. Tables report means over 100 replications; Monte Carlo standard errors (MCSEs) accompany the full results in the Supplementary Material, including the paired source-mismatch contrasts. DGP-based risk metrics and paired comparisons are detailed in the Supplementary Material.

\subsection{Predictive performance}
Tables~\ref{tab:loglogistic} and~\ref{tab:weibull} summarize the main simulation results. With training-selected GAM structures, SPT improves mean Brier score, AUC and truncated Uno concordance over T-GAM in all 24 setting--distribution--horizon combinations. Averaging paired relative Brier changes over the three horizons gives reductions of 3.31\%, 1.00\%, 3.07\% and 2.01\% in log-logistic settings A--D, respectively; the corresponding Weibull reductions are 3.18\%, 1.07\%, 3.09\% and 1.90\%. SPT also improves all three mean metrics over the matched prediction-regularized GAM in every combination, demonstrating the contribution of the source-centered smoothing penalty.

Under full agreement in A, SPT approaches the strong performance of Source GAM and Stratified GAM. In B, adaptation to the changed main effects yields the best means among the methods in the primary tables. In C, SPT combines improved probability accuracy with strong discrimination as the interaction attenuates. In D, SPT achieves the best means in both primary tables across all horizons and metrics, with the target interaction selected from the training data. The target selects an interaction in 99--100\% of full-training fits in A, 100\% in B, 94--99\% in C and 38--50\% in D; Supplementary Table~S15 gives the horizon-specific proportions.

In setting A, CoxTL, POTL-Cox and CoxKL improved discrimination across horizons, whereas Brier-score gains emerged at Q75, consistent with the distinction between ranking and horizon-specific probability accuracy under a misspecified linear Cox representation.

Additional experiments isolate the effects of target sample size and irrelevant predictors using prespecified working structures. With $n_T=200$ or 13 irrelevant predictors, SPT improves all three means over T-GAM in each of the 12 tasks. At $n_T=1000$, SPT improves Brier in 8/12 tasks and both discrimination measures in 7/12.

\input{table/main_table_01.tex}


\input{table/main_table_02.tex}


\subsection{Component contributions under source and structural mismatch}\label{sec:source_mismatch}
We examine four forms of mismatch to identify the roles of calibration, adaptive centering and source projection. The first three experiments hold the source and target interaction formulas fixed to isolate changes in the source risk function. Starting from setting A, we retain the target predictor $g_0=a+0.8u$ and change the source predictor to $g_0+\sigma$ (level shift), $g_0+0.8(x_4^2-1)/\sqrt{0.8}-0.8x_5$ (shape change), or $-g_0$ (direction reversal), where $\sigma$ is the corresponding distribution's error scale. Both event-time distributions use 100 replications, with target samples and prediction horizons held fixed across the source changes.

The component comparisons isolate the borrowing mechanisms. Prediction-regularized GAM (PR-GAM) fixes $\rho=0$, retaining the ordinary target penalty and calibrated soft-label loss. It tests the contribution of the source-projected center; fixing $\rho=1$ tests adaptive scaling. The uncalibrated variant uses the original source probabilities as soft labels, and the soft-label-free variant sets $\tau=0$ while retaining the adaptive center. All variants share the source score, target basis, smoothing penalties and internal folds. Except for the soft-label-free variant, they use the same cross-validation grid and selection rule for $\tau$.

\begin{figure}[!htbp]
\centering
\includegraphics[width=\textwidth]{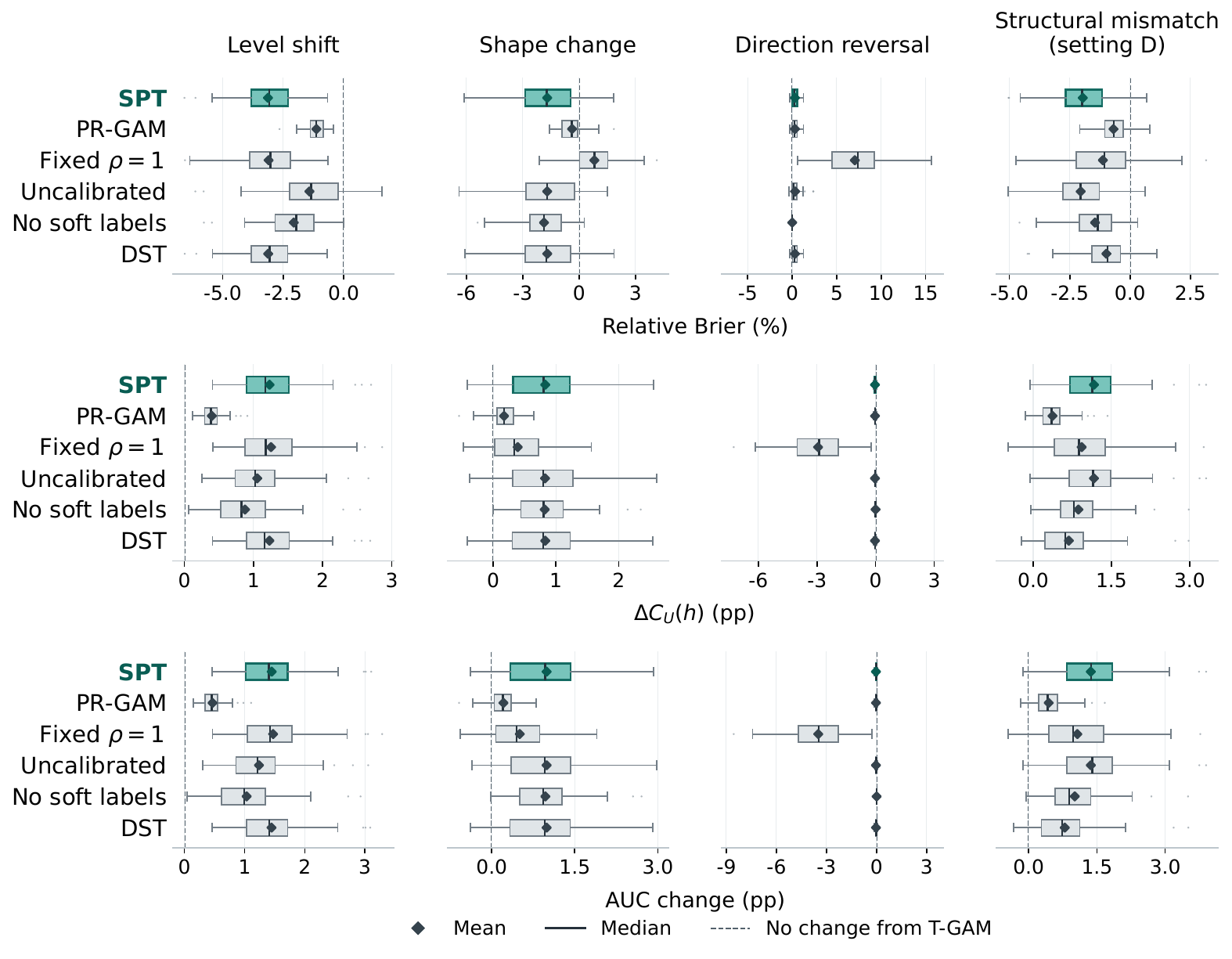}
\caption{Component contributions under source and structural mismatch. Columns show level shift, shape change, direction reversal and structural mismatch (setting D); rows show relative IPCW Brier changes (\%), truncated Uno concordance differences and AUC differences (percentage points), all relative to the matched T-GAM. Each panel compares SPT, PR-GAM, the fixed-center variant, uncalibrated soft labels, no soft labels and direct-score transfer (DST). The first three columns use fixed interaction formulas to isolate changes in source effects; the fourth uses training-only GCV structure selection followed by REML fitting. Each box summarizes 100 replication-level contrasts, averaging both distributions and three horizons within each replication. Boxes span the interquartile range, central lines mark medians, diamonds mark means, whiskers extend to the most extreme values within 1.5 interquartile ranges, and points show observations beyond the whiskers. Dashed lines mark zero change. SPT is highlighted. Horizontal ranges vary across panels. Means and paired MCSEs are given in Supplementary Table~S1.}\label{fig:source_mismatch}
\par\smallskip
\begin{minipage}{\textwidth}\small\raggedright
\textbf{Alt text:} Twelve horizontal boxplots compare six methods across level shift, shape change, direction reversal and structural mismatch. Rows show Brier, concordance and AUC changes relative to T-GAM. Calibration improves performance under level shift, and adaptive centering reduces harm under shape change and reversal. Under structural mismatch, SPT improves all three mean metrics over direct-score transfer. Each box summarizes 100 simulation replications.
\end{minipage}
\end{figure}

Figure~\ref{fig:source_mismatch} displays the distribution of paired changes relative to T-GAM; Supplementary Table~S1 reports the means and paired MCSEs. Under a source level shift, SPT reduces Brier error by 3.13\%, compared with 1.41\% using uncalibrated soft labels and 2.06\% without soft labels. Under shape change, the fixed-center variant increases Brier error by 0.82\%, whereas SPT reduces it by 1.70\%. SPT also improves on PR-GAM under both level and shape changes, supporting the contribution of source-informed smoothing beyond probability regularization. 

Under direction reversal, SPT limits the mean Brier increase to 0.31\%, compared with 7.04\% for the fixed-center variant. Removing soft labels matches T-GAM to the reported precision. The adaptive center accommodates shape mismatch, while calibrated soft labels add value when source probabilities remain informative after level adjustment.

The fourth experiment uses setting D to assess structural mismatch: the source predictor contains an interaction absent from the target generating predictor, and both fitted structures are selected by GCV before REML estimation. We compare SPT with DST as defined in \eqref{eq:dst}, matching the source fit, target basis and penalty, internal folds and $\tau$ selection rule. Comparing \eqref{eq:projected_score_form} with \eqref{eq:dst} isolates the value of replacing the complete source score by its target-space projection.

The fourth column of Figure~\ref{fig:source_mismatch} shows the resulting gains. SPT improves all three mean metrics over DST at every horizon in both distributions. Across the six combinations, Brier decreases by 0.00095--0.00303, AUC increases by 0.435--0.729 percentage points and concordance increases by 0.400--0.598 percentage points. SPT also improves on PR-GAM throughout D. Together, these comparisons show the value of borrowing a source-informed shape within the selected target representation. The calibration benefit is concentrated in the level-shift experiment; calibrated and uncalibrated variants have nearly identical mean performance under shape change and similar performance under structural mismatch (Supplementary Table~S1). Supplementary Table~S16 reports the paired contrasts.

\begin{figure}[htbp]
\centering
\includegraphics[width=\textwidth]{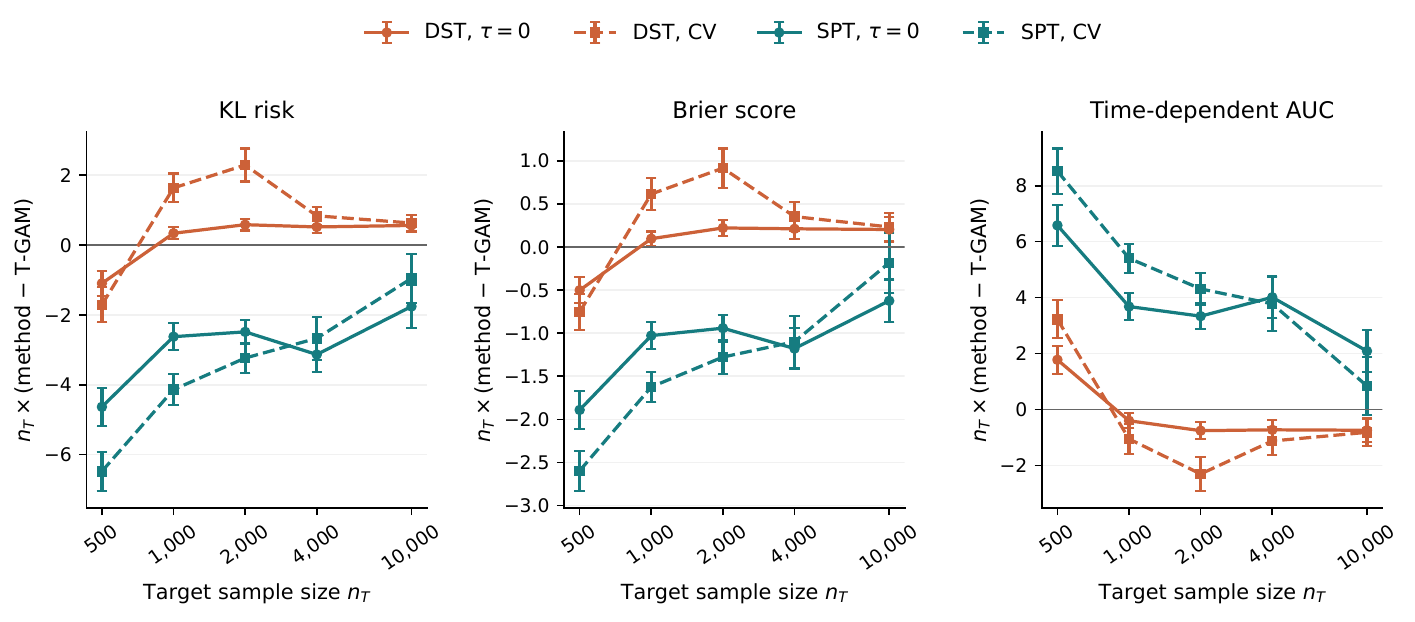}
\caption{Projected and direct-score borrowing across target sample sizes in log-logistic setting D with a fixed additive target representation. Panels show KL risk, IPCW Brier score and time-dependent AUC. Four curves compare DST and SPT with $\tau=0$ or cross-validation (CV), using differences from the matched T-GAM multiplied by $n_T$. Q25, Q50 and Q75 differences are averaged within each replication before calculating means and error bars of $\pm1.96$ paired Monte Carlo standard errors over 100 replications. Negative KL/Brier and positive AUC differences favor transfer; AUC differences use the original 0--1 scale. CV uses the paired one-standard-error rule with reference $\tau=0.5$. The KL panel concerns the theoretical log-risk comparison; Brier and AUC provide complementary predictive assessments.}\label{fig:sample_size_risk}
\par\smallskip
\begin{minipage}{\textwidth}\small\raggedright
\textbf{Alt text:} Three line plots show target-sample-size-scaled differences from T-GAM in KL risk, Brier score and AUC. Four curves represent SPT and direct-score transfer, each with or without cross-validated soft labels. SPT improves the three mean metrics at every sample size. Direct-score transfer has higher KL risk than T-GAM from sample size 1000 onward. Error bars represent Monte Carlo uncertainty.
\end{minipage}
\end{figure}

\paragraph{Projected borrowing across target sample sizes.}
To connect the structural comparison with the fixed-representation analysis in Theorem~\ref{thm:risk}, we retain log-logistic setting D and fit a fixed additive target GAM. The source GAM retains its interaction, with \texttt{ts}/$k=10$ source smooths and \texttt{ts}/$k=5$ target smooths. We vary $n_T$ over $500,1000,2000,4000,10000$, holding $n_S=3000$ and the test sample size at 5000, with 100 replications per sample size. Figure~\ref{fig:sample_size_risk} compares SPT and DST with $\tau=0$ or automatic soft-label selection against the matched T-GAM. KL risk uses the known DGP probabilities; Brier and AUC use IPCW evaluation of observed test outcomes.

SPT improves all three horizon-averaged metrics at every sample size under both rules. Direct-score borrowing initially benefits from the source information, but its KL risk exceeds T-GAM's from $n_T=1000$ onward. This contrast illustrates the value of retaining the projected source shape while excluding the additional direction. The automatic DST curve approaches its $\tau=0$ counterpart as target information increases: at $n_T=10000$, cross-validation selects $\tau=0$ in 99\%, 100\% and 98\% of replications at Q25, Q50 and Q75. These patterns support the risk distinction and adaptive rejection mechanisms described by the theoretical analysis. Supplementary Figures~S1 and S2 examine the fitted penalty magnitude and plug-in risk interpolation.

\clearpage


\section{Mortality prediction in a reduced-kidney-function population}\label{sec:realdata}
\subsection{Clinical question and cohorts}
We apply SPT to predict all-cause mortality in SPRINT standard-treatment participants with reduced kidney function, using UK Biobank as the source cohort. SPRINT compared intensive with standard blood-pressure treatment in adults with hypertension and increased cardiovascular risk, including a prespecified chronic-kidney-disease subgroup \citep{SPRINT2015}. Our clinical question concerns prognosis under standard treatment: whether information from a larger external cohort can improve identification of participants at higher mortality risk over the next two to three years. Such predictions could support follow-up prioritization and prognostic discussions after external validation. The analysis estimates mortality risk within the standard-treatment population.

UK Biobank provides a larger population-based source for learning nonlinear associations between routine baseline measurements and mortality \citep{Sudlow2015}. We apply a common kidney-function criterion to both cohorts: baseline eGFR from 20 to less than 60 mL/min/1.73 m$^2$, calculated using the race-free CKD-EPI 2021 creatinine equation \citep{Inker2021}. Eligibility is based on a single baseline measurement. After outcome, follow-up and complete-case filtering, the analysis includes 7059 source participants and 1126 target participants. Follow-up begins at UK Biobank baseline assessment or SPRINT randomization and ends at death or censoring. Detailed eligibility and harmonization procedures are given in the Supplementary Material.

The cohorts differ in clinically relevant baseline characteristics despite the common eGFR criterion. Mean age is 62.7 years in UK Biobank and 72.5 years in the SPRINT target cohort; mean eGFR is 51.5 and 47.2 mL/min/1.73 m$^2$, respectively. Men comprise 46.7\% and 58.7\% of the two cohorts.

\subsection{Prediction models and evaluation}
We use seven harmonized baseline predictors: smoking history, age, body mass index, eGFR, total cholesterol, creatinine and sex. Source and target GAMs include additive effects without interactions. The primary GAM-source implementation, denoted SPT-GAM in the clinical displays, uses shrinkage thin-plate smooths with source basis size 10 and target basis size 5. The soft-label weight is selected from $\{0,0.1,0.5,1\}$ by three-fold target cross-validation using the one-standard-error rule with reference $\tau=0.5$. Projection, calibration and target fitting are repeated within the relevant training samples. For source-model sensitivity analyses, we evaluate SPT-Cox and SPT-NN, which supply source predictions from a Cox model and an IPCW-trained neural classifier, respectively. Both retain the same target GAM and transfer-weight selection. The source Cox fit supplies event probabilities at all three horizons; the neural classifier is tuned using source-only cross-validation and fitted separately at each horizon. Source-learner specifications are given in the Supplementary Material.

Predictions are evaluated at 24, 30 and 36 months. The target cohort has 39, 53 and 67 observed deaths by these horizons, respectively, with 1052, 965 and 710 observed event-free controls. The corresponding estimated censoring survival probabilities are 0.968, 0.900 and 0.674. The event counts and increasing censoring motivate borrowing external information for horizon-specific mortality prediction.

We use 100 repeated five-fold target partitions shared by all methods. Each participant is evaluated in a held-out fold, and predictions are pooled across folds within each repetition. The source GAM is fitted using the external cohort at each horizon; target-dependent tuning uses only the outer-training participants. Performance is summarized by IPCW Brier score, time-dependent AUC and truncated Uno concordance. Figure~\ref{fig:clinical_boxplots} compares the three SPT implementations with T-GAM, Source GAM, Stratified GAM, intercept-plus-slope score recalibration, CoxTL and CoxKL. These comparisons assess direct use of source predictions, pooling, model updating and adaptive transfer within the same evaluation design. Additional survival-transfer and machine-learning results are provided in the supplement.

\subsection{Prediction results}
\begin{figure}[p]
\centering
\includegraphics[width=0.92\textwidth]{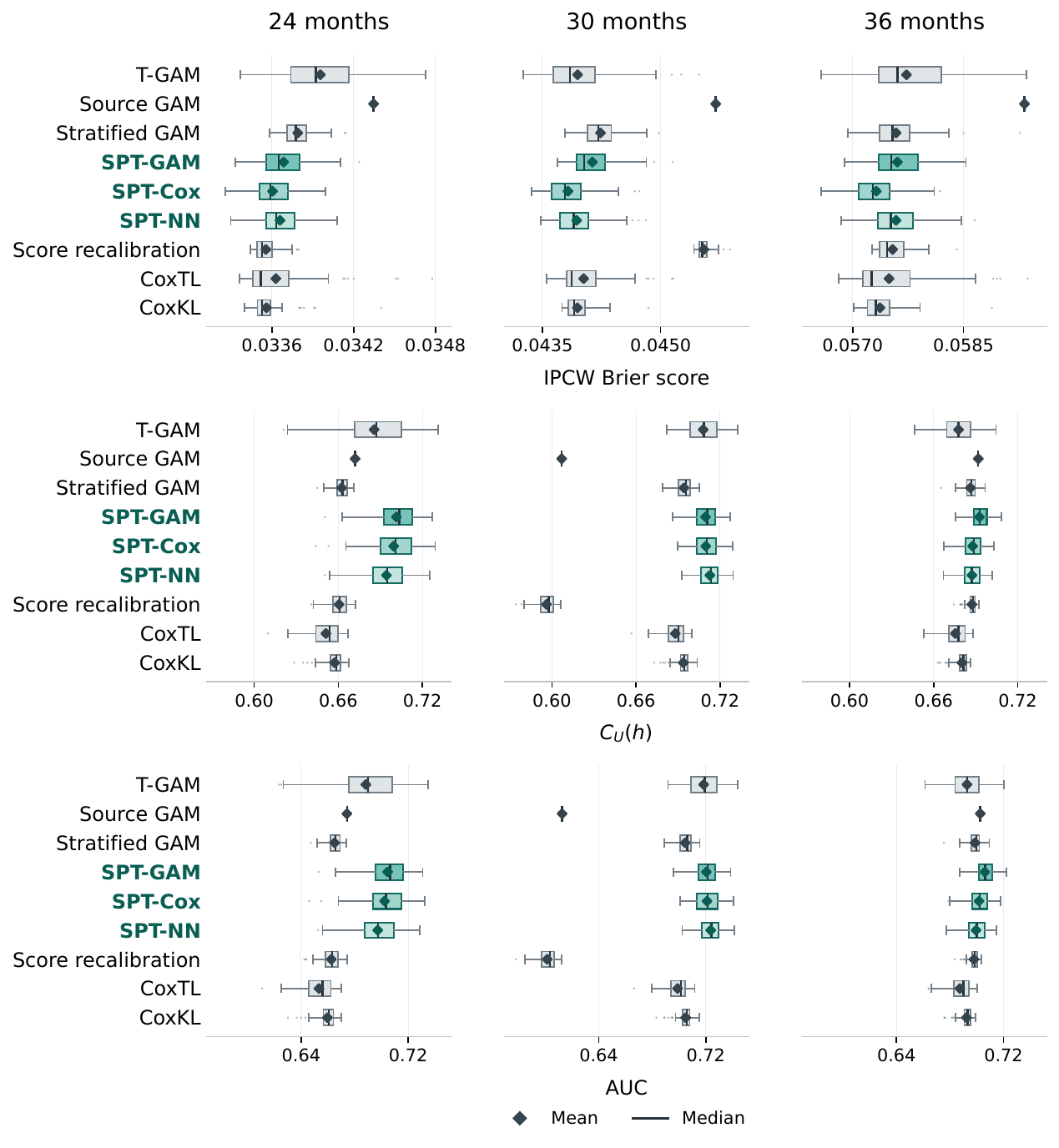}
\caption{Mortality prediction in SPRINT standard-treatment participants using UK Biobank as the source cohort. Columns show 24, 30 and 36 months; rows show IPCW Brier score, horizon-truncated Uno concordance and time-dependent AUC. Each box summarizes 100 repeated five-fold evaluations, with a pooled out-of-fold metric calculated within each repetition. Boxes span the interquartile range, lines mark medians, diamonds mark means, whiskers extend to the most extreme values within 1.5 interquartile ranges, and points show observations beyond the whiskers. The SPT variants are highlighted: SPT-GAM, SPT-Cox and SPT-NN use GAM, Cox and neural-network source predictions, respectively, while retaining the same target GAM. Source GAM predictions are fixed across target partitions. The distributions describe training-partition variability within the same cohort. Brier axes vary across horizons; concordance and AUC use common ranges within their rows. Means and partition SDs for these and additional comparators are reported in Supplementary Table~S3.}\label{fig:clinical_boxplots}
\par\smallskip
\begin{minipage}{\textwidth}\small\raggedright
\textbf{Alt text:} Nine horizontal boxplot panels compare mortality prediction methods at 24, 30 and 36 months. Rows show Brier score, truncated concordance and AUC. SPT-GAM improves mean discrimination over T-GAM at all three horizons; Brier scores improve at 24 and 36 months and increase slightly at 30 months. Cox and neural-network source variants are also shown. Boxes describe variation across repeated training partitions.
\end{minipage}
\end{figure}

SPT-GAM, the primary implementation, improves mean AUC and truncated Uno concordance over T-GAM at all three horizons (Figure~\ref{fig:clinical_boxplots}). At 24, 30 and 36 months, AUC increases from 0.6884, 0.7184 and 0.6928 to 0.7047, 0.7204 and 0.7061, respectively. The corresponding concordance values increase from 0.6855, 0.7080 and 0.6777 to 0.7014, 0.7097 and 0.6927.

Mean Brier scores for SPT-GAM versus T-GAM are 0.03369 versus 0.03396 at 24 months, 0.04413 versus 0.04395 at 30 months, and 0.05761 versus 0.05773 at 36 months. Thus SPT-GAM improves Brier score at 24 and 36 months, while the 30-month score increases by approximately 0.4\%. At 30 months, the gains in risk ranking are accompanied by a small increase in squared probability error. Calibration slopes provide a complementary summary: SPT-GAM's mean held-out slopes are 0.731, 0.945 and 0.823, compared with 0.636, 0.813 and 0.807 for T-GAM.

The direct-source and pooling comparisons illustrate the role of target adaptation. Source GAM has lower AUC than T-GAM at 24 and 30 months, but higher AUC at 36 months. Stratified GAM improves on direct source prediction at 30 months but remains below T-GAM in discrimination. SPT-GAM improves discrimination over T-GAM at all three horizons.

As source-model sensitivity analyses, SPT-Cox and SPT-NN replace the source learner while retaining the same target GAM and selection rule. Both improve mean AUC and truncated Uno concordance over T-GAM at each horizon. SPT-Cox has lower mean Brier scores than SPT-GAM at all three horizons; discrimination favors SPT-GAM at 24 and 36 months and SPT-NN at 30 months. The full horizon-specific results for each source learner are reported in Supplementary Table~S3.

\clearpage


\section{Discussion}\label{sec:discussion}
The experiments show why source effect shapes and source probabilities merit separate adaptation. Under a source level shift, SPT reduces Brier error relative to T-GAM by 3.13\%, compared with 1.41\% using uncalibrated soft labels and 2.06\% without soft labels. Under shape change, estimating the penalty center turns a 0.82\% Brier increase with a fixed center into a 1.70\% reduction; removing soft labels yields a slightly larger reduction of 1.87\%, showing that useful source shapes need not justify equally strong probability borrowing. Projection matters when source structure does not transfer: in setting D, SPT reduces Brier error by 1.96\%, compared with 0.98\% for direct-score transfer and 0.69\% for probability regularization alone. Under direction reversal, adaptation limits the mean Brier increase to 0.31\%, compared with 7.04\% for the fixed-center variant. These matched comparisons identify distinct benefits from calibration, adaptive centering and restricting borrowed information to the target representation (Figure~\ref{fig:source_mismatch}).

SPT is most useful when a flexible source model contains informative covariate relationships but the target population calls for a simpler model or different effect shapes. When the two populations agree closely, direct source prediction and pooling are strong alternatives, as setting A illustrates. When the main difference is overall risk level, intercept recalibration offers a simple starting point; SPT adds target-specific smooth corrections when changing the level alone is insufficient. When a source interaction has little target relevance, projection allows the target model to borrow the representable part of the source score, as demonstrated in D. With ample target information, a target-only fit becomes increasingly competitive. In practice, SPT should be assessed alongside T-GAM, source prediction, pooling, simple recalibration and probability regularization using the intended prediction horizon and loss. The clinical analysis retains SPT-GAM as the primary implementation and treats Cox and neural-network sources as sensitivity analyses; source-learner choice should follow target validation rather than the best observed result at each horizon.

Under a correctly specified fixed target representation and the stated local smoothing conditions, the risk comparison explains the benefit of retaining the projected source shape while avoiding the positive cost of estimating an uninformative additional direction; the rejection theorem describes how fixed-grid validation and adaptive centering discard persistently adverse information. The fitted penalties remain substantial relative to $\sqrt{n_T}$ over the sample sizes studied and decrease with $n_T$ (Supplementary Figure~S1), making finite-sample smoothing relevant to interpreting the centering gains in Figure~\ref{fig:sample_size_risk}. The asymptotic formulas explain the risk mechanisms, while the plug-in discrepancies in Supplementary Figure~S2 motivate a finite-penalty analysis that retains the fitted curvature and accounts for training-selected representations.

Several choices delimit the present evidence. SPT holds the T-GAM penalty matrix $\Omega$ fixed within each training split, and its cross-validation rule favors $\tau=0.5$ when the loss difference falls within one paired standard error; jointly estimating smoothing and borrowing, or changing this reference, may alter the fitted model. Marginal censoring weights assume independent censoring, and separate horizon fits do not enforce a monotone survival curve. The clinical study uses one target cohort, with reduced kidney function defined by a single baseline measurement. Some target ages exceed the source age range. Source recalibration also does not ensure calibration of the final predictor: SPT-GAM improves clinical discrimination at all three horizons, while its 30-month Brier score increases by approximately 0.4\%. Repeated partitions measure training stability within the observed cohort; independent populations are needed to assess transportability.

A natural extension is transfer from several source models. If $c_j$ denotes the projection of source $j$, the penalty center could be a weighted combination $c(\boldsymbol\rho)=\sum_{j=1}^J\rho_jc_j$, with $\rho_j\geq0$ and $\sum_j\rho_j\leq1$. This permits selective borrowing across sources while retaining zero as an admissible center, and replaces scalar profiling with a small constrained quadratic problem. Separately calibrated source probabilities could be combined through target validation, allowing different sources to contribute to smooth shapes and predicted risks. Further work could incorporate covariate-dependent censoring weights, link horizon-specific fits to enforce temporal coherence, and extend the risk analysis to selected bases and jointly estimated smoothing parameters. These extensions would broaden SPT's use while preserving its central aim: learning useful structure from flexible external models within a target model suited to the available data.


\bibliographystyle{plainnat}\bibliography{reference}
\end{document}


\maketitle
\begin{abstract}
This supplement provides proofs of the exact center-rejection condition, local expected log-risk comparisons and asymptotic rejection of mismatched borrowing channels, together with implementation details, data-generating mechanisms, comparator tuning and clinical cohort construction. Additional comparator, sample-size and irrelevant-variable experiments assess predictive performance and robustness using observed-outcome and DGP-based evaluation. SPT is the calibrated, profiled-center estimator with no residual predictor and the one-standard-error rule with reference weight 0.5. Matched component ablations are evaluated in simulation.
\end{abstract}

\section{Estimation algorithm and numerical specification}
\subsection{Target representation and penalty matrices}
For an additive target model, collect the intercept, $d_0$ unpenalized parametric terms $z$ and centered smooth bases in
\begin{equation}
 B(x)=\{1,z^\mathsf T,b_1(x_1)^\mathsf T,\ldots,b_J(x_J)^\mathsf T\}^\mathsf T,
 \qquad d=1+d_0+\sum_jd_j,
 \label{eq:basis_vector}
\end{equation}
where $b_j\in\mathbb R^{d_j}$. Shrinkage thin-plate bases (\texttt{bs="ts"}) use $k_T=5$, giving $d_j=4$ after centering. The clinical target has two binary and five continuous predictors, hence $d=23$; interactions append basis blocks. Standardization and basis construction use training covariates and are retained for prediction.

For $f_j(u)=b_j(u)^\mathsf T\theta_j$, the ordinary roughness penalty is \citep{Wood2003}
\begin{equation}
 \int\{f_j''(u)\}^2du=\theta_j^\mathsf TR_j\theta_j,
 \qquad (R_j)_{ab}=\int b_{ja}''(u)b_{jb}''(u)du.
 \label{eq:roughness_matrix}
\end{equation}
The shrinkage construction replaces zero penalty eigenvalues by small positive values, allowing whole smooths to shrink toward zero \citep{Wood2017}. Denote the centered, scaled local penalty by $\widetilde R_j\in\mathbb R^{d_j\times d_j}$. If $E_j\in\mathbb R^{d\times d_j}$ inserts term $j$ into the full coefficient vector, then
\begin{equation}
 \Omega_j=E_j\widetilde R_jE_j^\mathsf T,
 \qquad \Omega=\sum_j\widehat\lambda_j\Omega_j,
 \qquad \tfrac12\theta^\mathsf T\Omega\theta
       =\tfrac12\sum_j\widehat\lambda_j\theta_j^\mathsf T\widetilde R_j\theta_j.
 \label{eq:gam_penalty}
\end{equation}
The intercept and parametric blocks are zero. T-GAM estimates the smoothing strengths by REML using an IPCW-weighted quasibinomial logit fit with \texttt{gamma=1.4} \citep{Wood2011}. In \texttt{mgcv}, \texttt{lpmatrix} supplies $B$; \texttt{first.para}/\texttt{last.para} locate the local penalties, and \texttt{fit\$sp} supplies their multipliers.

\subsection{Source-model preparation}
SPT accepts a fitted source predictor evaluated on harmonized target covariates; source fitting is a separate preparation step that can be performed by the source-data holder. We fitted the source models ourselves in the simulations and UK Biobank analysis. The SPT-GAM implementation uses $\eta_S(x)=B_S(x)^\mathsf T\widehat\gamma_S$, the same fitting convention as T-GAM, and $k_S=10$ (nine coefficients per centered smooth; clinical source dimension 48). Source standardization uses source training data. In the main simulation, source-only GCV chooses the additive or interaction formula before REML fitting. Cox and neural-network source specifications appear in Section~\ref{sec:clinical_supp}.

\subsection{SPT algorithm and soft-label weight selection}
At each horizon, keep the fitted source model fixed and proceed using target training data:
\begin{enumerate}
\item Form three internal folds. Within each training fold, estimate censoring survival and fit T-GAM to obtain $B$ and $\Omega$. In the main simulation, select the interaction by training-only GCV before REML fitting.
\item Project source scores onto $B$ by Gaussian REML, obtaining $c$ and the separate projection penalty $P$; estimate the source calibration intercept on the same training fold. For each $\tau\in\mathcal T=\{0,0.1,0.5,1\}$, fit the profiled SPT objective with this $\Omega$ and evaluate held-out IPCW log loss.
\item Combine the out-of-fold losses $L_i(t)$ over all $n=n_T$ training participants. Let $t_*$ minimize their mean, breaking ties toward smaller weights. Define $d_i=L_i(0.5)-L_i(t_*)$, $\bar d=n^{-1}\sum_i d_i$ and $s_d=\{\sum_i(d_i-\bar d)^2/(n-1)\}^{1/2}$, and use the paired one-standard-error rule
\begin{equation}
 \widehat\tau=\begin{cases}
 0.5,&\bar d\leq s_d/\sqrt n,\\
 t_*,&\text{otherwise}.
 \end{cases}\label{eq:tau}
\end{equation}
\item Re-estimate T-GAM, projection and calibration on the full target training sample, fit the selected candidate, and predict the outer test sample.
\end{enumerate}

\subsection{Numerical details}\label{sec:numerical_details}
Censoring survival is floored at $10^{-3}$; fitting IPCW weights are capped at 20, whereas validation and simulation evaluation weights are uncapped. Probabilities and soft labels are clipped to $[10^{-3},1-10^{-3}]$. The reported routines evaluate the right-continuous censoring KM at $\min(Y_i,h)$; the theoretical event weight uses its left limit, differing only at an event time coinciding with a censoring-KM jump.

We profile $\rho$ analytically and optimize target coefficients by preconditioned L-BFGS-B, checking convergence and projected gradients at tolerance $2\times10^{-6}$. If the original boundary GAM fit fails, its working scale is fixed at one for the fallback fit. Further implementation details are provided in the code.

\section{Theoretical properties and proofs}
This section proves the exact center-rejection condition and the two risk results. The first theorem isolates projected-center borrowing at $\tau=0$; the second addresses automatic soft-label selection, center rejection and the full-sample refit.
Background on logistic-risk expansions and constrained $M$-estimation is provided by \citet{Bach2010} and \citet{Geyer1994}, respectively. The arguments below establish the results for the matched SPT and DST objectives under the stated conditions.
\subsection{Conditional setup, estimators and notation}

We condition on an independently fitted source model and a fixed target-basis construction. The source score $\eta_S$ is therefore a given function, and the probability statements below concern target samples with the target design treated as fixed.
Write $b(x)=B(x)\in\R^d$ for a fixed, identifiable target design vector with an intercept, $b(x)^\top e_0=1$. At a fixed horizon, write
\[
 p_\theta(x)=\expit\{b(x)^\top\theta\},\qquad
 \pst(x)=p_{\thst}(x),\qquad \expit(z)=(1+e^{-z})^{-1}.
\]
The observed horizon label is $y_i$ and the fitted IPCW weight is $\widehat w_i$. Define the average weighted outcome loss and the normalized smoothing matrix by
\[
 \ell_n(\eta)=\frac1n\sum_i\widehat w_i\CE\{y_i,\expit(\eta(X_i))\},
 \qquad \Lambda_n=\Omega_n/n.
\]
$\Omega_n$ is the effective penalty in the target-only \emph{sum-loss} criterion, including any working-scale convention. It may be selected by REML on the same outcomes and is inherited unchanged by the matched estimators.

For the fitted source logit $\eta_S$, define
\[
 c_n=(B^\top B+P_n)^{-1}B^\top\eta_S,\qquad
 r_n(x)=\eta_S(x)-b(x)^\top c_n.
\]
$P_n$ is the Gaussian projection penalty; $\Omega_n$ is the target outcome penalty. For example, with bounded $b$ and $\eta_S$, $\E[bb^\top]\succ0$ and $P_n/n\to_p\Pi\succeq0$, the law of large numbers gives
\[
 c_n\longrightarrow_p c=(\E[bb^\top]+\Pi)^{-1}\E[b\eta_S],
 \qquad r=\eta_S-b^\top c.
\]
The subsequent conditions allow any projection sequence with the stated convergence. The center-only comparisons use $\tau=0$:
\begin{align}
 \widehat\theta_\tg
 &=\argmin_\theta\{\ell_n(b^\top\theta)+\tfrac12\theta^\top\Lambda_n\theta\},
 \label{eq:tg}\\
 (\widehat\theta_{\spt,0},\widehat\rho_0)
 &\in\argmin_{\theta,\,0\le\rho\le1}
 \{\ell_n(b^\top\theta)+\tfrac12(\theta-\rho c_n)^\top\Lambda_n(\theta-\rho c_n)\},
 \label{eq:spt0}\\
 (\widehat\theta_\dst,\widehat v)
 &\in\argmin_{\theta,\,0\le v\le1}
 \{\ell_n(b^\top\theta+v r_n)+\tfrac12(\theta-vc_n)^\top\Lambda_n(\theta-vc_n)\}.
 \label{eq:dst}
\end{align}
Write $\widehat\eta_\tg=b^\top\widehat\theta_\tg$ and $\widehat\eta_\dst=b^\top\widehat\theta_\dst+\widehat v r_n$. The third estimator is direct-score transfer (DST). Substituting $\theta=\beta+vc_n$ gives its exact alternative form
\[
 \eta_\dst=b^\top\beta+v\eta_S,\qquad \text{penalty}=\tfrac12\beta^\top\Lambda_n\beta.
\]
For SPT, the substitution $\beta=\theta-\rho c_n$ instead gives
\begin{equation}
 \eta_\spt=b^\top\beta+\rho b^\top c_n,\qquad
 \tfrac12(\theta-\rho c_n)^\top\Lambda_n(\theta-\rho c_n)
 =\tfrac12\beta^\top\Lambda_n\beta.
 \label{eq:spt-equivalent}
\end{equation}
Thus SPT borrows the projected source shape and estimates a smooth target correction.

For a candidate $\tau>0$, SPT adds $\tau n^{-1}\sum_i\CE(q_i,p_\theta(X_i))$ to \eqref{eq:spt0}. The actual calibrated, clipped source probability $q_i$ is considered in Theorem~\ref{thm:safety}; it is absent from Theorem~\ref{thm:risk}.

For a fitted linear predictor $\eta$, define the target excess log risk
\begin{equation}
 \mathcal K(\eta)=\E_X\operatorname{KL}\{\operatorname{Bern}(p^\star(X))\,\Vert\,
                  \operatorname{Bern}(\expit\{\eta(X)\})\}.
 \label{eq:krisk}
\end{equation}
An outer $\E$ averages over target training samples conditional on the fitted source model and the fixed target basis. When the argument is a coefficient vector, its predictor is $b^\top\theta$. With $H=\E[p^\star(1-p^\star)bb^\top]$, the local expansion is
\begin{equation}
 \mathcal K(\theta^\star+\Delta)=\tfrac12\Delta^\top H\Delta+o(\|\Delta\|^2).
 \label{eq:risk-taylor}
\end{equation}

\paragraph{Notation.}
\begin{center}
\small
\begin{tabular}{@{}lp{0.69\linewidth}@{}}
\toprule
Symbol & Meaning\\
\midrule
$\Omega_n,\ P_n$ & Outcome and Gaussian projection penalty matrices.\\
$\Lambda_n,\ A_n$ & $\Omega_n/n$ and $\Omega_n/\sqrt n$, respectively.\\
$\mathbb P_n f$ & Empirical average $n^{-1}\sum_{i=1}^n f(O_i)$.\\
$\mathcal K(\eta),\ \ell(y,p)$ & Target excess log risk and Bernoulli cross-entropy loss.\\
$q_i,\ q_\infty$ & Calibrated soft labels and their limiting function.\\
$H,\ \mathbf h_r,\ h_{rr}$ & Target curvature, basis--residual cross-curvature and residual curvature.\\
$s,\ \nu$ & Residual Schur complement and efficient-score variance.\\
$\rho,\ v$ & Projected-center coefficient and direct-score coefficient.\\
$G_K,\ D(k)$ & Centering gain and extra-direction risk cost.\\
\bottomrule
\end{tabular}
\end{center}

\subsection{Profile penalty and exact center rejection}
\begin{lemma}[Profile penalty]\label{lem:profile}
For $\Lambda\succeq0$ and fixed $c$,
\[
 J_c(\theta)=\tfrac12\min_{0\le\rho\le1}\norm{\theta-\rho c}_\Lambda^2
\]
is convex and continuously differentiable, with
\[
 \nabla J_c(\theta)=\Lambda\{\theta-\rho(\theta)c\},\quad
 \rho(\theta)=\Pi_{[0,1]}\frac{c^\top\Lambda\theta}{c^\top\Lambda c},
\]
and gradient Lipschitz constant at most $\norm{\Lambda}_{\rm op}$. Set $\rho=0$ when $\Lambda c=0$.
\end{lemma}
\begin{proof}
The profile penalty is $\tfrac12\operatorname{dist}^2(\Lambda^{1/2}\theta,[0,\Lambda^{1/2}c])$. The residual of Euclidean projection onto a closed convex set is nonexpansive. Composition with $\Lambda^{1/2}$ proves the gradient formula and Lipschitz bound, including at endpoint switches.
\end{proof}

\begin{remark}[One additional direction at fixed curvature]
On an interior profile region with $c_n^\top\Lambda_nc_n>0$, minimizing over $\rho$ replaces $\Lambda_n$ by
\[
 \Lambda_{n,c}=\Lambda_n-
 \frac{\Lambda_nc_nc_n^\top\Lambda_n}{c_n^\top\Lambda_nc_n}.
\]
For the same fixed loss curvature $H_0\succ0$, define
\[
 F_0=H_0^{1/2}(H_0+\Lambda_n)^{-1}H_0^{1/2},\qquad
 F_c=H_0^{1/2}(H_0+\Lambda_{n,c})^{-1}H_0^{1/2}.
\]
Then $F_c-F_0\succeq0$ has rank at most one and
\[
 0\le\tr(F_c)-\tr(F_0)\le1.
\]
Profiling introduces at most one additional effective direction for this fixed-curvature smoother.
\end{remark}
\begin{proof}
Cauchy--Schwarz in the $\Lambda_n$ seminorm gives
$0\preceq\Lambda_{n,c}\preceq\Lambda_n$. Inversion reverses the positive-definite order, so $0\preceq F_0\preceq F_c\preceq I$. The resolvent identity gives $\operatorname{rank}(F_c-F_0)\le1$. Its sole possible nonzero eigenvalue lies in $[0,1]$, proving the trace bound. One also has
$0\le\tr(F_c^2)-\tr(F_0^2)\le1$: decreasingly ordered eigenvalues satisfy
$\lambda_i(F_c)\ge\lambda_i(F_0)\ge\lambda_{i+1}(F_c)$ for $i<d$ by rank-one interlacing. Squaring these nonnegative eigenvalues and telescoping bounds the difference by $\lambda_1(F_c)^2-\lambda_d(F_0)^2\le1$.
\end{proof}

For SPT at fixed $\tau$, the loss $F$ in the following proposition is
\[
 F(\theta)=\sum_iw_i\ell(y_i,p_i)+\tau\sum_i\ell(q_i,p_i),
 \qquad p_i=\expit(B_i^\mathsf T\theta).
\]
It includes the target outcome and calibrated soft-label losses and excludes the quadratic smoothness penalty; it is denoted $F_\tau$ in the main text.

\begin{proposition}[Exact rejection of the penalty center]\label{prop:direction}
Let $F(\theta)$ be a differentiable, strictly convex loss with a finite minimizer after adding $\theta^\top \Omega\theta/2$, where $\Omega\succeq0$. Write that zero-centered minimizer as $\widehat\theta_{\rm PR}$, the prediction-regularized fit with $\rho=0$. If $c^\top \Omega c>0$, the joint minimizer of
\[
 F(\theta)+\tfrac12(\theta-\rho c)^\top \Omega(\theta-\rho c),\quad 0\le\rho\le1
\]
is unique and satisfies
\[
 \widehat\rho=0\quad\Longleftrightarrow\quad c^\top \Omega\widehat\theta_{\rm PR}\le0.
\]
On this event $\widehat\theta=\widehat\theta_{\rm PR}$. If $c^\top\Omega c=0$, use the convention $\widehat\rho=0$.
\end{proposition}
\begin{proof}
At $(\widehat\theta_{\rm PR},0)$ the coefficient gradient is zero and the inward derivative in $\rho$ is $-c^\top \Omega\widehat\theta_{\rm PR}$. The KKT conditions for this convex problem are therefore equivalent to the displayed inequality. Strict convexity of $F$ makes the fitted coefficient vector unique; $c^\top \Omega c>0$ makes the center coefficient unique for that vector. Conversely, any optimum with $\rho=0$ must use the unique zero-centered coefficient and satisfy the same KKT condition.
\end{proof}
At $\tau=0$, the zero-centered fit is T-GAM; at positive $\tau$, it is the matched prediction-regularized fit.

\subsection{Regularity for actual risk comparisons}

Conditions C1--C4 concern the statistical minimizers in \eqref{eq:tg}--\eqref{eq:dst}.
\begin{enumerate}\renewcommand{\labelenumi}{C\arabic{enumi}.}
\item \textbf{Correct target and identifiable design.}
The observations are iid; $d$ is fixed; $b$ is bounded; $\E(bb^\top)\succ0$; and
$\varepsilon+\zeta\le\pst\le1-\varepsilon-\zeta$ for some $\zeta>0$. The target coefficient space is $\R^d$. For the DST comparison, $r_n\to r$ in a bounded-function regime supporting the empirical limits below, and $\widetilde b=(b^\top,r)^\top$ has positive-definite population Gram matrix. The limiting residual is fixed and does not vanish with $n$.
\item \textbf{Actual weighted loss and scores.}
The fitted weighted outcome loss, its local Hessian, and the directional functions in Lemma~\ref{lem:localization} satisfy uniform laws with the limits of correct target weighting. In particular, $\Pn\widehat w\to_p1$ and, using the actual fitted weights,
\[
 Z_{b,n}=\frac1{\sqrt n}\sum_i\widehat w_i b_i(y_i-\pst_i)
       =\frac1{\sqrt n}\sum_i\psi_b(O_i)+o_{L^2}(1),
 \quad \E\psi_b=0,\quad \E\psi_b\psi_b^\top=V.
\]
For DST the augmented score
\[
 \widetilde Z_n=(Z_{b,n}^\top,Z_{r,n})^\top,\qquad
 Z_{r,n}=\frac1{\sqrt n}\sum_i\widehat w_i r_n(X_i)(y_i-\pst_i),
\]
has a corresponding mean-zero asymptotically linear expansion with covariance $\widetilde V$; the local Hessian converges to
$\widetilde H=\E[\pst(1-\pst)\widetilde b\widetilde b^\top]$.
The influence functions include estimation of the censoring weights. Projection-score stability follows from the identity below; the displayed expansions and C4 provide the moment control used in the expected-risk calculations.
\item \textbf{Common smoothing and center.}
$\Omega_n\succeq0$, $\Omega_ne_0=0$, $A_n=\Omega_n/\sqrt n=O_p(1)$, and $c_n\to_p c$. The center comparison in Theorem~\ref{thm:risk}(ii) additionally assumes $A_n\to_p A$ for a deterministic $A\succeq0$. The DST comparison does not require this additional convergence or independence of $A_n$ and the scores.
\item \textbf{Uniform integrability.}
For the estimators being compared, their root-$n$ coefficient errors (including $\sqrt n\widehat v$ for DST) have uniformly bounded $2+\kappa$ moments for some $\kappa>0$. For DST, require the same moment control for $\sqrt n\|\widehat\eta_\dst-b^\top\thst\|_\infty$; bounded augmented basis functions and joint coefficient moment control are a sufficient route. The same bound holds for the score vectors and the linear penalty-score quantities appearing in their expansions, including $A_n\thst$ and $c_n^\top A_n\thst$. This ensures uniform integrability of the quadratic approximations and negligible expected scaled-risk remainders. Any measurable fallback on exceptional nonexistence or numerical-failure events is required to satisfy the same moment condition.
\end{enumerate}

The score and smoothing conditions apply to the actual fitted weights and realized smoothing matrices. The weighting interpretation is specified at the end of this section.

\paragraph{Projection-score stability.}
Conditional on the fitted source model, $r_n-r=-b^\top(c_n-c)$. Hence
\[
 \frac1{\sqrt n}\sum_i\widehat w_i\{r_n(X_i)-r(X_i)\}(y_i-\pst_i)
 =-(c_n-c)^\top Z_{b,n}=o_p(1),
\]
using $c_n\to_p c$ and $Z_{b,n}=O_p(1)$. Thus estimating the projection leaves the limiting residual score unchanged, including when the projection and outcome scores use the same target sample.

\begin{lemma}[Localization and consistency]\label{lem:localization}
Under the correct-weight uniform laws in C1--C2, $\Omega_m/m\to_p0$, $c_m=O_p(1)$, and uniform convergence of any bounded soft-label function, finite-grid SPT fits are localized in a fixed ball with probability tending to one and converge to the unique minimizers of their limiting objectives. The soft-label convergence condition is unnecessary for target-only and $\tau=0$ fits. For DST, a bounded residual and positive-definite augmented design give consistency for $(\thst,0)$.
\end{lemma}
\begin{proof}
For a unit vector $u$ and $t\ge0$, logistic loss obeys
\[
 \ell_m(tb^\top u)\ge t\,\mathbb P_m\widehat w
 \{y(-b^\top u)_++(1-y)(b^\top u)_+\}.
\]
The infimum of the population right-hand coefficient over unit $u$ is positive: correct weighting and overlap bound it below by
$\zeta\inf_{\norm u=1}\E|b^\top u|>0$. The directional uniform law makes this bound uniform empirically. Comparing the full nonnegative criterion with its value at $(0,0)$ therefore localizes the fits. A bounded residual offset $vr_m$, $v\in[0,1]$, only changes this coercivity bound by a bounded additive term. On compact sets the normalized penalty vanishes. Uniform convergence and strict convexity identify the limiting coefficient; positive-definite augmented design identifies $(\thst,0)$ for DST. Positive local curvature, a root-$n$ score, and C3 then give root-$n$ localization for the comparisons below.
\end{proof}

\subsection{Local decomposition for direct-score transfer}
The DST objective in \eqref{eq:dst} uses the exact reparameterization $\theta=\beta+vc_n$ for any projection coefficient $c_n$. If $r_n$ is identically zero on the prediction domain, DST and SPT coincide when their loss, penalty matrix and coefficient constraints agree.

For the augmented design $\widetilde b=(b^\top,r)^\top$, write
\[
 \widetilde H=
 \begin{pmatrix}H&\mathbf h_r\\\mathbf h_r^\top&h_{rr}\end{pmatrix},
 \qquad
 \mathbf h_r=\E[\pst(1-\pst)br],\qquad
 s=h_{rr}-\mathbf h_r^\top H^{-1}\mathbf h_r>0.
\]
Here $h_{rr}=\E[\pst(1-\pst)r^2]$. Target overlap and
$r\notin\operatorname{span}(b)$, interpreted in $L^2(P_X)$, ensure
the positive Schur complement. Define
\[
 Z_{{\rm eff},n}=Z_{r,n}-\mathbf h_r^\top H^{-1}Z_{b,n},\qquad
 k_n=(c_n+H^{-1}\mathbf h_r)^\top A_n\thst.
\]
C2 implies $Z_{{\rm eff},n}\Rightarrow W\sim N(0,\nu)$; we assume $\nu>0$ for the risk comparison.

If $c$ is the population $L^2(P_X)$ projection coefficient, then $\E[br]=0$, while the logistic curvature cross-term $\mathbf h_r=\E[\pst(1-\pst)br]$ can be nonzero. The positive Schur complement describes a stable, identifiable additional direction.

\begin{lemma}[Local decomposition for direct-score transfer]\label{lem:dst}
Under C1--C4, put
\[
 a_n=A_n\thst,\qquad u_n=H^{-1}(Z_{b,n}-a_n),\qquad
 e=(-\mathbf h_r^\top H^{-1},1)^\top,\qquad
 t_n=\frac{(Z_{{\rm eff},n}+k_n)_+}{s}.
\]
Then
\begin{align}
 \sqrt n(\widehat\theta_{\tg}-\thst)&=u_n+o_p(1),\label{eq:dst-tg-expansion}\\
 \sqrt n\begin{pmatrix}\widehat\theta_{\dst}-\thst\\
                         \widehat v\end{pmatrix}
 &=\begin{pmatrix}u_n\\0\end{pmatrix}+e\,t_n+o_p(1).
 \label{eq:dst-cone-expansion}
\end{align}
These expansions imply the expected log-risk identity
\eqref{eq:DSTgap}, including when $A_n$ is random and score-dependent.
\end{lemma}

\begin{proof}
The true augmented parameter is $(\thst,0)$, because $s>0$.
For bounded local coordinates $(u,t)$, the sum objective at
$(\thst+u/\sqrt n,t/\sqrt n)$, minus its value at $(\thst,0)$,
equals, uniformly on compact sets,
\[
 -Z_{b,n}^\top u-Z_{r,n}t+
 \frac12\begin{pmatrix}u\\t\end{pmatrix}^{\!\top}
 \widetilde H\begin{pmatrix}u\\t\end{pmatrix}
 +a_n^\top u-c_n^\top a_n t+o_p(1).
\]
Indeed the remaining quadratic penalty is bounded by a constant times
$\|A_n\|(\|u\|+|t|\|c_n\|)^2/\sqrt n=o_p(1)$.
The local constraint is $0\leq t\leq\sqrt n$; its limiting cone is
$t\geq0$.  Tightness of the local minimizers makes the upper endpoint
irrelevant.  Completing the square in $u$ gives
\[
 u=u_n-H^{-1}\mathbf h_r\,t,
 \qquad
 t=\arg\min_{z\geq0}
       \left\{\frac{s}{2}z^2-(Z_{{\rm eff},n}+k_n)z\right\}
   =t_n.
\]
Strict positive definiteness and the uniform quadratic approximation
give \eqref{eq:dst-tg-expansion}--\eqref{eq:dst-cone-expansion}.
The expansion holds jointly for the realized $A_n$ and scores, allowing their dependence.

The key orthogonality is
\[
 \widetilde H e=(0,\ldots,0,s)^\top,
 \qquad
 \begin{pmatrix}u_n\\0\end{pmatrix}^{\!\top}\widetilde H e=0.
\]
It yields the pathwise identity
\[
 \left\|\begin{pmatrix}u_n\\0\end{pmatrix}+e\,t_n\right\|_{\widetilde H}^2
 -\left\|\begin{pmatrix}u_n\\0\end{pmatrix}\right\|_{\widetilde H}^2
 =s t_n^2.
\]
The local population excess log-risk is one half of the squared
$\widetilde H$ norm. Its expectation expansion and uniform
integrability give
\begin{equation}
 n\{\E\mathcal K(\widehat\eta_\dst)
       -\E\mathcal K(\widehat\eta_\tg)\}
 =\frac{\E[(Z_{{\rm eff},n}+k_n)_+^2]}{2s}+o(1).
 \label{eq:DSTgap}
\end{equation}
Finally, under
information equality, $Z_{{\rm eff},n}\Rightarrow N(0,s)$; if
$A_n\to_p0$, then $k_n\to_p0$ and
$\E[(Z_{{\rm eff},n}+k_n)_+^2]\to s/2$.
\end{proof}

\subsection{Expected log-risk comparison}

Whenever $A_n\to_p A$ for a deterministic $A\succeq0$, set
\[
 M=AH^{-1}A,\qquad k=(c+H^{-1}\mathbf h_r)^\top A\thst,\qquad
 D(k)=\frac{\E[(W+k)_+^2]}{2s}.
\]
If $c^\top Ac>0$, define
\[
 \widetilde\rho=\Pi_{[0,1]}
      \frac{c^\top A\thst}{c^\top Ac},\qquad
 G_K=\widetilde\rho\,c^\top M\thst
             -\frac{\widetilde\rho^2}{2}c^\top Mc.
\]
If $c^\top Ac=0$, set $\widetilde\rho=0$ and $G_K=0$ as
risk conventions. In that case $Ac=0$, so the leading centering contribution is zero for any limit of the bounded fitted center coefficient.

\begin{theorem}[Expected log risk of projected and direct-score borrowing]\label{thm:risk}
Suppose $\tau=0$, the target model is correctly specified,
$r\notin\operatorname{span}(b)$, and $\nu>0$. Assume C1--C4 for all
three estimators, which use the same realized $\Omega_n$.
\begin{enumerate}\renewcommand{\labelenumi}{(\roman{enumi})}
\item For any common smoothing sequence satisfying
$A_n=\Omega_n/\sqrt n=O_p(1)$ under these conditions,
\begin{equation}
 \liminf_{n\to\infty}
 n\{\E\mathcal K(\widehat\eta_\dst)
           -\E\mathcal K(\widehat\eta_\tg)\}\ge0.
 \label{eq:dst-liminf}
\end{equation}
This conclusion also covers random, outcome-dependent $A_n$ with no deterministic limit.

\item If additionally $A_n\to_p A$ for deterministic $A\succeq0$, then
\begin{align}
 n\{\E\mathcal K(\widehat\eta_\tg)
       -\E\mathcal K(\widehat\theta_{\spt,0})\}
      &\longrightarrow G_K,\label{eq:centerK}\\
 n\{\E\mathcal K(\widehat\eta_\dst)
       -\E\mathcal K(\widehat\eta_\tg)\}
      &\longrightarrow D(k)
       =\frac{\E[(W+k)_+^2]}{2s}>0.\label{eq:dst-limit}
\end{align}
Consequently,
\begin{equation}
 n\{\E\mathcal K(\widehat\eta_\dst)
       -\E\mathcal K(\widehat\theta_{\spt,0})\}
       \longrightarrow G_K+D(k).
 \label{eq:combinedrisk}
\end{equation}
When $c^\top Ac>0$,
\begin{equation}
 G_K>0
 \quad\Longleftrightarrow\quad
 0<\widetilde\rho<
 \frac{2c^\top M\thst}{c^\top Mc}.
 \label{eq:gain-condition}
\end{equation}
If $G_K\ge0$, SPT has strictly smaller actual expected target
log-risk than DST for all sufficiently large $n$.

\item If $A_n\to_p0$ (so $A=0$), then $G_K=k=0$, SPT and T-GAM have the same
first-order expected log-risk, and
\begin{equation}
 n\{\E\mathcal K(\widehat\eta_\dst)
       -\E\mathcal K(\widehat\theta_{\spt,0})\}
       \longrightarrow \frac{\nu}{4s}>0.
 \label{eq:weakbenefit}
\end{equation}
For the correctly specified, uncensored, unit-weight Bernoulli
likelihood, $\nu=s$ and this risk difference is
$1/(4n)+o(n^{-1})$.
\end{enumerate}
\end{theorem}

\paragraph{Local smoothing interpretation.}
Part (ii) describes a local smoothing asymptotic regime:
\[
 \Lambda_n=\frac{\Omega_n}{n}
 =\frac{A}{\sqrt n}+o_p(n^{-1/2}).
\]
At this scale, the leading smoothing bias and the sampling error can
both contribute to the root-$n$ coefficient expansion. The result
provides a local approximation to finite-sample settings in which
smoothing bias is comparable to sampling uncertainty, analogous in
purpose to local alternatives in testing theory. When the fitted
penalty satisfies $\Omega_n=O_p(1)$, including a REML fit in this
bounded-smoothing regime, $A_n\to_p0$ and part (iii) is the relevant
conclusion: the centering contribution is $o(n^{-1})$, whereas
excluding an identifiable source residual retains a strictly positive
$1/n$-order log-risk benefit. The rate conditions are imposed on the realized smoothing matrix.

\begin{proof}
\textbf{Part (i).}
Lemma~\ref{lem:dst} gives the actual expected-risk identity
\eqref{eq:DSTgap}. Its leading term is nonnegative for every $n$,
which proves \eqref{eq:dst-liminf}; the liminf accommodates smoothing sequences whose induced residual costs vary with $n$.

\textbf{Part (ii): centering.}
Lemma~\ref{lem:localization} gives consistency.
If $c^\top Ac>0$, the exact profile formula and continuous mapping
imply $\widehat\rho_0\to_p\widetilde\rho$. The first-order optimality
equations, local Hessian convergence, and C3 give
\begin{align}
 \sqrt n(\widehat\theta_\tg-\thst)
 &=H^{-1}\{Z_{b,n}-A\thst\}+o_p(1),\label{eq:TGexp}\\
 \sqrt n(\widehat\theta_{\spt,0}-\thst)
 &=H^{-1}\{Z_{b,n}-A(\thst-\widetilde\rho c)\}+o_p(1).
 \label{eq:SPTexp}
\end{align}
Consistency of $\widehat\rho_0$ and the Lipschitz profile gradient, including at endpoint switches, suffice for these expansions.

If $c^\top Ac=0$, positive semidefiniteness gives $Ac=0$.
Since $0\le\widehat\rho_0\le1$ and $A_nc_n\to_p0$, both fits have the
same first-order expansion \eqref{eq:TGexp}, irrespective of any
limit of $\widehat\rho_0$.

In the nondegenerate case put $a=A\thst$ and
$a_c=A(\thst-\widetilde\rho c)$. C2 gives $\E Z_{b,n}\to0$.
C4 makes the probability remainders negligible in expected quadratic
risks. In \eqref{eq:risk-taylor}, the common score covariance cancels:
\[
 n(\E\mathcal K_\tg-\E\mathcal K_{\spt,0})
 \longrightarrow
 \tfrac12(a^\top H^{-1}a-a_c^\top H^{-1}a_c)=G_K.
\]
This also holds with $G_K=0$ in the degenerate case.

\textbf{Part (ii): the residual and strict improvement.}
Here $k_n\to_p k$ and $Z_{{\rm eff},n}\Rightarrow W$.
Slutsky's theorem, the continuous positive-part map, and C4 give
\[
 \E[(Z_{{\rm eff},n}+k_n)_+^2]\longrightarrow\E[(W+k)_+^2].
\]
Inserting this into \eqref{eq:DSTgap} proves \eqref{eq:dst-limit};
adding the two risk differences proves \eqref{eq:combinedrisk}.
Since $\nu>0$, a normal $W$ has $\Prb(W+k>0)>0$ for every finite
$k$, and hence $D(k)>0$. If $c^\top Ac>0$, then $c^\top Mc>0$ and
\[
 G_K=\frac{\widetilde\rho\,c^\top Mc}{2}
      \left\{\frac{2c^\top M\thst}{c^\top Mc}-\widetilde\rho\right\},
\]
which proves \eqref{eq:gain-condition}. For $G_K\ge0$ the limit
in \eqref{eq:combinedrisk} is strictly positive, implying the asserted
expected-risk inequality for all sufficiently large $n$.

\textbf{Part (iii).}
$A=0$ implies $G_K=k=0$. Symmetry gives
$\E[(W_+)^2]=\E[W^2]/2=\nu/2$, and therefore $D(0)=\nu/(4s)$.
Under information equality $\widetilde V=\widetilde H$, the
efficient residual score variance is $\nu=s$.
\end{proof}

\begin{remark}[Explicit residual cost and the scope of the ordering]
For $z=k/\sqrt\nu$, normal integration gives
\[
 D(k)=
 \frac{(\nu+k^2)\Phi(z)+k\sqrt\nu\,\phi(z)}{2s},
\]
where $\phi,\Phi$ denote the standard normal density and distribution
function. The condition $G_K\ge0$ is sufficient for SPT to improve on
DST. The exact condition for a strictly positive first-order improvement
is $G_K+D(k)>0$, which also permits a mildly negative $G_K$.
When this sum is zero, the ordering depends on smaller-order risk terms. In part (i), a bounded deterministic
$A_n$ can alternate between two matrices with different induced
residual costs; this is why the statement uses $\liminf$.
\end{remark}

\subsection{Automatic rejection of mismatched borrowing channels}
\label{sec:safety}

This result concerns the implemented fixed-grid selection and full-sample
refit. It separates rejection of the soft-label weight from rejection of
the penalty center. The argument uses uniform convergence of the calibrated soft-label function and validation risks. Let $\widehat\theta_{m,\tau}$ denote the
SPT fit on a training sample of size $m$, and let
$\widehat\rho_0$ denote the full-sample center coefficient for the
$\tau=0$ candidate. The matched target-only fit uses the same target
basis, fitted IPCW weights, and smoothing matrix as SPT.

\begin{enumerate}\renewcommand{\labelenumi}{N\arabic{enumi}.}
\item \textbf{Target model, weights, and folds.}
The target model is correctly specified, $\pst=p_{\thst}$, with fixed
dimension, bounded basis functions, identifiable design, and
$\varepsilon+\zeta\le\pst\le1-\varepsilon-\zeta$ for some $\zeta>0$.
The compact and directional weighted laws of large numbers in
Lemma~\ref{lem:localization} hold on every training sample and on the full
sample, with the population limits of the correctly weighted target
loss. The validation censoring estimates converge uniformly to the
correct survival function on the relevant time range, where this
function is bounded away from zero. Clipped validation losses obey
$0\le L_i(\tau)\le C$ for a fixed constant $C$. The number of folds is
fixed and each fold has a nonvanishing sample fraction. All
training-dependent fitting steps are recomputed within each fold and
again for the full-sample refit.

\item \textbf{Vanishing normalized smoothing.}
On each relevant training sample of size $m$, including the full sample,
\[
 \|\Omega_m\|_{\mathrm{op}}/m\longrightarrow_p0,
 \qquad c_m=O_p(1).
\]
The smoothing matrices may be chosen using the same training outcomes;
different smoothing blocks need not have a common rate.

\item \textbf{A persistent soft-label discrepancy visible to the target basis.}
The actual calibrated and clipped soft-label function converges uniformly, on each
training sample, to the same function
$q_\infty(x)\in[\varepsilon,1-\varepsilon]$, and
\[
 b_q:=\E\big[b(X)\{q_\infty(X)-\pst(X)\}\big]\ne0.
\]
This condition concerns a source discrepancy visible to the target
basis, rather than pointwise soft-label error alone.

\item \textbf{A strictly adverse center direction.}
For a fixed $\gamma>0$, the full-sample quantities satisfy
\[
 \Prb\left\{
 \Omega_nc_n=0\ \text{or}\
 \frac{(\Omega_nc_n)^\top\thst}{\|\Omega_nc_n\|}\le-\gamma
 \right\}\longrightarrow1,
\]
where the ratio is evaluated only on $\{\Omega_nc_n\ne0\}$. When
$\Omega_nc_n=0$, the analytic profile uses the convention $\widehat\rho=0$.
A sufficient nondegenerate condition is
$\Omega_n/\xi_n\to_p \Omega_0\succeq0$, $c_n\to_p c$, and
$c^\top \Omega_0\thst<0$ for some positive sequence $\xi_n$.

\item \textbf{The reference-based CV rule.}
The candidate grid $\Tset\subset[0,\infty)$ is fixed and finite,
contains $0$, and contains a fixed positive reference $\tau_r$.
Let
\[
 \widehat C_n(\tau)=\frac1n\sum_{i=1}^n L_i(\tau),
 \qquad
 \tau_n^{\min}\in\arg\min_{\tau\in\Tset}\widehat C_n(\tau),
\]
with ties resolved toward smaller $\tau$. For the paired individual
loss differences
$d_i=L_i(\tau_r)-L_i(\tau_n^{\min})$, let $s_d$ be their empirical
standard deviation. Select $\widehat\tau=\tau_r$ if
\[
 \widehat C_n(\tau_r)-\widehat C_n(\tau_n^{\min})
 \le s_d/\sqrt n,
\]
and otherwise select $\widehat\tau=\tau_n^{\min}$. Refit the selected
candidate on the full sample. The implemented choice
$\Tset=\{0,0.1,0.5,1\}$ and $\tau_r=0.5$ is included.
\end{enumerate}

\begin{theorem}[Asymptotic rejection of mismatched borrowing channels]
\label{thm:safety}
The two borrowing channels satisfy the following conclusions.
\begin{enumerate}\renewcommand{\labelenumi}{(\alph{enumi})}
\item Under N1--N3 and N5,
\[
 \Prb(\widehat\tau=0)\longrightarrow1.
\]
This conclusion does not require N4.
\item Under the full-sample conditions in N1--N2 and N4,
\[
 \Prb\{\widehat\rho_0=0,\
 \widehat\theta_{n,0}=\widehat\theta_{\tg}\}
 \longrightarrow1.
\]
This conclusion concerns the $\tau=0$ candidate and does not require
N3 or CV selection.
\item Under N1--N5,
\[
 \Prb\{\widehat\tau=0,\ \widehat\rho_{\widehat\tau}=0,\
 \widehat\theta_{n,\widehat\tau}=\widehat\theta_{\tg}\}
 \longrightarrow1.
\]
Thus the final refitted estimator coincides with the matched target-only
estimator with probability tending to one.
\end{enumerate}
\end{theorem}

\begin{proof}
\textbf{Part (a): limiting candidate fits.}
Define
\[
 R(\theta)=\E\,\CE\{\pst(X),p_\theta(X)\},
 \qquad
 Q_\tau(\theta)=R(\theta)
       +\tau\E\,\CE\{q_\infty(X),p_\theta(X)\}.
\]
By N2, the profiled penalty converges uniformly to zero on compact
coefficient sets: evaluating its minimum at $\rho=0$ bounds it above by
$\|\Omega_m\|_{\mathrm{op}}\|\theta\|^2/(2m)$. N1--N3 and
Lemma~\ref{lem:localization} therefore give, on every fold training
sample and for the full refit,
\[
 \widehat\theta_{m,\tau}\longrightarrow_p
 \theta_\tau:=\arg\min_\theta Q_\tau(\theta).
\]
The target risk is coercive and strictly convex under the interior
truth and identifiable design. The added soft-label loss is nonnegative
and convex, so each $Q_\tau$ has a finite unique minimizer. For $\tau=0$,
$\theta_0=\thst$; for each positive grid value,
\[
 \nabla Q_\tau(\thst)=-\tau b_q\ne0,
 \qquad \theta_\tau\ne\thst.
\]

\textbf{A fixed positive validation gap.}
Write
\[
 \bar p_\theta(x)=
 \min\{1-\varepsilon,\max\{\varepsilon,p_\theta(x)\}\},
 \qquad
 R_\varepsilon(\theta)=\E\,\CE\{\pst(X),\bar p_\theta(X)\}.
\]
Since the truth is strictly inside the clipping interval,
$\bar p_\theta=\pst$ can hold only where $p_\theta=\pst$.
Identifiability and $\theta_\tau\ne\thst$ consequently imply
\[
 \Delta_\tau:=R_\varepsilon(\theta_\tau)
                    -R_\varepsilon(\thst)
 =\E\,\KL\{\operatorname{Bern}(\pst),
                  \operatorname{Bern}(\bar p_{\theta_\tau})\}>0
 \quad(\tau>0).
\]
The finite fixed grid gives
$\Delta_{\min}=\min_{\tau\in\Tset,\tau>0}\Delta_\tau>0$.

For a fixed fold and candidate, replace the fitted validation nuisance
weights by their correct limits. Uniform nuisance consistency,
positivity, and bounded clipped loss make the resulting error
$o_p(1)$. Replace the fitted coefficient by $\theta_\tau$; consistency,
bounded design, and the Lipschitz property of clipped log-loss again
give an $o_p(1)$ error. The remaining held-out average is an average of
a fixed integrable function and obeys the law of large numbers. There
are finitely many folds and candidates, hence
\[
 \max_{\tau\in\Tset}
 |\widehat C_n(\tau)-R_\varepsilon(\theta_\tau)|
 \longrightarrow_p0.
\]
The strict gap implies $\Prb(\tau_n^{\min}=0)\to1$, establishing CV consistency through the limiting fold fits and their held-out risks.

\textbf{The reference preference.}
By N1, $|d_i|\le C$, even though the paired out-of-fold losses need not
be independent. Deterministically,
\[
 \frac{s_d}{\sqrt n}\le\frac{C}{\sqrt{n-1}}\longrightarrow0.
\]
On the event $\{\tau_n^{\min}=0\}$, the estimated improvement over
the positive reference converges to $\Delta_{\tau_r}>0$. The reference
rule therefore retains zero with probability tending to one, proving
part (a). This includes the implemented reference $\tau_r=0.5$.

\textbf{Part (b): exact rejection of the center.}
The full-sample conditions in N1--N2 and
Lemma~\ref{lem:localization} imply
$\widehat\theta_{\tg}\to_p\thst$. On the nonzero-direction event in N4,
\[
 \frac{c_n^\top \Omega_n\widehat\theta_{\tg}}{\|\Omega_nc_n\|}
 \le-\gamma+\|\widehat\theta_{\tg}-\thst\|<0
\]
with probability tending to one. Proposition~\ref{prop:direction}
then shows that $(\widehat\theta_{\tg},0)$ is the joint SPT minimizer
at $\tau=0$. The weighted design has full column rank with probability
tending to one by N1, so the outcome loss is strictly convex in
$\theta$. When $\Omega_nc_n\ne0$, positive semidefiniteness gives
$c_n^\top \Omega_nc_n>0$; together these facts ensure uniqueness of the
joint minimizer. If $\Omega_nc_n=0$, the penalty does not depend on $\rho$,
the coefficient fit equals the target-only fit, and the stated profile
convention selects zero. This proves part (b).

\textbf{Part (c): combining the channels.}
The full-data candidate $\widehat\theta_{n,0}$ is precisely the full
refit used when $\widehat\tau=0$. By the union bound,
\begin{align*}
 &\Prb\{\widehat\tau=0,\ \widehat\rho_{\widehat\tau}=0,\
              \widehat\theta_{n,\widehat\tau}=\widehat\theta_{\tg}\}\\
 &\quad\ge1-\Prb(\widehat\tau\ne0)
       -\Prb\{\widehat\rho_0\ne0\ \text{or}\
              \widehat\theta_{n,0}\ne\widehat\theta_{\tg}\}
 \longrightarrow1.
\end{align*}
\end{proof}

\paragraph{Censoring weights and projection.}
Under marginal independent censoring, event contributions use $1/G(Y-)$ and event-free contributions use $1/G(h)$. Uniform consistency and positivity supply the weighted laws for Theorem~\ref{thm:safety}; Theorem~\ref{thm:risk} additionally uses the score expansion including censoring estimation. The implemented cap of 20 and floor of $10^{-3}$ are asymptotically inactive when $\inf_{t\le h}G(t)>1/20+\delta_G$ for a fixed $\delta_G>0$ and the censoring estimator is uniformly consistent. For the penalized Gaussian projection, $(B^\top B+P_n)c_n=B^\top\eta_S$ implies $B^\top r_n=P_nc_n$. The Schur complement in the risk comparison accommodates this dependence between the target basis and residual direction.

\section{Simulation design, comparison methods and evaluation}
\subsection{Simulation design and model specifications}
Covariate distributions, the additive function $a$, the interaction $u$, and source--target predictors for settings A--D are defined in Section~3.1 of the main text. The main design has common covariate support. The target sample-size experiments use B and D with $n_T=200$ or 1000. The nuisance-variable experiments append 13 independent uniform variables to both cohorts and include all 20 predictors in every fitted model; these added variables have no effect on outcomes. They retain $n_T=500$ and settings B and D. Both survival distributions are evaluated in every group.

In the main A--D experiment, source and target each compare an additive formula with a formula including the $X_1$--$X_3$ interaction, retaining every main effect. GCV selects the formula using IPCW-weighted quasibinomial fitting with $\mathtt{gamma}=1.4$; the selected outcome GAM is then fitted by REML. By design, the source and target parameterize the interaction differently: the source uses a varying-coefficient smooth of $X_3$ by numeric $X_1$, while the target uses an identifiable ordered-factor-by smooth with the binary main effect retained. Source smooths use ts/$k=10$; target and projection smooths use ts/$k=5$. Each inner target training fold selects its own formula, basis and outcome penalty; the full-training fit repeats this procedure. Gaussian projection uses the selected target space and its own REML smoothing parameters. The candidate transfer models share these fitted quantities within each training sample. Table~\ref{tab:structure_selection} reports full-training interaction selection rates.

Pooled GAMs retain the shared-interaction formula with $k=5$ in A--D, reflecting the interaction represented in the pooled source observations. The sample-size and irrelevant-variable experiments use prespecified interaction formulas for B and additive target formulas for D, with the source interaction included in both settings. The level, shape and reversal experiments retain the setting-A interaction formulas in both cohorts to isolate changes in source effects. These additional designs use REML outcome fitting and Gaussian REML projection.

The event-time generators and conditional survival functions are given in Section~3.1 of the main text. Their error scales are $\sigma_{\mathrm{LL}}=1.245686424145402$ and $\sigma_{\mathrm{W}}=1.761666635490472$, with $\gamma_E=0.5772156649015329$. The Oracle event risk is $1-S_D(h\mid x)$ under the corresponding survival function.

The independent exponential censoring rates are calibrated separately for each setting and distribution to expected censoring proportions of 40\% in the source sample, 50\% in the target training sample and 30\% in the target test sample; realized proportions vary across replications.

\subsection{Comparison methods}\label{sec:comparison_methods}
In simulations, all methods use the same source/target covariates, target test sets and replication-specific horizons. They retain their own model classes and tuning criteria. Matched component comparisons in Section~3.3 of the main text isolate the transfer mechanisms within a common model class; implementation details are given in Section~\ref{sec:ablations}.

\paragraph{GAM references.}
T-GAM uses only target outcomes, with GCV-selected structure in the main A--D experiment and the prespecified working structure in the additional designs. Source GAM directly applies the source ts/$k=10$ predictor without target calibration. Pooled GAM combines both training cohorts with shared smooths and an intercept. Stratified GAM adds a cohort indicator to the pooled GAM, allowing cohort-specific intercepts while retaining shared smooth effects. Separate within-cohort censoring estimates determine pooled fitting weights. GAM references use REML, \texttt{gamma=1.4} and ts/$k=5$ for target or pooled smooths.

\paragraph{CoxTL and POTL-Cox.}
CoxTL uses the authors' implementation of density-ratio adjustment and calibrated source contribution \citep{Lu2026}. The transfer grid is $\{0.001,0.005,0.01,0.05,0.1,1,2,\ldots,10\}$, with five-fold internal target tuning. The implementation scales the nominal source-contribution parameter by $n_T/n_S$. POTL-Cox uses the authors' Cox implementation \citep{Gu2026}, source Cox predictions and the target baseline estimate, with the grid $\{0\}\cup\{2^j:j=-15,\ldots,15\}$ and five-fold target tuning. The prediction-oriented objective provides a reference for borrowing on the probability scale; SPT additionally adapts the target smoothness-penalty center.

\paragraph{Additional transfer methods.}
TransCox uses the local implementation of \citet{Li2023}, selecting two penalty constants over $\{0.1,0.5,1,2\}^2$ by BIC. CoxKL \citep{CoxKL2026} uses the implementation's cross-validated partial-likelihood criterion; the simulation grid extends its ten-point 0--5 grid with 10, 20 and 50. DiscreteKL uses the authors' discrete-time KL framework \citep{Wang2025}. Simulation intervals subdivide each span between consecutive horizons into four equal parts, yielding 12 intervals through Q75. Its borrowing grid is 0 to 10 by 0.1. Trans-AFT uses the authors' weighted two-stage accelerated failure-time implementation \citep{TransAFT2025}, with its ten constants from 1 to 0.1 and eight-fold tuning. A locally implemented residual-survival calculation converts its fitted time score into risks at the requested horizons.  Target Cox is an ordinary target-only linear Cox fit with a target baseline hazard.

\paragraph{Flexible target-only references.}
RSF \citep{Ishwaran2008} uses 500 trees, log-rank splitting, minimum node sizes $\{10,20,40\}$, and $m_{\mathrm{try}}\in\{\lfloor\sqrt p\rfloor,\lceil p/2\rceil,p\}$. Three-fold target cross-validation selects the combination using average IPCW log-loss across the prediction horizons. IPCW-XGBoost \citep{Chen2016} fits a horizon-specific weighted binary learner, with depths $\{1,2,3\}$ and boosting rounds $\{100,300\}$ selected by three-fold target cross-validation. Learning rate is 0.05, row and column subsampling fractions are 0.8, and minimum child weight is 5. These methods serve as flexible prediction benchmarks without imposing an additive structure.

\paragraph{Clinical-analysis settings.}
The clinical CoxKL comparison uses the ten-point 0--5 grid. DiscreteKL uses six-month intervals through 48 months. RSF tuning uses 24/36/48-month risks; the same fitted survival forest supplies the 30-month prediction. Score recalibration fits $p_h(x)=\expit\{\alpha_h+\beta_h\eta_{S,h}(x)\}$ by unpenalized IPCW logistic regression on target training observations, with an unrestricted intercept and slope \citep{Steyerberg2004}.

\subsection{Evaluation metrics and Monte Carlo uncertainty}\label{sec:evaluation_metrics}
For simulation evaluation, the censoring distribution is estimated from the independent target test sample, which is shared across methods within each replication. Write $\widehat G$ for its censoring Kaplan--Meier estimate. For a test prediction $p_i=p_h(X_i)$, the IPCW Brier score estimate \citep{Graf1999} is
\[
 \widehat{\mathrm{BS}}(h)=\frac1{n}\sum_i
 \left\{\frac{\delta_i I(Y_i\leq h)}{\widehat G(Y_i)}(1-p_i)^2+
 \frac{I(Y_i>h)}{\widehat G(h)}p_i^2\right\}.
\]
Here $\widehat G(Y_i)$ is evaluated using the right-continuous KM estimate, following the convention in Section~\ref{sec:numerical_details}.
The time-dependent AUC at $h$ compares individuals with $T^*\leq h$ against those with $T^*>h$. Its IPCW estimate compares observed cases $Y_i\leq h,\delta_i=1$ with controls $Y_j>h$, using weights $1/\widehat G(Y_i)$ and $1/\widehat G(h)$ and half credit for tied scores \citep{Uno2007,Blanche2013}.

Truncated Uno concordance uses the horizon-specific risk $p_h$ to rank event times through $h$ \citep{Uno2011}. In the absence of event-time ties,
\[
 \widehat C_U(h)=
 \frac{\sum_{i,j}\delta_i I(Y_i\leq h)I(Y_i<Y_j)\widehat G(Y_i)^{-2}
 [I(p_i>p_j)+\tfrac12I(p_i=p_j)]}
 {\sum_{i,j}\delta_i I(Y_i\leq h)I(Y_i<Y_j)\widehat G(Y_i)^{-2}}.
\]

Let $\pi_i=p_{T,h}^{\mathrm{true}}(X_i)$. The conditional expected Brier score, averaged over test covariates, is
\[
 \mathrm{BS}_{\mathrm{DGP}}(h)=\frac1n\sum_i\{(p_i-\pi_i)^2+\pi_i(1-\pi_i)\}.
\]
The DGP-risk AUC uses case and control weights $\pi_i$ and $1-\pi_j$ in the same weighted rank calculation. DGP-based Brier scores and AUCs use the known event probabilities at each horizon. Concordance is evaluated from observed survival outcomes using IPCW.

For horizon-specific summaries, MCSE is the replication SD divided by $\sqrt{100}$. For the paired contrasts in Table~\ref{tab:source_mismatch} and Figure~1 of the main text, the same calculation is applied to the 100 replication-level contrast summaries. Horizon-specific SPT-minus-DST contrasts in Table~\ref{tab:structural_contrasts} use differences within each replication before calculating means and MCSEs.

\section{Implementation of matched component comparisons}\label{sec:ablations}
Section~3.3 and Figure~1 of the main text assess component contributions under source-effect changes and structural mismatch. Table~\ref{tab:source_mismatch} reports the corresponding means and paired MCSEs. Each variant uses the same source score, target design $B$, target penalty $\Omega$, projection $c$ and internal folds. The PR-GAM, fixed-center, uncalibrated and soft-label-free comparisons retain the predictor $B^\mathsf T\theta$. PR-GAM uses the penalty $\theta^\mathsf T\Omega \theta/2$, whereas the fixed-center variant uses $(\theta-c)^\mathsf T\Omega (\theta-c)/2$. The remaining variants retain SPT's profiled penalty center. Uncalibrated soft labels replace $q_i$ by clipped $\expit\{\eta_S(X_i)\}$; setting $\tau=0$ removes the soft-label loss. The projection-centered penalty remains active in this $\tau=0$ comparison.

\paragraph{Direct-score transfer.}
DST borrows the complete source score with one coefficient and an ordinary penalty on the target correction; see Section~2.2 of the main text. At each $\tau$, it jointly estimates the correction and source-score coefficient using the same outcome and soft-label losses as SPT; both methods share the source fit, target space, smoothing penalty, calibration, folds and $\tau$ selection rule. The source-effect experiments retain fixed interaction formulas, while the main A--D experiments use GCV-selected formulas followed by REML fitting; Table~\ref{tab:structural_contrasts} reports the paired SPT-minus-DST differences in D.

For each source-mismatch replication, target training and test observations, latent risks and prediction horizons are retained from the matched setting-A replication. Source censoring is recalibrated to 40\% under each changed source predictor. Paired contrasts are first averaged over the two distributions and three horizons within each replication, then summarized over the 100 independent replication blocks, as described in Section~\ref{sec:evaluation_metrics}.

\input{table/supp_table_01.tex}

\clearpage

\section{Clinical cohort construction and supplementary results}\label{sec:clinical_supp}
\subsection{Eligibility, follow-up and predictors}
The UK Biobank source is defined by the race-free CKD-EPI 2021 creatinine eGFR range $[20,60)$ and outcome and complete-case eligibility criteria. Among 7368 source participants initially below eGFR 60, exclusions comprise one participant without usable exact death information and 191 below eGFR 20; 7176 have valid positive follow-up and 7059 meet the complete-case requirement. Among 2311 SPRINT participants below eGFR 60, exclusions comprise 15 with invalid time/outcome information and 23 with missing values in the eight-variable eligibility set, leaving 2273 participants across treatment arms. The target cohort comprises the 1126 standard-treatment participants. Table~\ref{tab:cohorts} summarizes the cohorts and horizon-specific outcome support.

UK Biobank follow-up begins at baseline assessment and ends at recorded death or censoring, including the administrative cutoff of 30 November 2020 or earlier recorded loss to follow-up. SPRINT follow-up begins at randomization. Follow-up days are divided by 30.4375 to obtain months. Complete-case eligibility uses the eight-variable set including cardiovascular disease history; the prediction model uses the seven remaining predictors. Smoking histories and laboratory variables use harmonized definitions across cohorts. Baseline eGFR is recomputed consistently across cohorts \citep{Inker2021}. 

The target outer folds are stratified on the recorded 36-month event indicator and share the same 100 partitions across methods. The source GAM is fitted once at each horizon on all eligible source observations. Internal target tuning is repeated within each outer training sample, including fresh projection and calibration. The resulting 100 out-of-fold prediction vectors are different fits on the same participants. Table~\ref{tab:realdata} reports means and partition SDs to quantify sensitivity to training/test allocation within this cohort.

\input{table/supp_table_02.tex}

\clearpage

\paragraph{Alternative source learners.}
SPT-GAM uses the source ts/$k=10$ GAM. SPT-Cox and SPT-NN replace this source learner while retaining the target additive ts/$k=5$ GAM, Gaussian projection, calibrated soft labels, profiled $\rho\in[0,1]$ and the same $\tau$ grid and selection rule. All variants use the same participants, outcomes, outer five-fold partitions and inner three-fold partitions.

The source Cox model includes the seven linear main effects and uses Efron's handling of tied event times. One fit supplies the baseline cumulative hazard and event probabilities $\widehat p_{S,h}(x)=1-\exp\{-\widehat H_{0,S}(h)\exp(x^\mathsf T\widehat\beta_S)\}$ at all three horizons. SPT uses the logit of these absolute event probabilities as its source score. The source neural classifier has one logistic hidden layer and a logistic output, fitted by IPCW-weighted Bernoulli loss with weight decay. Source-only five-fold cross-validation selects hidden-layer size from $\{4,8\}$ and decay from $\{0.1,1,10\}$ using IPCW log-loss. Each configuration averages the probabilities from three fixed initializations. Standardization and censoring estimation use the corresponding source training folds. A final source ensemble is fitted separately at each horizon and shared across target partitions. Both learners supply probabilities clipped to $[10^{-3},1-10^{-3}]$ before the logit transformation.

\input{table/supp_table_03.tex}

\clearpage
\section{Additional simulation results}
Tables~\ref{tab:allG2500loglogistic0}--\ref{tab:allG2500weibull1} provide the full main-simulation comparisons. Tables~\ref{tab:allG3200loglogistic0}--\ref{tab:allG3200weibull0} and~\ref{tab:allG31000loglogistic0}--\ref{tab:allG31000weibull0} report the smaller and larger target samples, respectively. Tables~\ref{tab:allG5500loglogistic0}--\ref{tab:allG5500weibull0} cover irrelevant predictors, and Table~\ref{tab:dgp_check} compares observed-outcome and DGP-risk evaluation.

\subsection{Projected borrowing across target sample sizes}\label{sec:fixed_additive_D}
The fixed-additive comparison in main-text Figure~2 uses log-logistic setting D, $g_S=a+0.8u$ and $g_T=a$, at $n_T=500,1000,2000,4000,10000$. The source GAM includes the original interaction with \texttt{ts}/$k=10$ smooths; the target uses a fixed additive \texttt{ts}/$k=5$ basis. We retain $n_S=3000$, 5000 target test observations, the same generating coefficients and covariate shift, and expected source, target-training and test censoring proportions of 40\%, 50\% and 30\%. All five methods share source fits, target data, smoothing matrices, horizons and validation folds within each replication. Horizons are the target-training observed-time quantiles and are recomputed at each sample size. Source and test observations are reused by replication across sample sizes; the target samples are not nested.

The KL evaluation averages $\operatorname{KL}\{\operatorname{Bern}(p_i^\star)\,\Vert\,\operatorname{Bern}(\widehat p_i)\}$ over the test covariates. Subtracting T-GAM's KL risk gives the same contrast as subtracting its DGP expected logistic loss. For each metric, differences at Q25, Q50 and Q75 are averaged within replication and multiplied by $n_T$; the resulting 100 paired summaries supply the plotted mean and MCSE.

\begin{figure}[p]\centering
\includegraphics[width=\textwidth]{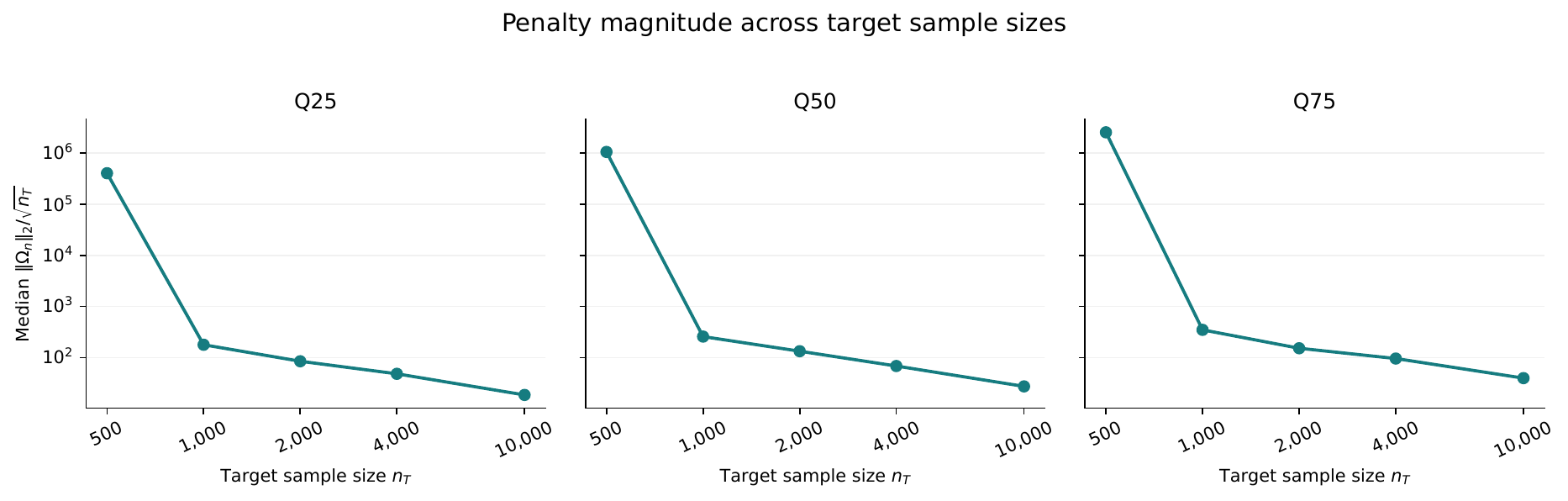}
\caption{Median scaled target penalty across sample sizes. Each point is the median of $\|\Omega_n\|_2/\sqrt{n_T}$ over 100 replications, using the complete fitted penalty matrix, including its estimated smoothing parameters. Columns correspond to Q25, Q50 and Q75; the vertical axis is logarithmic.}\label{fig:penalty_norm}
\end{figure}

Figure~\ref{fig:penalty_norm} records the median spectral norm $\|A_n\|_2=\|\Omega_n\|_2/\sqrt{n_T}$ of the fitted target penalty. It decreases at every horizon, accompanying the reduction in the scaled SPT--T-GAM KL contrast between the smallest and largest target samples. At Q50, the median falls from 258 at $n_T=1000$ to 133, 69 and 27 at the three larger sample sizes. This diagnostic tracks the fitted penalty in the working basis.

\begin{figure}[p]\centering
\includegraphics[width=\textwidth]{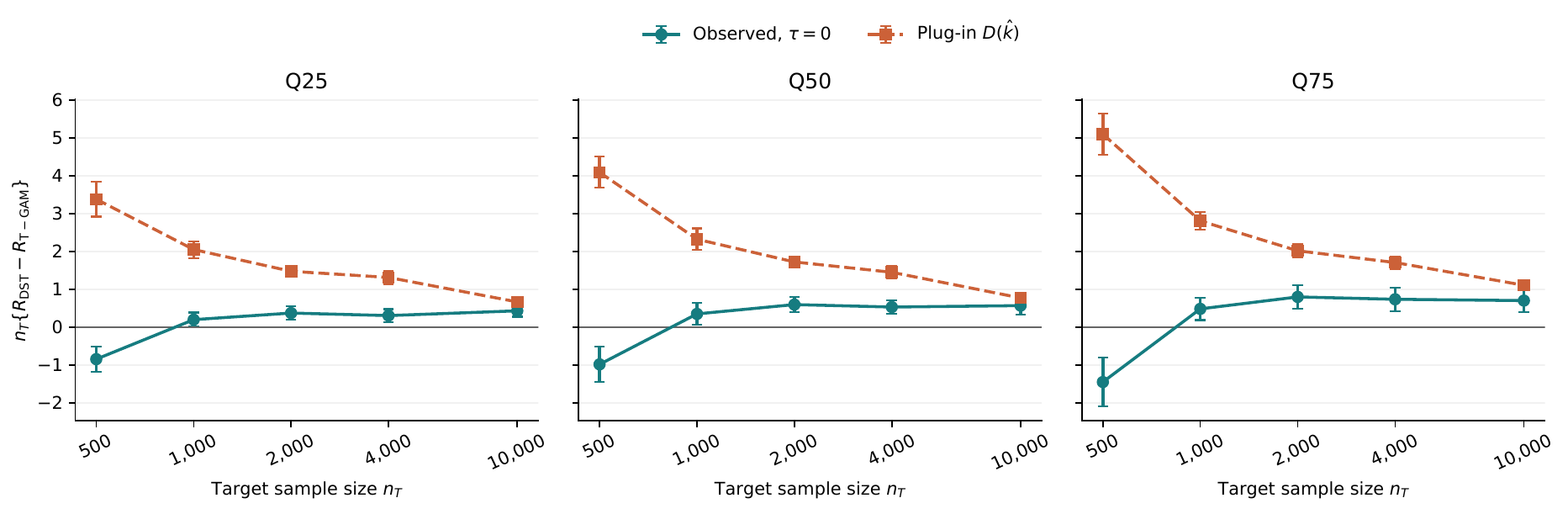}
\caption{Plug-in interpolation and observed scaled log-risk differences. Solid curves show the mean $n_T\{\mathcal K(\widehat\eta_{\mathrm{DST}})-\mathcal K(\widehat\eta_{\mathrm{TG}})\}$ at $\tau=0$; dashed curves show the mean replication-specific $\widehat D$. Error bars are $\pm1.96$ MCSE. Each horizon is evaluated separately over 100 replications at each target sample size.}\label{fig:risk_interpolation}
\end{figure}

For Figure~\ref{fig:risk_interpolation}, we estimate the local quantities in Theorem~\ref{thm:risk} from each fitted model. The plug-in target coefficient is the T-GAM estimate. We estimate $H$, $\mathbf h_r$ and $s$ using the independent test covariates and the corresponding fitted Bernoulli curvature, and set $\widehat k=(c+\widehat H^{-1}\widehat{\mathbf h}_r)^\top A_n\widehat\theta_{\mathrm{TG}}$. The estimated effective-score variance $\widehat\nu$ includes the influence of estimating the target censoring Kaplan--Meier function, conditional on the fitted source score and projection. The normal interpolation is
\[
 \widehat D=
 \frac{(\widehat\nu+\widehat k^2)\Phi(\widehat k/\sqrt{\widehat\nu})
       +\widehat k\sqrt{\widehat\nu}\,\phi(\widehat k/\sqrt{\widehat\nu})}
      {2\widehat s},
\]
where $\Phi$ and $\phi$ are the standard normal distribution and density functions. We average these replication-specific predictions and compare them with the observed $n_T$-scaled DST-minus-T-GAM KL contrasts at $\tau=0$. The interpolation approaches the observed contrast as sample size grows. At $n_T=10000$, predicted values are 0.663, 0.768 and 1.103, compared with observed values of 0.428, 0.565 and 0.698. The paired prediction-minus-observation differences are 0.235 (0.095), 0.203 (0.138) and 0.405 (0.182), with MCSEs in parentheses. The finite-sample interpretation is discussed in Section~5 of the main text.

\clearpage

\input{table/supp_table_04.tex}

\input{table/supp_table_05.tex}

\input{table/supp_table_06.tex}

\input{table/supp_table_07.tex}

\input{table/supp_table_08.tex}

\input{table/supp_table_09.tex}

\input{table/supp_table_10.tex}

\input{table/supp_table_11.tex}

\input{table/supp_table_12.tex}

\input{table/supp_table_13.tex}

\input{table/supp_table_14.tex}

\input{table/supp_table_15.tex}

\input{table/supp_table_16.tex}


\clearpage


\clearpage\bibliographystyle{plainnat}\bibliography{reference}

%% file: table/main_table_01.tex
\begin{table}[!p]
\centering
\caption{Log-logistic simulation results: $n_S=3000$, $n_T=500$, $p=7$. Means over 100 replications; corresponding MCSEs are reported in the Supplementary Material. The best fitted-method mean in each setting and column is bold; Oracle is excluded.}\label{tab:loglogistic}
\begingroup\fontsize{9}{11}\selectfont\setlength{\tabcolsep}{2.5pt}\renewcommand{\arraystretch}{1.12}
\sbox{\maintablebox}{%
\begin{tabular}{@{}l*{9}{c}@{}}\toprule
 & \multicolumn{3}{c}{Q25} & \multicolumn{3}{c}{Q50} & \multicolumn{3}{c}{Q75}\\
\cmidrule(lr){2-4}\cmidrule(lr){5-7}\cmidrule(lr){8-10}
Method & Brier & $C_U(h)$ & AUC & Brier & $C_U(h)$ & AUC & Brier & $C_U(h)$ & AUC\\
\midrule
\multicolumn{10}{l}{\textit{Setting A}}\\[2pt]
T-GAM & 0.1238 & 0.786 & 0.811 & 0.1721 & 0.756 & 0.793 & 0.1933 & 0.730 & 0.775\\
Source GAM & \textbf{0.1197} & \textbf{0.799} & \textbf{0.825} & \textbf{0.1658} & \textbf{0.769} & \textbf{0.808} & \textbf{0.1839} & \textbf{0.747} & \textbf{0.795}\\
Stratified GAM & 0.1200 & 0.799 & 0.825 & 0.1663 & 0.769 & 0.808 & 0.1844 & 0.747 & 0.795\\
\textbf{SPT} & 0.1203 & 0.798 & 0.824 & 0.1669 & 0.768 & 0.807 & 0.1853 & 0.746 & 0.794\\
CoxTL & 0.1323 & 0.769 & 0.793 & 0.1829 & 0.734 & 0.768 & 0.2030 & 0.711 & 0.750\\
POTL-Cox & 0.1324 & 0.768 & 0.792 & 0.1826 & 0.734 & 0.767 & 0.2031 & 0.711 & 0.749\\
RSF & 0.1298 & 0.770 & 0.794 & 0.1820 & 0.739 & 0.773 & 0.2025 & 0.716 & 0.757\\
\textit{Oracle} & \textit{0.1187} & \textit{0.802} & \textit{0.829} & \textit{0.1645} & \textit{0.772} & \textit{0.811} & \textit{0.1822} & \textit{0.750} & \textit{0.799}\\
\midrule
\multicolumn{10}{l}{\textit{Setting B}}\\[2pt]
T-GAM & 0.1330 & 0.749 & 0.772 & 0.1848 & 0.726 & 0.763 & 0.2042 & 0.703 & 0.748\\
Source GAM & 0.1418 & 0.728 & 0.749 & 0.1969 & 0.702 & 0.734 & 0.2168 & 0.681 & 0.719\\
Stratified GAM & 0.1367 & 0.737 & 0.758 & 0.1916 & 0.710 & 0.744 & 0.2117 & 0.689 & 0.730\\
\textbf{SPT} & \textbf{0.1318} & \textbf{0.754} & \textbf{0.777} & \textbf{0.1829} & \textbf{0.730} & \textbf{0.768} & \textbf{0.2020} & \textbf{0.708} & \textbf{0.753}\\
CoxTL & 0.1405 & 0.715 & 0.735 & 0.1981 & 0.689 & 0.718 & 0.2191 & 0.669 & 0.704\\
POTL-Cox & 0.1405 & 0.716 & 0.735 & 0.1980 & 0.689 & 0.718 & 0.2189 & 0.669 & 0.704\\
RSF & 0.1385 & 0.731 & 0.752 & 0.1939 & 0.709 & 0.743 & 0.2120 & 0.690 & 0.732\\
\textit{Oracle} & \textit{0.1275} & \textit{0.770} & \textit{0.795} & \textit{0.1761} & \textit{0.745} & \textit{0.785} & \textit{0.1921} & \textit{0.727} & \textit{0.777}\\
\midrule
\multicolumn{10}{l}{\textit{Setting C}}\\[2pt]
T-GAM & 0.1315 & 0.755 & 0.779 & 0.1830 & 0.731 & 0.767 & 0.2025 & 0.708 & 0.753\\
Source GAM & 0.1287 & 0.769 & 0.793 & 0.1784 & 0.742 & 0.780 & 0.1946 & 0.724 & 0.772\\
Stratified GAM & 0.1281 & \textbf{0.769} & \textbf{0.794} & 0.1784 & 0.743 & 0.781 & \textbf{0.1945} & \textbf{0.724} & \textbf{0.773}\\
\textbf{SPT} & \textbf{0.1279} & 0.769 & 0.794 & \textbf{0.1780} & \textbf{0.743} & \textbf{0.781} & 0.1947 & 0.724 & 0.773\\
CoxTL & 0.1353 & 0.749 & 0.772 & 0.1875 & 0.721 & 0.756 & 0.2045 & 0.702 & 0.746\\
POTL-Cox & 0.1356 & 0.748 & 0.771 & 0.1877 & 0.721 & 0.756 & 0.2045 & 0.702 & 0.746\\
RSF & 0.1368 & 0.740 & 0.762 & 0.1918 & 0.715 & 0.750 & 0.2098 & 0.696 & 0.739\\
\textit{Oracle} & \textit{0.1259} & \textit{0.775} & \textit{0.800} & \textit{0.1748} & \textit{0.749} & \textit{0.788} & \textit{0.1905} & \textit{0.730} & \textit{0.781}\\
\midrule
\multicolumn{10}{l}{\textit{Setting D}}\\[2pt]
T-GAM & 0.1399 & 0.718 & 0.738 & 0.1955 & 0.699 & 0.734 & 0.2116 & 0.681 & 0.728\\
Source GAM & 0.1477 & 0.712 & 0.732 & 0.2033 & 0.692 & 0.726 & 0.2167 & 0.679 & 0.724\\
Stratified GAM & 0.1425 & 0.717 & 0.737 & 0.1991 & 0.697 & 0.732 & 0.2128 & 0.683 & 0.730\\
\textbf{SPT} & \textbf{0.1378} & \textbf{0.728} & \textbf{0.750} & \textbf{0.1920} & \textbf{0.709} & \textbf{0.746} & \textbf{0.2056} & \textbf{0.695} & \textbf{0.745}\\
CoxTL & 0.1411 & 0.713 & 0.734 & 0.1967 & 0.695 & 0.730 & 0.2104 & 0.682 & 0.730\\
POTL-Cox & 0.1408 & 0.714 & 0.734 & 0.1963 & 0.695 & 0.730 & 0.2099 & 0.682 & 0.731\\
RSF & 0.1441 & 0.698 & 0.717 & 0.2024 & 0.683 & 0.715 & 0.2174 & 0.670 & 0.714\\
\textit{Oracle} & \textit{0.1344} & \textit{0.740} & \textit{0.762} & \textit{0.1868} & \textit{0.720} & \textit{0.758} & \textit{0.1997} & \textit{0.705} & \textit{0.758}\\
\bottomrule\end{tabular}}
\typeout{MAIN TABLE tab:loglogistic NATURAL WIDTH=\the\wd\maintablebox; TEXT WIDTH=\the\textwidth}
\ifdim\wd\maintablebox>\textwidth\resizebox{\textwidth}{!}{\usebox{\maintablebox}}\else\usebox{\maintablebox}\fi
\endgroup
\par\smallskip\begin{minipage}{\textwidth}\footnotesize
Lower Brier and higher $C_U(h)$/AUC are better. Q25/Q50/Q75 denote quantiles of observed target-training follow-up $Y=\min(T^*,C)$, chosen separately in each replication. All metrics use censored test outcomes. Target-training/test censoring rates are 50\%/30\%. Source and target GAM structures are selected by training-only GCV and fitted by REML. SPT uses the one-standard-error rule with reference $\tau=0.5$. The cohort-intercept pooled GAM shares smooth effects.
\end{minipage}\end{table}

%% file: table/main_table_02.tex
\begin{table}[!p]
\centering
\caption{Weibull simulation results: $n_S=3000$, $n_T=500$, $p=7$. Means over 100 replications; corresponding MCSEs are reported in the Supplementary Material. The best fitted-method mean in each setting and column is bold; Oracle is excluded.}\label{tab:weibull}
\begingroup\fontsize{9}{11}\selectfont\setlength{\tabcolsep}{2.5pt}\renewcommand{\arraystretch}{1.12}
\sbox{\maintablebox}{%
\begin{tabular}{@{}l*{9}{c}@{}}\toprule
 & \multicolumn{3}{c}{Q25} & \multicolumn{3}{c}{Q50} & \multicolumn{3}{c}{Q75}\\
\cmidrule(lr){2-4}\cmidrule(lr){5-7}\cmidrule(lr){8-10}
Method & Brier & $C_U(h)$ & AUC & Brier & $C_U(h)$ & AUC & Brier & $C_U(h)$ & AUC\\
\midrule
\multicolumn{10}{l}{\textit{Setting A}}\\[2pt]
T-GAM & 0.1336 & 0.747 & 0.774 & 0.1748 & 0.736 & 0.781 & 0.1879 & 0.724 & 0.785\\
Source GAM & \textbf{0.1294} & 0.760 & 0.788 & \textbf{0.1685} & \textbf{0.748} & \textbf{0.796} & \textbf{0.1794} & \textbf{0.738} & \textbf{0.802}\\
Stratified GAM & 0.1296 & \textbf{0.760} & \textbf{0.788} & 0.1690 & 0.748 & 0.796 & 0.1800 & 0.737 & 0.802\\
\textbf{SPT} & 0.1299 & 0.759 & 0.787 & 0.1694 & 0.747 & 0.795 & 0.1808 & 0.736 & 0.801\\
CoxTL & 0.1380 & 0.735 & 0.761 & 0.1833 & 0.719 & 0.761 & 0.1984 & 0.706 & 0.760\\
POTL-Cox & 0.1380 & 0.735 & 0.760 & 0.1831 & 0.719 & 0.760 & 0.1984 & 0.706 & 0.759\\
RSF & 0.1385 & 0.733 & 0.758 & 0.1835 & 0.722 & 0.764 & 0.1978 & 0.710 & 0.767\\
\textit{Oracle} & \textit{0.1281} & \textit{0.763} & \textit{0.791} & \textit{0.1666} & \textit{0.751} & \textit{0.799} & \textit{0.1773} & \textit{0.741} & \textit{0.806}\\
\midrule
\multicolumn{10}{l}{\textit{Setting B}}\\[2pt]
T-GAM & 0.1418 & 0.708 & 0.731 & 0.1888 & 0.704 & 0.747 & 0.2007 & 0.695 & 0.755\\
Source GAM & 0.1484 & 0.694 & 0.715 & 0.1991 & 0.685 & 0.724 & 0.2123 & 0.677 & 0.730\\
Stratified GAM & 0.1439 & 0.701 & 0.723 & 0.1940 & 0.693 & 0.733 & 0.2074 & 0.684 & 0.740\\
\textbf{SPT} & \textbf{0.1405} & \textbf{0.713} & \textbf{0.737} & \textbf{0.1867} & \textbf{0.709} & \textbf{0.752} & \textbf{0.1982} & \textbf{0.700} & \textbf{0.761}\\
CoxTL & 0.1460 & 0.684 & 0.704 & 0.1995 & 0.674 & 0.710 & 0.2154 & 0.665 & 0.713\\
POTL-Cox & 0.1459 & 0.684 & 0.704 & 0.1993 & 0.675 & 0.710 & 0.2151 & 0.665 & 0.714\\
RSF & 0.1460 & 0.693 & 0.714 & 0.1967 & 0.691 & 0.730 & 0.2091 & 0.684 & 0.741\\
\textit{Oracle} & \textit{0.1362} & \textit{0.730} & \textit{0.755} & \textit{0.1797} & \textit{0.723} & \textit{0.769} & \textit{0.1881} & \textit{0.716} & \textit{0.784}\\
\midrule
\multicolumn{10}{l}{\textit{Setting C}}\\[2pt]
T-GAM & 0.1410 & 0.714 & 0.738 & 0.1872 & 0.709 & 0.751 & 0.1989 & 0.700 & 0.761\\
Source GAM & 0.1380 & 0.729 & 0.754 & 0.1823 & 0.720 & 0.765 & 0.1915 & 0.713 & 0.778\\
Stratified GAM & \textbf{0.1373} & \textbf{0.730} & \textbf{0.755} & 0.1820 & 0.721 & 0.766 & \textbf{0.1914} & 0.714 & 0.778\\
\textbf{SPT} & 0.1374 & 0.729 & 0.754 & \textbf{0.1817} & \textbf{0.721} & \textbf{0.766} & 0.1915 & \textbf{0.714} & \textbf{0.779}\\
CoxTL & 0.1416 & 0.713 & 0.737 & 0.1893 & 0.704 & 0.745 & 0.2009 & 0.695 & 0.754\\
POTL-Cox & 0.1417 & 0.713 & 0.736 & 0.1895 & 0.703 & 0.744 & 0.2010 & 0.695 & 0.753\\
RSF & 0.1447 & 0.703 & 0.725 & 0.1942 & 0.698 & 0.738 & 0.2066 & 0.691 & 0.748\\
\textit{Oracle} & \textit{0.1351} & \textit{0.735} & \textit{0.761} & \textit{0.1780} & \textit{0.727} & \textit{0.773} & \textit{0.1866} & \textit{0.720} & \textit{0.787}\\
\midrule
\multicolumn{10}{l}{\textit{Setting D}}\\[2pt]
T-GAM & 0.1482 & 0.674 & 0.694 & 0.1998 & 0.677 & 0.716 & 0.2098 & 0.673 & 0.733\\
Source GAM & 0.1544 & 0.676 & 0.696 & 0.2082 & 0.672 & 0.709 & 0.2164 & 0.668 & 0.727\\
Stratified GAM & 0.1494 & 0.681 & 0.700 & 0.2033 & 0.676 & 0.714 & 0.2122 & 0.673 & 0.733\\
\textbf{SPT} & \textbf{0.1457} & \textbf{0.689} & \textbf{0.710} & \textbf{0.1961} & \textbf{0.687} & \textbf{0.728} & \textbf{0.2051} & \textbf{0.683} & \textbf{0.747}\\
CoxTL & 0.1474 & 0.679 & 0.699 & 0.2000 & 0.676 & 0.714 & 0.2093 & 0.673 & 0.733\\
POTL-Cox & 0.1472 & 0.680 & 0.700 & 0.1995 & 0.676 & 0.715 & 0.2086 & 0.673 & 0.734\\
RSF & 0.1505 & 0.663 & 0.681 & 0.2056 & 0.664 & 0.700 & 0.2165 & 0.663 & 0.720\\
\textit{Oracle} & \textit{0.1425} & \textit{0.701} & \textit{0.723} & \textit{0.1907} & \textit{0.698} & \textit{0.740} & \textit{0.1978} & \textit{0.695} & \textit{0.762}\\
\bottomrule\end{tabular}}
\typeout{MAIN TABLE tab:weibull NATURAL WIDTH=\the\wd\maintablebox; TEXT WIDTH=\the\textwidth}
\ifdim\wd\maintablebox>\textwidth\resizebox{\textwidth}{!}{\usebox{\maintablebox}}\else\usebox{\maintablebox}\fi
\endgroup
\par\smallskip\begin{minipage}{\textwidth}\footnotesize
Lower Brier and higher $C_U(h)$/AUC are better. Q25/Q50/Q75 denote quantiles of observed target-training follow-up $Y=\min(T^*,C)$, chosen separately in each replication. All metrics use censored test outcomes. Target-training/test censoring rates are 50\%/30\%. Source and target GAM structures are selected by training-only GCV and fitted by REML. SPT uses the one-standard-error rule with reference $\tau=0.5$. The cohort-intercept pooled GAM shares smooth effects.
\end{minipage}\end{table}

%% file: table/supp_table_01.tex
\begin{table}[p]\centering
\caption{Component comparisons under source and structural mismatch: paired changes relative to T-GAM.}\label{tab:source_mismatch}
\small\setlength{\tabcolsep}{4pt}\renewcommand{\arraystretch}{1.1}
\begin{tabular}{llrrr}\toprule
Setting & Method & Brier (\%) & $\Delta C_U(h)$ (pp) & $\Delta$AUC (pp)\\\midrule
Level & SPT & -3.13 (0.11) & +1.23 (0.05) & +1.45 (0.06)\\
Level & PR-GAM & -1.13 (0.04) & +0.40 (0.02) & +0.47 (0.02)\\
Level & Fixed center ($\rho=1$) & -3.10 (0.12) & +1.25 (0.05) & +1.47 (0.06)\\
Level & Uncalibrated & -1.41 (0.14) & +1.05 (0.05) & +1.24 (0.05)\\
Level & No soft labels ($\tau=0$) & -2.06 (0.11) & +0.88 (0.05) & +1.03 (0.05)\\
Level & DST & -3.12 (0.11) & +1.23 (0.05) & +1.45 (0.06)\\
\addlinespace[3pt]
Shape & SPT & -1.70 (0.17) & +0.83 (0.06) & +0.99 (0.07)\\
Shape & PR-GAM & -0.37 (0.06) & +0.17 (0.02) & +0.21 (0.03)\\
Shape & Fixed center ($\rho=1$) & +0.82 (0.11) & +0.39 (0.05) & +0.51 (0.06)\\
Shape & Uncalibrated & -1.70 (0.17) & +0.83 (0.06) & +0.99 (0.07)\\
Shape & No soft labels ($\tau=0$) & -1.87 (0.12) & +0.82 (0.05) & +0.97 (0.05)\\
Shape & DST & -1.69 (0.17) & +0.83 (0.06) & +0.99 (0.07)\\
\addlinespace[3pt]
Reverse & SPT & +0.31 (0.03) & -0.03 (0.00) & -0.03 (0.00)\\
Reverse & PR-GAM & +0.31 (0.03) & -0.03 (0.00) & -0.03 (0.00)\\
Reverse & Fixed center ($\rho=1$) & +7.04 (0.34) & -2.94 (0.15) & -3.49 (0.18)\\
Reverse & Uncalibrated & +0.32 (0.05) & -0.03 (0.00) & -0.04 (0.01)\\
Reverse & No soft labels ($\tau=0$) & +0.00 (0.00) & +0.00 (0.00) & +0.00 (0.00)\\
Reverse & DST & +0.31 (0.03) & -0.03 (0.00) & -0.03 (0.00)\\
\addlinespace[3pt]
Structural (D) & SPT & -1.96 (0.12) & +1.16 (0.06) & +1.37 (0.07)\\
Structural (D) & PR-GAM & -0.69 (0.06) & +0.37 (0.03) & +0.44 (0.03)\\
Structural (D) & Fixed center ($\rho=1$) & -1.13 (0.15) & +0.93 (0.07) & +1.07 (0.09)\\
Structural (D) & Uncalibrated & -2.04 (0.12) & +1.16 (0.06) & +1.37 (0.07)\\
Structural (D) & No soft labels ($\tau=0$) & -1.45 (0.10) & +0.87 (0.05) & +1.01 (0.06)\\
Structural (D) & DST & -0.98 (0.10) & +0.68 (0.06) & +0.79 (0.07)\\
\bottomrule\end{tabular}
\par\smallskip\begin{minipage}{0.97\linewidth}\footnotesize
Entries are mean (paired MCSE). Within each replication, average the six distribution--horizon contrasts, then summarize the 100 replication blocks. Brier changes are $100(\widehat{\mathrm{BS}}_m/\widehat{\mathrm{BS}}_{\mathrm{TG}}-1)$; discrimination changes are percentage points. The first three scenarios retain fixed interaction formulas. Structural mismatch uses setting D with training-only GCV formula selection followed by REML fitting. Within each scenario, all methods share their source fit and target working space. Automatic variants use the same reference-weight rule. The soft-label-free variant fixes $\tau=0$; uncalibrated soft labels omit intercept recalibration; DST borrows the complete source score.
\end{minipage}
\end{table}

%% file: table/supp_table_02.tex
\begin{table}[htbp]\centering\caption{Baseline characteristics and horizon-specific outcome support.}\label{tab:cohorts}
\small\begin{tabular}{lrr}\toprule
Characteristic & UK Biobank & SPRINT standard\\\midrule
Participants & 7059 & 1126\\
Male (\%) & 46.7 & 58.7\\
Ever smoked (\%) & 51.6 & 54.4\\
Age (years) & 62.7 (6.0) & 72.5 (9.5)\\
BMI (kg/m$^2$) & 29.4 (5.4) & 29.4 (5.8)\\
eGFR (mL/min/1.73 m$^2$) & 51.5 (8.1) & 47.2 (9.4)\\
Total cholesterol (mg/dL) & 203.6 (49.3) & 184.3 (39.0)\\
Creatinine (mg/dL) & 1.4 (0.3) & 1.5 (0.3)\\
\midrule\multicolumn{3}{l}{24 months}\\
Observed deaths & 140 & 39\\
Observed event-free controls & 6913 & 1052\\
Censored before horizon & 6 & 35\\
$\widehat G(h)$ & 0.999 & 0.968\\
\midrule\multicolumn{3}{l}{30 months}\\
Observed deaths & 178 & 53\\
Observed event-free controls & 6875 & 965\\
Censored before horizon & 6 & 108\\
$\widehat G(h)$ & 0.999 & 0.900\\
\midrule\multicolumn{3}{l}{36 months}\\
Observed deaths & 231 & 67\\
Observed event-free controls & 6822 & 710\\
Censored before horizon & 6 & 349\\
$\widehat G(h)$ & 0.999 & 0.674\\
\bottomrule\end{tabular}\par\smallskip\footnotesize Continuous variables: mean (SD). Eligibility uses a single baseline eGFR measurement.\end{table}

%% file: table/supp_table_03.tex
\begin{sidewaystable}[p]\centering
\caption{UK Biobank to SPRINT standard treatment: mortality prediction at 24, 30 and 36 months.}\label{tab:realdata}
\scriptsize\setlength{\tabcolsep}{1.8pt}\renewcommand{\arraystretch}{1.2}
\sbox{\maintablebox}{\begin{tabular}{l*{9}{r}}\toprule
 & \multicolumn{3}{c}{24 months} & \multicolumn{3}{c}{30 months} & \multicolumn{3}{c}{36 months}\\
\cmidrule(lr){2-4}\cmidrule(lr){5-7}\cmidrule(lr){8-10}
Method & Brier & $C_U(h)$ & AUC & Brier & $C_U(h)$ & AUC & Brier & $C_U(h)$ & AUC\\\midrule
T-GAM & 0.03396 (0.00028) & 0.686 (0.02440) & 0.688 (0.02454) & 0.04395 (0.00046) & 0.708 (0.01174) & 0.718 (0.01211) & 0.05773 (0.00061) & 0.678 (0.01199) & 0.693 (0.01268)\\
Source GAM & 0.03435 (0.00000) & 0.672 (0.00000) & 0.674 (0.00000) & 0.04570 (0.00000) & 0.607 (0.00000) & 0.613 (0.00000) & 0.05933 (0.00000) & 0.692 (0.00000) & 0.703 (0.00000)\\
Pooled GAM & 0.03378 (0.00008) & 0.664 (0.00461) & 0.667 (0.00472) & 0.04456 (0.00025) & 0.691 (0.00610) & 0.702 (0.00622) & 0.05781 (0.00037) & 0.685 (0.00503) & 0.695 (0.00518)\\
Stratified GAM & 0.03379 (0.00012) & 0.662 (0.00521) & 0.665 (0.00533) & 0.04424 (0.00023) & 0.694 (0.00581) & 0.705 (0.00598) & 0.05759 (0.00034) & 0.686 (0.00493) & 0.699 (0.00515)\\
SPT-GAM & 0.03369 (0.00018) & \textbf{0.701 (0.01495)} & \textbf{0.705 (0.01511)} & 0.04413 (0.00030) & 0.710 (0.00917) & 0.720 (0.00949) & 0.05761 (0.00039) & \textbf{0.693 (0.00727)} & \textbf{0.706 (0.00762)}\\
SPT-Cox & 0.03361 (0.00015) & 0.700 (0.01704) & 0.702 (0.01719) & \textbf{0.04383 (0.00028)} & 0.710 (0.00932) & 0.721 (0.00955) & \textbf{0.05732 (0.00034)} & 0.688 (0.00811) & 0.702 (0.00853)\\
SPT-NN & 0.03366 (0.00016) & 0.694 (0.01515) & 0.697 (0.01533) & 0.04394 (0.00029) & \textbf{0.713 (0.00841)} & \textbf{0.724 (0.00869)} & 0.05759 (0.00037) & 0.687 (0.00822) & 0.699 (0.00859)\\
Score recalibration & \textbf{0.03356 (0.00009)} & 0.661 (0.00661) & 0.663 (0.00674) & 0.04555 (0.00009) & 0.596 (0.00664) & 0.601 (0.00689) & 0.05754 (0.00022) & 0.687 (0.00331) & 0.698 (0.00358)\\
Target Cox & 0.03416 (0.00023) & 0.631 (0.01135) & 0.632 (0.01155) & 0.04453 (0.00033) & 0.674 (0.00994) & 0.685 (0.01022) & 0.05829 (0.00046) & 0.660 (0.00864) & 0.675 (0.00899)\\
CoxTL & 0.03363 (0.00026) & 0.651 (0.01084) & 0.653 (0.01114) & 0.04402 (0.00033) & 0.688 (0.00801) & 0.699 (0.00822) & 0.05749 (0.00053) & 0.675 (0.00863) & 0.687 (0.00849)\\
POTL-Cox & 0.03375 (0.00014) & 0.658 (0.00804) & 0.660 (0.00823) & 0.04404 (0.00020) & 0.694 (0.00552) & 0.705 (0.00576) & 0.05762 (0.00029) & 0.679 (0.00500) & 0.691 (0.00506)\\
CoxKL & 0.03356 (0.00013) & 0.657 (0.00656) & 0.660 (0.00671) & 0.04394 (0.00016) & 0.694 (0.00499) & 0.705 (0.00513) & 0.05737 (0.00025) & 0.680 (0.00474) & 0.692 (0.00462)\\
TransCox & 0.03407 (0.00025) & 0.634 (0.01204) & 0.636 (0.01230) & 0.04451 (0.00033) & 0.675 (0.01040) & 0.685 (0.01074) & 0.05826 (0.00049) & 0.660 (0.00929) & 0.675 (0.00969)\\
DiscreteKL & 0.03374 (0.00013) & 0.660 (0.00658) & 0.663 (0.00671) & 0.04400 (0.00017) & 0.697 (0.00535) & 0.708 (0.00550) & 0.05759 (0.00027) & 0.681 (0.00513) & 0.692 (0.00520)\\
Trans-AFT & 0.03405 (0.00011) & 0.544 (0.02805) & 0.545 (0.02843) & 0.04576 (0.00013) & 0.572 (0.02122) & 0.575 (0.02208) & 0.05988 (0.00017) & 0.574 (0.01904) & 0.569 (0.02024)\\
RSF & 0.03432 (0.00020) & 0.590 (0.01894) & 0.592 (0.01937) & 0.04493 (0.00037) & 0.647 (0.01357) & 0.655 (0.01388) & 0.05879 (0.00055) & 0.654 (0.01341) & 0.670 (0.01436)\\
IPCW-XGBoost & 0.03402 (0.00021) & 0.638 (0.02354) & 0.639 (0.02393) & 0.04508 (0.00038) & 0.671 (0.01522) & 0.680 (0.01582) & 0.05871 (0.00053) & 0.645 (0.01648) & 0.659 (0.01766)\\
\bottomrule\end{tabular}}
\ifdim\wd\maintablebox>\linewidth
\resizebox{\linewidth}{!}{\usebox{\maintablebox}}
\else\usebox{\maintablebox}\fi
\par\smallskip\begin{minipage}{0.95\linewidth}\footnotesize All metrics use IPCW. Mean (partition SD) over 100 repeated five-fold evaluations on the same 1126 participants. SDs describe partition variability on this cohort. Bold denotes the best mean among displayed methods. Source GAM is fitted once per horizon and shared across target partitions, giving zero partition SD. SPT-GAM, SPT-Cox and SPT-NN use GAM, Cox and neural-network source predictions, respectively. All retain the target ts/$k=5$ GAM and the one-standard-error rule with reference $\tau=0.5$; the source GAM uses ts/$k=10$. The cohort-intercept pooled GAM shares smooth effects.\end{minipage}\end{sidewaystable}

%% file: table/supp_table_04.tex
\begin{sidewaystable}[p]
\centering
\caption{Loglogistic: $n_T=500$, $p=7$, all reported comparators. Settings A/B.}\label{tab:allG2500loglogistic0}
\scriptsize\setlength{\tabcolsep}{1.3pt}\renewcommand{\arraystretch}{1.08}
\begin{tabular}{l*{9}{r}}\toprule
 & \multicolumn{3}{c}{Q25} & \multicolumn{3}{c}{Q50} & \multicolumn{3}{c}{Q75}\\
Method & Brier & $C_U(h)$ & AUC & Brier & $C_U(h)$ & AUC & Brier & $C_U(h)$ & AUC\\\midrule
\multicolumn{10}{l}{\textbf{Setting A}}\\[2pt]
T-GAM & 0.1238 (0.0006) & 0.786 (0.0010) & 0.811 (0.0011) & 0.1721 (0.0005) & 0.756 (0.0009) & 0.793 (0.0011) & 0.1933 (0.0005) & 0.730 (0.0010) & 0.775 (0.0013)\\
Source GAM & 0.1197 (0.0006) & \textbf{0.799 (0.0008)} & \textbf{0.825 (0.0008)} & \textbf{0.1658 (0.0004)} & \textbf{0.769 (0.0006)} & \textbf{0.808 (0.0007)} & \textbf{0.1839 (0.0002)} & \textbf{0.747 (0.0005)} & \textbf{0.795 (0.0005)}\\
Pooled GAM & \textbf{0.1197 (0.0006)} & 0.799 (0.0008) & 0.825 (0.0008) & 0.1660 (0.0004) & 0.769 (0.0006) & 0.808 (0.0007) & 0.1840 (0.0002) & 0.747 (0.0005) & 0.795 (0.0005)\\
Stratified GAM & 0.1200 (0.0006) & 0.799 (0.0008) & 0.825 (0.0008) & 0.1663 (0.0004) & 0.769 (0.0006) & 0.808 (0.0007) & 0.1844 (0.0002) & 0.747 (0.0005) & 0.795 (0.0005)\\
SPT & 0.1203 (0.0006) & 0.798 (0.0008) & 0.824 (0.0008) & 0.1669 (0.0004) & 0.768 (0.0006) & 0.807 (0.0007) & 0.1853 (0.0003) & 0.746 (0.0005) & 0.794 (0.0006)\\
Target Cox & 0.1306 (0.0007) & 0.767 (0.0009) & 0.790 (0.0010) & 0.1824 (0.0004) & 0.732 (0.0007) & 0.765 (0.0008) & 0.2050 (0.0004) & 0.708 (0.0006) & 0.746 (0.0008)\\
CoxTL & 0.1323 (0.0007) & 0.769 (0.0009) & 0.793 (0.0009) & 0.1829 (0.0004) & 0.734 (0.0006) & 0.768 (0.0007) & 0.2030 (0.0003) & 0.711 (0.0005) & 0.750 (0.0007)\\
POTL-Cox & 0.1324 (0.0008) & 0.768 (0.0009) & 0.792 (0.0010) & 0.1826 (0.0004) & 0.734 (0.0006) & 0.767 (0.0007) & 0.2031 (0.0003) & 0.711 (0.0005) & 0.749 (0.0007)\\
CoxKL & 0.1321 (0.0007) & 0.768 (0.0009) & 0.792 (0.0009) & 0.1829 (0.0004) & 0.734 (0.0006) & 0.767 (0.0007) & 0.2032 (0.0003) & 0.711 (0.0005) & 0.750 (0.0007)\\
TransCox & 0.1306 (0.0007) & 0.767 (0.0009) & 0.790 (0.0010) & 0.1823 (0.0004) & 0.732 (0.0007) & 0.765 (0.0008) & 0.2048 (0.0004) & 0.709 (0.0006) & 0.746 (0.0008)\\
DiscreteKL & 0.1316 (0.0007) & 0.769 (0.0009) & 0.792 (0.0010) & 0.1817 (0.0004) & 0.734 (0.0006) & 0.768 (0.0007) & 0.2028 (0.0003) & 0.711 (0.0005) & 0.750 (0.0007)\\
Trans-AFT & 0.1395 (0.0009) & 0.745 (0.0025) & 0.767 (0.0027) & 0.1940 (0.0008) & 0.712 (0.0022) & 0.743 (0.0024) & 0.2142 (0.0007) & 0.690 (0.0019) & 0.725 (0.0023)\\
RSF & 0.1298 (0.0007) & 0.770 (0.0009) & 0.794 (0.0010) & 0.1820 (0.0005) & 0.739 (0.0008) & 0.773 (0.0009) & 0.2025 (0.0005) & 0.716 (0.0008) & 0.757 (0.0011)\\
IPCW-XGBoost & 0.1287 (0.0007) & 0.771 (0.0010) & 0.795 (0.0011) & 0.1794 (0.0005) & 0.741 (0.0008) & 0.776 (0.0010) & 0.2048 (0.0006) & 0.707 (0.0013) & 0.748 (0.0016)\\
\textit{Oracle} & \textit{0.1187 (0.0006)} & \textit{0.802 (0.0008)} & \textit{0.829 (0.0008)} & \textit{0.1645 (0.0004)} & \textit{0.772 (0.0006)} & \textit{0.811 (0.0007)} & \textit{0.1822 (0.0002)} & \textit{0.750 (0.0005)} & \textit{0.799 (0.0005)}\\
\addlinespace[4pt]
\multicolumn{10}{l}{\textbf{Setting B}}\\[2pt]
T-GAM & 0.1330 (0.0008) & 0.749 (0.0012) & 0.772 (0.0013) & 0.1848 (0.0005) & 0.726 (0.0011) & 0.763 (0.0013) & 0.2042 (0.0005) & 0.703 (0.0011) & 0.748 (0.0013)\\
Source GAM & 0.1418 (0.0008) & 0.728 (0.0011) & 0.749 (0.0012) & 0.1969 (0.0005) & 0.702 (0.0008) & 0.734 (0.0009) & 0.2168 (0.0004) & 0.681 (0.0007) & 0.719 (0.0010)\\
Pooled GAM & 0.1377 (0.0008) & 0.737 (0.0011) & 0.758 (0.0011) & 0.1916 (0.0005) & 0.710 (0.0008) & 0.744 (0.0009) & 0.2111 (0.0004) & 0.690 (0.0007) & 0.730 (0.0009)\\
Stratified GAM & 0.1367 (0.0008) & 0.737 (0.0011) & 0.758 (0.0011) & 0.1916 (0.0005) & 0.710 (0.0008) & 0.744 (0.0009) & 0.2117 (0.0004) & 0.689 (0.0007) & 0.730 (0.0009)\\
SPT & \textbf{0.1318 (0.0008)} & \textbf{0.754 (0.0011)} & \textbf{0.777 (0.0012)} & \textbf{0.1829 (0.0005)} & \textbf{0.730 (0.0009)} & \textbf{0.768 (0.0012)} & \textbf{0.2020 (0.0005)} & \textbf{0.708 (0.0010)} & \textbf{0.753 (0.0012)}\\
Target Cox & 0.1405 (0.0008) & 0.715 (0.0009) & 0.735 (0.0010) & 0.1982 (0.0004) & 0.689 (0.0008) & 0.719 (0.0009) & 0.2191 (0.0004) & 0.670 (0.0007) & 0.705 (0.0009)\\
CoxTL & 0.1405 (0.0008) & 0.715 (0.0010) & 0.735 (0.0010) & 0.1981 (0.0004) & 0.689 (0.0008) & 0.718 (0.0010) & 0.2191 (0.0004) & 0.669 (0.0007) & 0.704 (0.0010)\\
POTL-Cox & 0.1405 (0.0008) & 0.716 (0.0009) & 0.735 (0.0010) & 0.1980 (0.0004) & 0.689 (0.0008) & 0.718 (0.0009) & 0.2189 (0.0004) & 0.669 (0.0007) & 0.704 (0.0010)\\
CoxKL & 0.1406 (0.0008) & 0.715 (0.0009) & 0.735 (0.0010) & 0.1981 (0.0004) & 0.688 (0.0008) & 0.718 (0.0009) & 0.2191 (0.0004) & 0.669 (0.0007) & 0.703 (0.0010)\\
TransCox & 0.1405 (0.0008) & 0.715 (0.0009) & 0.735 (0.0010) & 0.1981 (0.0004) & 0.689 (0.0008) & 0.719 (0.0009) & 0.2190 (0.0004) & 0.670 (0.0007) & 0.705 (0.0009)\\
DiscreteKL & 0.1402 (0.0008) & 0.714 (0.0009) & 0.733 (0.0010) & 0.1986 (0.0004) & 0.687 (0.0008) & 0.715 (0.0010) & 0.2214 (0.0004) & 0.667 (0.0007) & 0.700 (0.0011)\\
Trans-AFT & 0.1470 (0.0009) & 0.688 (0.0025) & 0.706 (0.0027) & 0.2081 (0.0007) & 0.665 (0.0022) & 0.691 (0.0025) & 0.2271 (0.0006) & 0.647 (0.0020) & 0.679 (0.0025)\\
RSF & 0.1385 (0.0008) & 0.731 (0.0012) & 0.752 (0.0013) & 0.1939 (0.0005) & 0.709 (0.0009) & 0.743 (0.0011) & 0.2120 (0.0004) & 0.690 (0.0009) & 0.732 (0.0012)\\
IPCW-XGBoost & 0.1375 (0.0008) & 0.730 (0.0013) & 0.751 (0.0014) & 0.1924 (0.0006) & 0.708 (0.0011) & 0.742 (0.0014) & 0.2150 (0.0006) & 0.678 (0.0015) & 0.718 (0.0018)\\
\textit{Oracle} & \textit{0.1275 (0.0007)} & \textit{0.770 (0.0008)} & \textit{0.795 (0.0008)} & \textit{0.1761 (0.0004)} & \textit{0.745 (0.0007)} & \textit{0.785 (0.0008)} & \textit{0.1921 (0.0003)} & \textit{0.727 (0.0005)} & \textit{0.777 (0.0007)}\\
\addlinespace[4pt]
\bottomrule\end{tabular}
\par\smallskip\begin{minipage}{0.97\linewidth}\footnotesize
Entries are mean (MCSE), using 100 replications for every method and horizon. Bold identifies the best fitted-method mean within the displayed set; Oracle is excluded. Q25/Q50/Q75 are target-training observed-follow-up quantiles. SPT uses calibrated soft labels and the one-standard-error rule with reference $\tau=0.5$.
GCV selects source and target GAM structures before REML fitting. All three metrics use censored test outcomes. Target-training/test censoring rates are 50\%/30\%.
\end{minipage}\end{sidewaystable}

%% file: table/supp_table_05.tex
\begin{sidewaystable}[p]
\centering
\caption{Loglogistic: $n_T=500$, $p=7$, all reported comparators. Settings C/D.}\label{tab:allG2500loglogistic1}
\scriptsize\setlength{\tabcolsep}{1.3pt}\renewcommand{\arraystretch}{1.08}
\begin{tabular}{l*{9}{r}}\toprule
 & \multicolumn{3}{c}{Q25} & \multicolumn{3}{c}{Q50} & \multicolumn{3}{c}{Q75}\\
Method & Brier & $C_U(h)$ & AUC & Brier & $C_U(h)$ & AUC & Brier & $C_U(h)$ & AUC\\\midrule
\multicolumn{10}{l}{\textbf{Setting C}}\\[2pt]
T-GAM & 0.1315 (0.0007) & 0.755 (0.0012) & 0.779 (0.0012) & 0.1830 (0.0005) & 0.731 (0.0010) & 0.767 (0.0011) & 0.2025 (0.0005) & 0.708 (0.0011) & 0.753 (0.0014)\\
Source GAM & 0.1287 (0.0007) & 0.769 (0.0008) & 0.793 (0.0009) & 0.1784 (0.0005) & 0.742 (0.0007) & 0.780 (0.0007) & 0.1946 (0.0003) & 0.724 (0.0005) & 0.772 (0.0006)\\
Pooled GAM & 0.1281 (0.0007) & 0.769 (0.0008) & 0.794 (0.0008) & \textbf{0.1779 (0.0004)} & 0.743 (0.0007) & 0.781 (0.0007) & \textbf{0.1941 (0.0003)} & \textbf{0.724 (0.0005)} & \textbf{0.773 (0.0006)}\\
Stratified GAM & 0.1281 (0.0007) & \textbf{0.769 (0.0008)} & \textbf{0.794 (0.0008)} & 0.1784 (0.0004) & 0.743 (0.0007) & 0.781 (0.0007) & 0.1945 (0.0003) & 0.724 (0.0005) & 0.773 (0.0006)\\
SPT & \textbf{0.1279 (0.0007)} & 0.769 (0.0008) & 0.794 (0.0009) & 0.1780 (0.0004) & \textbf{0.743 (0.0007)} & \textbf{0.781 (0.0008)} & 0.1947 (0.0003) & 0.724 (0.0006) & 0.773 (0.0007)\\
Target Cox & 0.1352 (0.0007) & 0.745 (0.0009) & 0.767 (0.0010) & 0.1885 (0.0004) & 0.717 (0.0007) & 0.752 (0.0008) & 0.2068 (0.0003) & 0.698 (0.0006) & 0.741 (0.0008)\\
CoxTL & 0.1353 (0.0007) & 0.749 (0.0008) & 0.772 (0.0009) & 0.1875 (0.0004) & 0.721 (0.0006) & 0.756 (0.0007) & 0.2045 (0.0003) & 0.702 (0.0005) & 0.746 (0.0007)\\
POTL-Cox & 0.1356 (0.0007) & 0.748 (0.0009) & 0.771 (0.0009) & 0.1877 (0.0004) & 0.721 (0.0007) & 0.756 (0.0008) & 0.2045 (0.0003) & 0.702 (0.0006) & 0.746 (0.0007)\\
CoxKL & 0.1357 (0.0008) & 0.748 (0.0009) & 0.771 (0.0009) & 0.1880 (0.0004) & 0.721 (0.0006) & 0.756 (0.0007) & 0.2047 (0.0003) & 0.702 (0.0005) & 0.746 (0.0007)\\
TransCox & 0.1352 (0.0007) & 0.745 (0.0009) & 0.767 (0.0010) & 0.1885 (0.0004) & 0.717 (0.0007) & 0.752 (0.0008) & 0.2067 (0.0003) & 0.699 (0.0006) & 0.742 (0.0008)\\
DiscreteKL & 0.1343 (0.0007) & 0.749 (0.0008) & 0.772 (0.0009) & 0.1866 (0.0004) & 0.721 (0.0006) & 0.756 (0.0007) & 0.2043 (0.0003) & 0.702 (0.0005) & 0.746 (0.0007)\\
Trans-AFT & 0.1408 (0.0008) & 0.728 (0.0021) & 0.749 (0.0022) & 0.1966 (0.0006) & 0.701 (0.0019) & 0.733 (0.0021) & 0.2143 (0.0006) & 0.683 (0.0017) & 0.723 (0.0020)\\
RSF & 0.1368 (0.0007) & 0.740 (0.0011) & 0.762 (0.0011) & 0.1918 (0.0005) & 0.715 (0.0009) & 0.750 (0.0010) & 0.2098 (0.0004) & 0.696 (0.0008) & 0.739 (0.0011)\\
IPCW-XGBoost & 0.1352 (0.0008) & 0.743 (0.0011) & 0.766 (0.0012) & 0.1891 (0.0005) & 0.718 (0.0010) & 0.752 (0.0012) & 0.2109 (0.0005) & 0.690 (0.0010) & 0.733 (0.0012)\\
\textit{Oracle} & \textit{0.1259 (0.0007)} & \textit{0.775 (0.0008)} & \textit{0.800 (0.0008)} & \textit{0.1748 (0.0004)} & \textit{0.749 (0.0006)} & \textit{0.788 (0.0007)} & \textit{0.1905 (0.0003)} & \textit{0.730 (0.0005)} & \textit{0.781 (0.0006)}\\
\addlinespace[4pt]
\multicolumn{10}{l}{\textbf{Setting D}}\\[2pt]
T-GAM & 0.1399 (0.0008) & 0.718 (0.0013) & 0.738 (0.0014) & 0.1955 (0.0005) & 0.699 (0.0012) & 0.734 (0.0014) & 0.2116 (0.0005) & 0.681 (0.0012) & 0.728 (0.0015)\\
Source GAM & 0.1477 (0.0008) & 0.712 (0.0010) & 0.732 (0.0010) & 0.2033 (0.0006) & 0.692 (0.0007) & 0.726 (0.0008) & 0.2167 (0.0004) & 0.679 (0.0006) & 0.724 (0.0008)\\
Pooled GAM & 0.1433 (0.0007) & 0.717 (0.0009) & 0.737 (0.0010) & 0.1987 (0.0005) & 0.697 (0.0007) & 0.732 (0.0008) & 0.2126 (0.0004) & 0.683 (0.0006) & 0.730 (0.0008)\\
Stratified GAM & 0.1425 (0.0007) & 0.717 (0.0009) & 0.737 (0.0010) & 0.1991 (0.0005) & 0.697 (0.0007) & 0.732 (0.0008) & 0.2128 (0.0004) & 0.683 (0.0006) & 0.730 (0.0008)\\
SPT & \textbf{0.1378 (0.0007)} & \textbf{0.728 (0.0010)} & \textbf{0.750 (0.0010)} & \textbf{0.1920 (0.0005)} & \textbf{0.709 (0.0009)} & \textbf{0.746 (0.0010)} & \textbf{0.2056 (0.0004)} & \textbf{0.695 (0.0007)} & \textbf{0.745 (0.0009)}\\
Target Cox & 0.1418 (0.0008) & 0.709 (0.0009) & 0.729 (0.0009) & 0.1984 (0.0004) & 0.691 (0.0008) & 0.725 (0.0009) & 0.2124 (0.0003) & 0.678 (0.0006) & 0.726 (0.0008)\\
CoxTL & 0.1411 (0.0008) & 0.713 (0.0008) & 0.734 (0.0008) & 0.1967 (0.0004) & 0.695 (0.0007) & 0.730 (0.0008) & 0.2104 (0.0003) & 0.682 (0.0006) & 0.730 (0.0008)\\
POTL-Cox & 0.1408 (0.0008) & 0.714 (0.0008) & 0.734 (0.0008) & 0.1963 (0.0004) & 0.695 (0.0007) & 0.730 (0.0008) & 0.2099 (0.0003) & 0.682 (0.0006) & 0.731 (0.0008)\\
CoxKL & 0.1412 (0.0008) & 0.714 (0.0008) & 0.734 (0.0008) & 0.1967 (0.0004) & 0.695 (0.0007) & 0.730 (0.0008) & 0.2102 (0.0003) & 0.682 (0.0006) & 0.731 (0.0008)\\
TransCox & 0.1418 (0.0008) & 0.709 (0.0009) & 0.729 (0.0009) & 0.1983 (0.0004) & 0.691 (0.0008) & 0.725 (0.0009) & 0.2123 (0.0003) & 0.678 (0.0006) & 0.726 (0.0008)\\
DiscreteKL & 0.1405 (0.0007) & 0.714 (0.0008) & 0.734 (0.0008) & 0.1963 (0.0004) & 0.695 (0.0007) & 0.730 (0.0007) & 0.2104 (0.0003) & 0.682 (0.0005) & 0.731 (0.0007)\\
Trans-AFT & 0.1459 (0.0009) & 0.686 (0.0027) & 0.704 (0.0030) & 0.2055 (0.0007) & 0.669 (0.0025) & 0.700 (0.0030) & 0.2206 (0.0008) & 0.658 (0.0024) & 0.701 (0.0031)\\
RSF & 0.1441 (0.0008) & 0.698 (0.0011) & 0.717 (0.0012) & 0.2024 (0.0004) & 0.683 (0.0010) & 0.715 (0.0011) & 0.2174 (0.0003) & 0.670 (0.0008) & 0.714 (0.0010)\\
IPCW-XGBoost & 0.1429 (0.0008) & 0.704 (0.0014) & 0.724 (0.0014) & 0.1998 (0.0005) & 0.689 (0.0011) & 0.722 (0.0013) & 0.2168 (0.0004) & 0.670 (0.0010) & 0.715 (0.0013)\\
\textit{Oracle} & \textit{0.1344 (0.0007)} & \textit{0.740 (0.0008)} & \textit{0.762 (0.0009)} & \textit{0.1868 (0.0004)} & \textit{0.720 (0.0006)} & \textit{0.758 (0.0007)} & \textit{0.1997 (0.0003)} & \textit{0.705 (0.0005)} & \textit{0.758 (0.0007)}\\
\addlinespace[4pt]
\bottomrule\end{tabular}
\par\smallskip\begin{minipage}{0.97\linewidth}\footnotesize
Entries are mean (MCSE), using 100 replications for every method and horizon. Bold identifies the best fitted-method mean within the displayed set; Oracle is excluded. Q25/Q50/Q75 are target-training observed-follow-up quantiles. SPT uses calibrated soft labels and the one-standard-error rule with reference $\tau=0.5$.
GCV selects source and target GAM structures before REML fitting. All three metrics use censored test outcomes. Target-training/test censoring rates are 50\%/30\%.
\end{minipage}\end{sidewaystable}

%% file: table/supp_table_06.tex
\begin{sidewaystable}[p]
\centering
\caption{Weibull: $n_T=500$, $p=7$, all reported comparators. Settings A/B.}\label{tab:allG2500weibull0}
\scriptsize\setlength{\tabcolsep}{1.3pt}\renewcommand{\arraystretch}{1.08}
\begin{tabular}{l*{9}{r}}\toprule
 & \multicolumn{3}{c}{Q25} & \multicolumn{3}{c}{Q50} & \multicolumn{3}{c}{Q75}\\
Method & Brier & $C_U(h)$ & AUC & Brier & $C_U(h)$ & AUC & Brier & $C_U(h)$ & AUC\\\midrule
\multicolumn{10}{l}{\textbf{Setting A}}\\[2pt]
T-GAM & 0.1336 (0.0006) & 0.747 (0.0011) & 0.774 (0.0012) & 0.1748 (0.0004) & 0.736 (0.0009) & 0.781 (0.0011) & 0.1879 (0.0005) & 0.724 (0.0009) & 0.785 (0.0011)\\
Source GAM & 0.1294 (0.0006) & 0.760 (0.0008) & 0.788 (0.0009) & \textbf{0.1685 (0.0004)} & \textbf{0.748 (0.0006)} & \textbf{0.796 (0.0007)} & \textbf{0.1794 (0.0003)} & \textbf{0.738 (0.0005)} & \textbf{0.802 (0.0006)}\\
Pooled GAM & \textbf{0.1293 (0.0006)} & 0.760 (0.0008) & 0.788 (0.0009) & 0.1686 (0.0004) & 0.748 (0.0006) & 0.796 (0.0007) & 0.1795 (0.0003) & 0.737 (0.0005) & 0.802 (0.0006)\\
Stratified GAM & 0.1296 (0.0006) & \textbf{0.760 (0.0008)} & \textbf{0.788 (0.0009)} & 0.1690 (0.0004) & 0.748 (0.0006) & 0.796 (0.0007) & 0.1800 (0.0003) & 0.737 (0.0005) & 0.802 (0.0006)\\
SPT & 0.1299 (0.0006) & 0.759 (0.0008) & 0.787 (0.0009) & 0.1694 (0.0004) & 0.747 (0.0006) & 0.795 (0.0007) & 0.1808 (0.0003) & 0.736 (0.0005) & 0.801 (0.0006)\\
Target Cox & 0.1372 (0.0007) & 0.733 (0.0009) & 0.759 (0.0010) & 0.1831 (0.0004) & 0.717 (0.0006) & 0.759 (0.0007) & 0.1999 (0.0003) & 0.704 (0.0005) & 0.757 (0.0007)\\
CoxTL & 0.1380 (0.0007) & 0.735 (0.0009) & 0.761 (0.0010) & 0.1833 (0.0003) & 0.719 (0.0006) & 0.761 (0.0007) & 0.1984 (0.0003) & 0.706 (0.0005) & 0.760 (0.0006)\\
POTL-Cox & 0.1380 (0.0007) & 0.735 (0.0009) & 0.760 (0.0010) & 0.1831 (0.0004) & 0.719 (0.0006) & 0.760 (0.0007) & 0.1984 (0.0003) & 0.706 (0.0005) & 0.759 (0.0006)\\
CoxKL & 0.1381 (0.0007) & 0.735 (0.0009) & 0.760 (0.0010) & 0.1835 (0.0004) & 0.719 (0.0006) & 0.761 (0.0007) & 0.1986 (0.0003) & 0.706 (0.0005) & 0.759 (0.0006)\\
TransCox & 0.1372 (0.0007) & 0.733 (0.0009) & 0.759 (0.0010) & 0.1830 (0.0004) & 0.718 (0.0006) & 0.759 (0.0007) & 0.1998 (0.0003) & 0.704 (0.0005) & 0.757 (0.0007)\\
DiscreteKL & 0.1380 (0.0007) & 0.735 (0.0009) & 0.760 (0.0010) & 0.1829 (0.0003) & 0.719 (0.0006) & 0.761 (0.0007) & 0.1982 (0.0003) & 0.706 (0.0005) & 0.760 (0.0006)\\
Trans-AFT & 0.1474 (0.0008) & 0.711 (0.0023) & 0.734 (0.0026) & 0.1972 (0.0008) & 0.695 (0.0021) & 0.732 (0.0026) & 0.2124 (0.0008) & 0.681 (0.0020) & 0.729 (0.0025)\\
RSF & 0.1385 (0.0006) & 0.733 (0.0009) & 0.758 (0.0009) & 0.1835 (0.0004) & 0.722 (0.0007) & 0.764 (0.0009) & 0.1978 (0.0004) & 0.710 (0.0007) & 0.767 (0.0010)\\
IPCW-XGBoost & 0.1384 (0.0007) & 0.732 (0.0012) & 0.757 (0.0013) & 0.1821 (0.0005) & 0.722 (0.0008) & 0.765 (0.0010) & 0.1992 (0.0006) & 0.706 (0.0009) & 0.762 (0.0013)\\
\textit{Oracle} & \textit{0.1281 (0.0006)} & \textit{0.763 (0.0008)} & \textit{0.791 (0.0009)} & \textit{0.1666 (0.0004)} & \textit{0.751 (0.0006)} & \textit{0.799 (0.0007)} & \textit{0.1773 (0.0003)} & \textit{0.741 (0.0005)} & \textit{0.806 (0.0006)}\\
\addlinespace[4pt]
\multicolumn{10}{l}{\textbf{Setting B}}\\[2pt]
T-GAM & 0.1418 (0.0008) & 0.708 (0.0012) & 0.731 (0.0014) & 0.1888 (0.0005) & 0.704 (0.0011) & 0.747 (0.0013) & 0.2007 (0.0005) & 0.695 (0.0011) & 0.755 (0.0014)\\
Source GAM & 0.1484 (0.0008) & 0.694 (0.0011) & 0.715 (0.0012) & 0.1991 (0.0005) & 0.685 (0.0008) & 0.724 (0.0009) & 0.2123 (0.0004) & 0.677 (0.0007) & 0.730 (0.0009)\\
Pooled GAM & 0.1448 (0.0008) & 0.701 (0.0010) & 0.723 (0.0012) & 0.1942 (0.0004) & 0.693 (0.0008) & 0.733 (0.0009) & 0.2068 (0.0003) & 0.684 (0.0006) & 0.740 (0.0008)\\
Stratified GAM & 0.1439 (0.0007) & 0.701 (0.0010) & 0.723 (0.0012) & 0.1940 (0.0004) & 0.693 (0.0008) & 0.733 (0.0009) & 0.2074 (0.0004) & 0.684 (0.0006) & 0.740 (0.0008)\\
SPT & \textbf{0.1405 (0.0007)} & \textbf{0.713 (0.0011)} & \textbf{0.737 (0.0012)} & \textbf{0.1867 (0.0004)} & \textbf{0.709 (0.0009)} & \textbf{0.752 (0.0011)} & \textbf{0.1982 (0.0005)} & \textbf{0.700 (0.0009)} & \textbf{0.761 (0.0012)}\\
Target Cox & 0.1460 (0.0008) & 0.684 (0.0009) & 0.704 (0.0010) & 0.1996 (0.0004) & 0.675 (0.0007) & 0.710 (0.0009) & 0.2152 (0.0004) & 0.665 (0.0006) & 0.714 (0.0009)\\
CoxTL & 0.1460 (0.0008) & 0.684 (0.0010) & 0.704 (0.0011) & 0.1995 (0.0004) & 0.674 (0.0008) & 0.710 (0.0010) & 0.2154 (0.0004) & 0.665 (0.0007) & 0.713 (0.0010)\\
POTL-Cox & 0.1459 (0.0008) & 0.684 (0.0009) & 0.704 (0.0010) & 0.1993 (0.0004) & 0.675 (0.0007) & 0.710 (0.0009) & 0.2151 (0.0003) & 0.665 (0.0006) & 0.714 (0.0009)\\
CoxKL & 0.1459 (0.0008) & 0.684 (0.0009) & 0.704 (0.0010) & 0.1993 (0.0004) & 0.675 (0.0007) & 0.710 (0.0009) & 0.2151 (0.0003) & 0.665 (0.0006) & 0.713 (0.0009)\\
TransCox & 0.1460 (0.0008) & 0.684 (0.0009) & 0.704 (0.0010) & 0.1995 (0.0004) & 0.675 (0.0007) & 0.711 (0.0009) & 0.2151 (0.0003) & 0.665 (0.0006) & 0.715 (0.0009)\\
DiscreteKL & 0.1461 (0.0008) & 0.682 (0.0009) & 0.702 (0.0010) & 0.2002 (0.0004) & 0.672 (0.0008) & 0.707 (0.0010) & 0.2175 (0.0004) & 0.662 (0.0007) & 0.710 (0.0010)\\
Trans-AFT & 0.1525 (0.0009) & 0.655 (0.0022) & 0.672 (0.0025) & 0.2109 (0.0007) & 0.646 (0.0021) & 0.676 (0.0026) & 0.2266 (0.0007) & 0.637 (0.0021) & 0.678 (0.0029)\\
RSF & 0.1460 (0.0008) & 0.693 (0.0012) & 0.714 (0.0013) & 0.1967 (0.0004) & 0.691 (0.0009) & 0.730 (0.0011) & 0.2091 (0.0005) & 0.684 (0.0008) & 0.741 (0.0011)\\
IPCW-XGBoost & 0.1466 (0.0009) & 0.686 (0.0014) & 0.706 (0.0015) & 0.1961 (0.0005) & 0.688 (0.0011) & 0.726 (0.0014) & 0.2117 (0.0006) & 0.674 (0.0013) & 0.727 (0.0018)\\
\textit{Oracle} & \textit{0.1362 (0.0007)} & \textit{0.730 (0.0009)} & \textit{0.755 (0.0009)} & \textit{0.1797 (0.0004)} & \textit{0.723 (0.0006)} & \textit{0.769 (0.0008)} & \textit{0.1881 (0.0003)} & \textit{0.716 (0.0005)} & \textit{0.784 (0.0006)}\\
\addlinespace[4pt]
\bottomrule\end{tabular}
\par\smallskip\begin{minipage}{0.97\linewidth}\footnotesize
Entries are mean (MCSE), using 100 replications for every method and horizon. Bold identifies the best fitted-method mean within the displayed set; Oracle is excluded. Q25/Q50/Q75 are target-training observed-follow-up quantiles. SPT uses calibrated soft labels and the one-standard-error rule with reference $\tau=0.5$.
GCV selects source and target GAM structures before REML fitting. All three metrics use censored test outcomes. Target-training/test censoring rates are 50\%/30\%.
\end{minipage}\end{sidewaystable}

%% file: table/supp_table_07.tex
\begin{sidewaystable}[p]
\centering
\caption{Weibull: $n_T=500$, $p=7$, all reported comparators. Settings C/D.}\label{tab:allG2500weibull1}
\scriptsize\setlength{\tabcolsep}{1.3pt}\renewcommand{\arraystretch}{1.08}
\begin{tabular}{l*{9}{r}}\toprule
 & \multicolumn{3}{c}{Q25} & \multicolumn{3}{c}{Q50} & \multicolumn{3}{c}{Q75}\\
Method & Brier & $C_U(h)$ & AUC & Brier & $C_U(h)$ & AUC & Brier & $C_U(h)$ & AUC\\\midrule
\multicolumn{10}{l}{\textbf{Setting C}}\\[2pt]
T-GAM & 0.1410 (0.0007) & 0.714 (0.0012) & 0.738 (0.0014) & 0.1872 (0.0004) & 0.709 (0.0010) & 0.751 (0.0012) & 0.1989 (0.0005) & 0.700 (0.0009) & 0.761 (0.0012)\\
Source GAM & 0.1380 (0.0007) & 0.729 (0.0009) & 0.754 (0.0009) & 0.1823 (0.0004) & 0.720 (0.0007) & 0.765 (0.0008) & 0.1915 (0.0003) & 0.713 (0.0005) & 0.778 (0.0007)\\
Pooled GAM & \textbf{0.1373 (0.0007)} & 0.730 (0.0008) & 0.755 (0.0009) & \textbf{0.1816 (0.0004)} & 0.721 (0.0006) & 0.766 (0.0008) & \textbf{0.1909 (0.0003)} & 0.714 (0.0005) & 0.778 (0.0007)\\
Stratified GAM & 0.1373 (0.0007) & \textbf{0.730 (0.0008)} & \textbf{0.755 (0.0009)} & 0.1820 (0.0004) & 0.721 (0.0006) & 0.766 (0.0008) & 0.1914 (0.0003) & 0.714 (0.0005) & 0.778 (0.0007)\\
SPT & 0.1374 (0.0007) & 0.729 (0.0009) & 0.754 (0.0010) & 0.1817 (0.0004) & \textbf{0.721 (0.0006)} & \textbf{0.766 (0.0008)} & 0.1915 (0.0003) & \textbf{0.714 (0.0005)} & \textbf{0.779 (0.0007)}\\
Target Cox & 0.1421 (0.0007) & 0.710 (0.0009) & 0.733 (0.0010) & 0.1905 (0.0004) & 0.700 (0.0007) & 0.741 (0.0008) & 0.2030 (0.0003) & 0.692 (0.0006) & 0.749 (0.0007)\\
CoxTL & 0.1416 (0.0007) & 0.713 (0.0009) & 0.737 (0.0009) & 0.1893 (0.0003) & 0.704 (0.0006) & 0.745 (0.0008) & 0.2009 (0.0003) & 0.695 (0.0005) & 0.754 (0.0006)\\
POTL-Cox & 0.1417 (0.0007) & 0.713 (0.0009) & 0.736 (0.0010) & 0.1895 (0.0004) & 0.703 (0.0006) & 0.744 (0.0008) & 0.2010 (0.0003) & 0.695 (0.0005) & 0.753 (0.0007)\\
CoxKL & 0.1419 (0.0007) & 0.713 (0.0009) & 0.736 (0.0010) & 0.1897 (0.0003) & 0.703 (0.0006) & 0.744 (0.0008) & 0.2012 (0.0003) & 0.695 (0.0005) & 0.753 (0.0007)\\
TransCox & 0.1420 (0.0007) & 0.710 (0.0009) & 0.733 (0.0010) & 0.1904 (0.0004) & 0.701 (0.0007) & 0.741 (0.0008) & 0.2028 (0.0003) & 0.692 (0.0005) & 0.750 (0.0007)\\
DiscreteKL & 0.1414 (0.0007) & 0.713 (0.0009) & 0.737 (0.0010) & 0.1891 (0.0003) & 0.704 (0.0006) & 0.745 (0.0008) & 0.2008 (0.0002) & 0.695 (0.0005) & 0.754 (0.0006)\\
Trans-AFT & 0.1493 (0.0009) & 0.687 (0.0024) & 0.708 (0.0027) & 0.2018 (0.0008) & 0.677 (0.0024) & 0.714 (0.0029) & 0.2145 (0.0008) & 0.669 (0.0022) & 0.720 (0.0029)\\
RSF & 0.1447 (0.0007) & 0.703 (0.0010) & 0.725 (0.0011) & 0.1942 (0.0004) & 0.698 (0.0008) & 0.738 (0.0009) & 0.2066 (0.0004) & 0.691 (0.0007) & 0.748 (0.0010)\\
IPCW-XGBoost & 0.1446 (0.0007) & 0.702 (0.0011) & 0.724 (0.0012) & 0.1927 (0.0005) & 0.698 (0.0009) & 0.738 (0.0011) & 0.2076 (0.0005) & 0.685 (0.0010) & 0.742 (0.0013)\\
\textit{Oracle} & \textit{0.1351 (0.0007)} & \textit{0.735 (0.0008)} & \textit{0.761 (0.0009)} & \textit{0.1780 (0.0004)} & \textit{0.727 (0.0006)} & \textit{0.773 (0.0007)} & \textit{0.1866 (0.0003)} & \textit{0.720 (0.0005)} & \textit{0.787 (0.0006)}\\
\addlinespace[4pt]
\multicolumn{10}{l}{\textbf{Setting D}}\\[2pt]
T-GAM & 0.1482 (0.0008) & 0.674 (0.0014) & 0.694 (0.0016) & 0.1998 (0.0005) & 0.677 (0.0012) & 0.716 (0.0014) & 0.2098 (0.0005) & 0.673 (0.0011) & 0.733 (0.0015)\\
Source GAM & 0.1544 (0.0009) & 0.676 (0.0010) & 0.696 (0.0010) & 0.2082 (0.0005) & 0.672 (0.0007) & 0.709 (0.0008) & 0.2164 (0.0004) & 0.668 (0.0006) & 0.727 (0.0008)\\
Pooled GAM & 0.1503 (0.0008) & 0.680 (0.0009) & 0.700 (0.0010) & 0.2030 (0.0005) & 0.676 (0.0007) & 0.714 (0.0008) & 0.2118 (0.0003) & 0.673 (0.0006) & 0.733 (0.0008)\\
Stratified GAM & 0.1494 (0.0008) & 0.681 (0.0009) & 0.700 (0.0010) & 0.2033 (0.0005) & 0.676 (0.0007) & 0.714 (0.0008) & 0.2122 (0.0004) & 0.673 (0.0006) & 0.733 (0.0008)\\
SPT & \textbf{0.1457 (0.0008)} & \textbf{0.689 (0.0010)} & \textbf{0.710 (0.0011)} & \textbf{0.1961 (0.0004)} & \textbf{0.687 (0.0009)} & \textbf{0.728 (0.0010)} & \textbf{0.2051 (0.0004)} & \textbf{0.683 (0.0008)} & \textbf{0.747 (0.0011)}\\
Target Cox & 0.1480 (0.0008) & 0.676 (0.0008) & 0.696 (0.0009) & 0.2013 (0.0004) & 0.672 (0.0007) & 0.710 (0.0008) & 0.2108 (0.0003) & 0.670 (0.0006) & 0.729 (0.0009)\\
CoxTL & 0.1474 (0.0008) & 0.679 (0.0008) & 0.699 (0.0009) & 0.2000 (0.0004) & 0.676 (0.0007) & 0.714 (0.0008) & 0.2093 (0.0003) & 0.673 (0.0006) & 0.733 (0.0008)\\
POTL-Cox & 0.1472 (0.0008) & 0.680 (0.0008) & 0.700 (0.0009) & 0.1995 (0.0004) & 0.676 (0.0007) & 0.715 (0.0008) & 0.2086 (0.0003) & 0.673 (0.0005) & 0.734 (0.0008)\\
CoxKL & 0.1473 (0.0008) & 0.680 (0.0008) & 0.700 (0.0008) & 0.1997 (0.0004) & 0.676 (0.0006) & 0.715 (0.0008) & 0.2088 (0.0003) & 0.673 (0.0005) & 0.734 (0.0007)\\
TransCox & 0.1480 (0.0008) & 0.676 (0.0008) & 0.696 (0.0009) & 0.2012 (0.0004) & 0.673 (0.0007) & 0.711 (0.0008) & 0.2107 (0.0003) & 0.670 (0.0006) & 0.729 (0.0009)\\
DiscreteKL & 0.1475 (0.0008) & 0.679 (0.0008) & 0.699 (0.0009) & 0.2000 (0.0004) & 0.676 (0.0006) & 0.714 (0.0008) & 0.2090 (0.0003) & 0.673 (0.0005) & 0.733 (0.0007)\\
Trans-AFT & 0.1521 (0.0009) & 0.651 (0.0023) & 0.668 (0.0026) & 0.2097 (0.0007) & 0.647 (0.0023) & 0.680 (0.0029) & 0.2219 (0.0008) & 0.644 (0.0024) & 0.695 (0.0032)\\
RSF & 0.1505 (0.0008) & 0.663 (0.0012) & 0.681 (0.0013) & 0.2056 (0.0004) & 0.664 (0.0009) & 0.700 (0.0011) & 0.2165 (0.0004) & 0.663 (0.0008) & 0.720 (0.0011)\\
IPCW-XGBoost & 0.1502 (0.0008) & 0.665 (0.0013) & 0.684 (0.0015) & 0.2039 (0.0005) & 0.667 (0.0012) & 0.704 (0.0014) & 0.2153 (0.0004) & 0.662 (0.0010) & 0.719 (0.0013)\\
\textit{Oracle} & \textit{0.1425 (0.0008)} & \textit{0.701 (0.0008)} & \textit{0.723 (0.0009)} & \textit{0.1907 (0.0004)} & \textit{0.698 (0.0006)} & \textit{0.740 (0.0007)} & \textit{0.1978 (0.0003)} & \textit{0.695 (0.0005)} & \textit{0.762 (0.0007)}\\
\addlinespace[4pt]
\bottomrule\end{tabular}
\par\smallskip\begin{minipage}{0.97\linewidth}\footnotesize
Entries are mean (MCSE), using 100 replications for every method and horizon. Bold identifies the best fitted-method mean within the displayed set; Oracle is excluded. Q25/Q50/Q75 are target-training observed-follow-up quantiles. SPT uses calibrated soft labels and the one-standard-error rule with reference $\tau=0.5$.
GCV selects source and target GAM structures before REML fitting. All three metrics use censored test outcomes. Target-training/test censoring rates are 50\%/30\%.
\end{minipage}\end{sidewaystable}

%% file: table/supp_table_08.tex
\begin{sidewaystable}[p]
\centering
\caption{Loglogistic: $n_T=200$, $p=7$, all reported comparators. Settings B/D.}\label{tab:allG3200loglogistic0}
\scriptsize\setlength{\tabcolsep}{1.3pt}\renewcommand{\arraystretch}{1.08}
\begin{tabular}{l*{9}{r}}\toprule
 & \multicolumn{3}{c}{Q25} & \multicolumn{3}{c}{Q50} & \multicolumn{3}{c}{Q75}\\
Method & Brier & $C_U(h)$ & AUC & Brier & $C_U(h)$ & AUC & Brier & $C_U(h)$ & AUC\\\midrule
\multicolumn{10}{l}{\textbf{Setting B}}\\[2pt]
T-GAM & 0.1410 (0.0011) & 0.726 (0.0016) & 0.747 (0.0017) & 0.1946 (0.0008) & 0.704 (0.0016) & 0.738 (0.0019) & 0.2166 (0.0009) & 0.681 (0.0018) & 0.721 (0.0022)\\
Source GAM & 0.1434 (0.0010) & 0.727 (0.0012) & 0.748 (0.0012) & 0.1974 (0.0007) & 0.701 (0.0009) & 0.733 (0.0010) & 0.2168 (0.0004) & 0.681 (0.0008) & 0.719 (0.0010)\\
Pooled GAM & 0.1415 (0.0010) & 0.730 (0.0012) & 0.752 (0.0012) & 0.1951 (0.0007) & 0.705 (0.0009) & 0.737 (0.0010) & 0.2143 (0.0004) & 0.685 (0.0008) & 0.723 (0.0010)\\
Stratified GAM & 0.1405 (0.0010) & 0.730 (0.0012) & 0.752 (0.0012) & 0.1953 (0.0007) & 0.705 (0.0009) & 0.737 (0.0010) & 0.2154 (0.0004) & 0.685 (0.0008) & 0.723 (0.0010)\\
SPT & \textbf{0.1371 (0.0011)} & \textbf{0.740 (0.0015)} & \textbf{0.763 (0.0015)} & \textbf{0.1888 (0.0007)} & \textbf{0.718 (0.0013)} & \textbf{0.754 (0.0015)} & \textbf{0.2081 (0.0006)} & \textbf{0.698 (0.0012)} & \textbf{0.740 (0.0016)}\\
Target Cox & 0.1447 (0.0011) & 0.703 (0.0014) & 0.722 (0.0016) & 0.2033 (0.0007) & 0.678 (0.0012) & 0.707 (0.0015) & 0.2256 (0.0007) & 0.661 (0.0011) & 0.694 (0.0014)\\
CoxTL & 0.1448 (0.0011) & 0.702 (0.0015) & 0.721 (0.0016) & 0.2034 (0.0007) & 0.676 (0.0013) & 0.704 (0.0016) & 0.2259 (0.0006) & 0.658 (0.0012) & 0.689 (0.0016)\\
POTL-Cox & 0.1441 (0.0011) & 0.705 (0.0015) & 0.724 (0.0016) & 0.2024 (0.0007) & 0.678 (0.0013) & 0.707 (0.0015) & 0.2244 (0.0006) & 0.660 (0.0012) & 0.692 (0.0015)\\
CoxKL & 0.1445 (0.0011) & 0.705 (0.0014) & 0.723 (0.0015) & 0.2025 (0.0007) & 0.678 (0.0013) & 0.706 (0.0015) & 0.2246 (0.0006) & 0.659 (0.0012) & 0.691 (0.0015)\\
TransCox & 0.1445 (0.0011) & 0.704 (0.0014) & 0.723 (0.0015) & 0.2031 (0.0007) & 0.679 (0.0012) & 0.707 (0.0014) & 0.2253 (0.0006) & 0.661 (0.0011) & 0.694 (0.0014)\\
DiscreteKL & 0.1440 (0.0010) & 0.701 (0.0014) & 0.719 (0.0015) & 0.2039 (0.0007) & 0.674 (0.0013) & 0.701 (0.0015) & 0.2283 (0.0006) & 0.655 (0.0012) & 0.685 (0.0015)\\
Trans-AFT & 0.1514 (0.0012) & 0.666 (0.0038) & 0.681 (0.0041) & 0.2143 (0.0010) & 0.643 (0.0035) & 0.666 (0.0041) & 0.2348 (0.0010) & 0.626 (0.0032) & 0.651 (0.0041)\\
RSF & 0.1469 (0.0011) & 0.701 (0.0019) & 0.720 (0.0020) & 0.2047 (0.0007) & 0.682 (0.0018) & 0.712 (0.0021) & 0.2219 (0.0006) & 0.668 (0.0016) & 0.705 (0.0019)\\
IPCW-XGBoost & 0.1482 (0.0011) & 0.690 (0.0020) & 0.707 (0.0022) & 0.2072 (0.0009) & 0.671 (0.0022) & 0.699 (0.0026) & 0.2307 (0.0010) & 0.645 (0.0024) & 0.677 (0.0029)\\
\textit{Oracle} & \textit{0.1288 (0.0009)} & \textit{0.770 (0.0009)} & \textit{0.795 (0.0009)} & \textit{0.1765 (0.0005)} & \textit{0.745 (0.0007)} & \textit{0.785 (0.0008)} & \textit{0.1920 (0.0003)} & \textit{0.727 (0.0006)} & \textit{0.777 (0.0007)}\\
\addlinespace[4pt]
\multicolumn{10}{l}{\textbf{Setting D}}\\[2pt]
T-GAM & 0.1483 (0.0012) & 0.687 (0.0023) & 0.705 (0.0025) & 0.2058 (0.0009) & 0.674 (0.0019) & 0.706 (0.0022) & 0.2235 (0.0008) & 0.657 (0.0018) & 0.699 (0.0023)\\
Source GAM & 0.1487 (0.0011) & 0.712 (0.0010) & 0.732 (0.0010) & 0.2034 (0.0007) & 0.692 (0.0008) & 0.726 (0.0008) & 0.2164 (0.0004) & 0.679 (0.0007) & 0.725 (0.0008)\\
Pooled GAM & 0.1467 (0.0011) & 0.713 (0.0010) & 0.734 (0.0010) & 0.2014 (0.0006) & 0.694 (0.0008) & 0.728 (0.0008) & 0.2146 (0.0004) & 0.680 (0.0007) & 0.727 (0.0008)\\
Stratified GAM & 0.1457 (0.0011) & 0.713 (0.0010) & 0.734 (0.0010) & 0.2022 (0.0007) & 0.694 (0.0008) & 0.728 (0.0008) & 0.2158 (0.0004) & 0.680 (0.0007) & 0.727 (0.0008)\\
SPT & \textbf{0.1401 (0.0010)} & \textbf{0.726 (0.0011)} & \textbf{0.747 (0.0012)} & \textbf{0.1927 (0.0006)} & \textbf{0.709 (0.0008)} & \textbf{0.746 (0.0009)} & \textbf{0.2067 (0.0004)} & \textbf{0.694 (0.0007)} & \textbf{0.745 (0.0009)}\\
Target Cox & 0.1451 (0.0011) & 0.699 (0.0013) & 0.718 (0.0014) & 0.2027 (0.0007) & 0.681 (0.0012) & 0.714 (0.0014) & 0.2178 (0.0006) & 0.669 (0.0011) & 0.715 (0.0014)\\
CoxTL & 0.1431 (0.0011) & 0.711 (0.0009) & 0.732 (0.0010) & 0.1982 (0.0006) & 0.693 (0.0008) & 0.728 (0.0009) & 0.2120 (0.0004) & 0.680 (0.0007) & 0.729 (0.0009)\\
POTL-Cox & 0.1423 (0.0010) & 0.712 (0.0010) & 0.732 (0.0011) & 0.1972 (0.0006) & 0.693 (0.0009) & 0.728 (0.0010) & 0.2107 (0.0004) & 0.681 (0.0008) & 0.729 (0.0009)\\
CoxKL & 0.1431 (0.0011) & 0.712 (0.0009) & 0.733 (0.0010) & 0.1978 (0.0006) & 0.694 (0.0008) & 0.729 (0.0009) & 0.2113 (0.0003) & 0.681 (0.0007) & 0.730 (0.0008)\\
TransCox & 0.1448 (0.0011) & 0.701 (0.0013) & 0.720 (0.0014) & 0.2021 (0.0007) & 0.683 (0.0012) & 0.716 (0.0014) & 0.2170 (0.0006) & 0.671 (0.0010) & 0.717 (0.0013)\\
DiscreteKL & 0.1417 (0.0010) & 0.713 (0.0009) & 0.733 (0.0009) & 0.1969 (0.0006) & 0.694 (0.0008) & 0.729 (0.0008) & 0.2108 (0.0003) & 0.682 (0.0006) & 0.730 (0.0008)\\
Trans-AFT & 0.1489 (0.0012) & 0.673 (0.0035) & 0.690 (0.0038) & 0.2093 (0.0010) & 0.657 (0.0032) & 0.686 (0.0037) & 0.2251 (0.0009) & 0.646 (0.0029) & 0.686 (0.0037)\\
RSF & 0.1499 (0.0011) & 0.675 (0.0016) & 0.691 (0.0017) & 0.2094 (0.0006) & 0.663 (0.0014) & 0.693 (0.0017) & 0.2237 (0.0006) & 0.653 (0.0013) & 0.694 (0.0018)\\
IPCW-XGBoost & 0.1499 (0.0012) & 0.676 (0.0020) & 0.693 (0.0022) & 0.2101 (0.0008) & 0.664 (0.0017) & 0.693 (0.0020) & 0.2267 (0.0008) & 0.647 (0.0017) & 0.687 (0.0021)\\
\textit{Oracle} & \textit{0.1353 (0.0010)} & \textit{0.739 (0.0008)} & \textit{0.762 (0.0008)} & \textit{0.1869 (0.0005)} & \textit{0.720 (0.0007)} & \textit{0.758 (0.0007)} & \textit{0.1995 (0.0002)} & \textit{0.705 (0.0005)} & \textit{0.758 (0.0006)}\\
\addlinespace[4pt]
\bottomrule\end{tabular}
\par\smallskip\begin{minipage}{0.97\linewidth}\footnotesize
Entries are mean (MCSE), using 100 replications for every method and horizon. Bold identifies the best fitted-method mean within the displayed set; Oracle is excluded. Q25/Q50/Q75 are target-training observed-follow-up quantiles. SPT uses calibrated soft labels and the one-standard-error rule with reference $\tau=0.5$.
All three metrics use censored test outcomes. Target-training/test censoring rates are 50\%/30\%.
\end{minipage}\end{sidewaystable}

%% file: table/supp_table_09.tex
\begin{sidewaystable}[p]
\centering
\caption{Weibull: $n_T=200$, $p=7$, all reported comparators. Settings B/D.}\label{tab:allG3200weibull0}
\scriptsize\setlength{\tabcolsep}{1.3pt}\renewcommand{\arraystretch}{1.08}
\begin{tabular}{l*{9}{r}}\toprule
 & \multicolumn{3}{c}{Q25} & \multicolumn{3}{c}{Q50} & \multicolumn{3}{c}{Q75}\\
Method & Brier & $C_U(h)$ & AUC & Brier & $C_U(h)$ & AUC & Brier & $C_U(h)$ & AUC\\\midrule
\multicolumn{10}{l}{\textbf{Setting B}}\\[2pt]
T-GAM & 0.1496 (0.0011) & 0.687 (0.0017) & 0.708 (0.0019) & 0.1979 (0.0007) & 0.685 (0.0014) & 0.724 (0.0018) & 0.2130 (0.0009) & 0.676 (0.0016) & 0.729 (0.0021)\\
Source GAM & 0.1501 (0.0010) & 0.694 (0.0011) & 0.715 (0.0012) & 0.1995 (0.0006) & 0.685 (0.0008) & 0.723 (0.0009) & 0.2121 (0.0004) & 0.677 (0.0007) & 0.730 (0.0009)\\
Pooled GAM & 0.1485 (0.0010) & 0.697 (0.0011) & 0.719 (0.0012) & 0.1973 (0.0006) & 0.688 (0.0008) & 0.727 (0.0009) & 0.2097 (0.0004) & 0.680 (0.0007) & 0.734 (0.0009)\\
Stratified GAM & 0.1476 (0.0010) & 0.697 (0.0011) & 0.718 (0.0012) & 0.1973 (0.0006) & 0.688 (0.0008) & 0.727 (0.0009) & 0.2108 (0.0004) & 0.680 (0.0007) & 0.734 (0.0009)\\
SPT & \textbf{0.1455 (0.0010)} & \textbf{0.702 (0.0014)} & \textbf{0.725 (0.0016)} & \textbf{0.1921 (0.0006)} & \textbf{0.698 (0.0011)} & \textbf{0.739 (0.0013)} & \textbf{0.2042 (0.0006)} & \textbf{0.691 (0.0011)} & \textbf{0.749 (0.0015)}\\
Target Cox & 0.1502 (0.0010) & 0.675 (0.0012) & 0.695 (0.0014) & 0.2044 (0.0006) & 0.666 (0.0011) & 0.700 (0.0014) & 0.2215 (0.0006) & 0.657 (0.0010) & 0.703 (0.0013)\\
CoxTL & 0.1504 (0.0010) & 0.673 (0.0013) & 0.692 (0.0015) & 0.2048 (0.0006) & 0.663 (0.0012) & 0.696 (0.0014) & 0.2223 (0.0006) & 0.654 (0.0011) & 0.698 (0.0015)\\
POTL-Cox & 0.1496 (0.0010) & 0.676 (0.0013) & 0.695 (0.0015) & 0.2035 (0.0006) & 0.666 (0.0012) & 0.699 (0.0014) & 0.2207 (0.0006) & 0.656 (0.0011) & 0.701 (0.0014)\\
CoxKL & 0.1499 (0.0011) & 0.675 (0.0013) & 0.695 (0.0014) & 0.2035 (0.0006) & 0.665 (0.0011) & 0.699 (0.0014) & 0.2207 (0.0006) & 0.656 (0.0010) & 0.701 (0.0014)\\
TransCox & 0.1500 (0.0010) & 0.676 (0.0012) & 0.696 (0.0014) & 0.2040 (0.0006) & 0.667 (0.0011) & 0.701 (0.0013) & 0.2211 (0.0006) & 0.658 (0.0010) & 0.704 (0.0013)\\
DiscreteKL & 0.1497 (0.0010) & 0.673 (0.0012) & 0.692 (0.0014) & 0.2048 (0.0006) & 0.662 (0.0011) & 0.695 (0.0013) & 0.2235 (0.0006) & 0.652 (0.0010) & 0.696 (0.0014)\\
Trans-AFT & 0.1576 (0.0012) & 0.629 (0.0040) & 0.643 (0.0045) & 0.2179 (0.0010) & 0.621 (0.0038) & 0.646 (0.0046) & 0.2354 (0.0011) & 0.612 (0.0036) & 0.645 (0.0048)\\
RSF & 0.1538 (0.0010) & 0.666 (0.0019) & 0.685 (0.0022) & 0.2071 (0.0007) & 0.667 (0.0016) & 0.701 (0.0020) & 0.2193 (0.0007) & 0.664 (0.0014) & 0.714 (0.0019)\\
IPCW-XGBoost & 0.1556 (0.0011) & 0.650 (0.0027) & 0.667 (0.0030) & 0.2096 (0.0009) & 0.655 (0.0020) & 0.687 (0.0025) & 0.2280 (0.0011) & 0.644 (0.0022) & 0.688 (0.0030)\\
\textit{Oracle} & \textit{0.1377 (0.0009)} & \textit{0.730 (0.0008)} & \textit{0.756 (0.0009)} & \textit{0.1800 (0.0005)} & \textit{0.723 (0.0006)} & \textit{0.769 (0.0007)} & \textit{0.1880 (0.0003)} & \textit{0.717 (0.0005)} & \textit{0.783 (0.0006)}\\
\addlinespace[4pt]
\multicolumn{10}{l}{\textbf{Setting D}}\\[2pt]
T-GAM & 0.1555 (0.0011) & 0.647 (0.0022) & 0.664 (0.0025) & 0.2101 (0.0008) & 0.653 (0.0018) & 0.687 (0.0023) & 0.2228 (0.0010) & 0.649 (0.0019) & 0.702 (0.0026)\\
Source GAM & 0.1557 (0.0011) & 0.676 (0.0009) & 0.696 (0.0010) & 0.2084 (0.0006) & 0.672 (0.0007) & 0.709 (0.0008) & 0.2160 (0.0004) & 0.668 (0.0006) & 0.727 (0.0008)\\
Pooled GAM & 0.1538 (0.0011) & 0.678 (0.0009) & 0.698 (0.0010) & 0.2060 (0.0006) & 0.673 (0.0007) & 0.711 (0.0008) & 0.2139 (0.0004) & 0.670 (0.0006) & 0.729 (0.0008)\\
Stratified GAM & 0.1527 (0.0011) & 0.678 (0.0009) & 0.698 (0.0010) & 0.2066 (0.0006) & 0.673 (0.0007) & 0.711 (0.0008) & 0.2155 (0.0004) & 0.670 (0.0006) & 0.729 (0.0008)\\
SPT & \textbf{0.1480 (0.0010)} & \textbf{0.688 (0.0011)} & \textbf{0.709 (0.0012)} & \textbf{0.1972 (0.0005)} & \textbf{0.687 (0.0008)} & \textbf{0.728 (0.0010)} & \textbf{0.2060 (0.0005)} & \textbf{0.683 (0.0007)} & \textbf{0.747 (0.0010)}\\
Target Cox & 0.1515 (0.0011) & 0.668 (0.0011) & 0.687 (0.0012) & 0.2055 (0.0006) & 0.665 (0.0010) & 0.701 (0.0013) & 0.2160 (0.0006) & 0.662 (0.0009) & 0.719 (0.0014)\\
CoxTL & 0.1497 (0.0011) & 0.678 (0.0009) & 0.698 (0.0010) & 0.2016 (0.0006) & 0.674 (0.0008) & 0.713 (0.0010) & 0.2108 (0.0004) & 0.671 (0.0007) & 0.731 (0.0010)\\
POTL-Cox & 0.1490 (0.0010) & 0.679 (0.0009) & 0.699 (0.0009) & 0.2004 (0.0005) & 0.675 (0.0007) & 0.714 (0.0009) & 0.2090 (0.0004) & 0.672 (0.0006) & 0.733 (0.0009)\\
CoxKL & 0.1494 (0.0011) & 0.679 (0.0009) & 0.699 (0.0010) & 0.2009 (0.0005) & 0.675 (0.0007) & 0.714 (0.0009) & 0.2098 (0.0004) & 0.672 (0.0006) & 0.733 (0.0009)\\
TransCox & 0.1512 (0.0011) & 0.670 (0.0011) & 0.689 (0.0012) & 0.2050 (0.0006) & 0.666 (0.0010) & 0.703 (0.0013) & 0.2153 (0.0006) & 0.663 (0.0009) & 0.720 (0.0014)\\
DiscreteKL & 0.1491 (0.0010) & 0.679 (0.0008) & 0.699 (0.0009) & 0.2006 (0.0005) & 0.675 (0.0007) & 0.714 (0.0009) & 0.2091 (0.0003) & 0.672 (0.0006) & 0.733 (0.0008)\\
Trans-AFT & 0.1555 (0.0011) & 0.637 (0.0033) & 0.653 (0.0037) & 0.2140 (0.0009) & 0.634 (0.0032) & 0.664 (0.0039) & 0.2273 (0.0011) & 0.631 (0.0031) & 0.678 (0.0041)\\
RSF & 0.1561 (0.0011) & 0.641 (0.0015) & 0.657 (0.0016) & 0.2124 (0.0007) & 0.646 (0.0014) & 0.679 (0.0017) & 0.2234 (0.0007) & 0.646 (0.0013) & 0.697 (0.0018)\\
IPCW-XGBoost & 0.1570 (0.0012) & 0.639 (0.0021) & 0.654 (0.0023) & 0.2134 (0.0008) & 0.645 (0.0015) & 0.677 (0.0019) & 0.2272 (0.0009) & 0.639 (0.0016) & 0.688 (0.0023)\\
\textit{Oracle} & \textit{0.1437 (0.0010)} & \textit{0.701 (0.0008)} & \textit{0.723 (0.0009)} & \textit{0.1909 (0.0005)} & \textit{0.698 (0.0006)} & \textit{0.741 (0.0007)} & \textit{0.1974 (0.0003)} & \textit{0.695 (0.0005)} & \textit{0.762 (0.0007)}\\
\addlinespace[4pt]
\bottomrule\end{tabular}
\par\smallskip\begin{minipage}{0.97\linewidth}\footnotesize
Entries are mean (MCSE), using 100 replications for every method and horizon. Bold identifies the best fitted-method mean within the displayed set; Oracle is excluded. Q25/Q50/Q75 are target-training observed-follow-up quantiles. SPT uses calibrated soft labels and the one-standard-error rule with reference $\tau=0.5$.
All three metrics use censored test outcomes. Target-training/test censoring rates are 50\%/30\%.
\end{minipage}\end{sidewaystable}

%% file: table/supp_table_10.tex
\begin{sidewaystable}[p]
\centering
\caption{Loglogistic: $n_T=1000$, $p=7$, all reported comparators. Settings B/D.}\label{tab:allG31000loglogistic0}
\scriptsize\setlength{\tabcolsep}{1.3pt}\renewcommand{\arraystretch}{1.08}
\begin{tabular}{l*{9}{r}}\toprule
 & \multicolumn{3}{c}{Q25} & \multicolumn{3}{c}{Q50} & \multicolumn{3}{c}{Q75}\\
Method & Brier & $C_U(h)$ & AUC & Brier & $C_U(h)$ & AUC & Brier & $C_U(h)$ & AUC\\\midrule
\multicolumn{10}{l}{\textbf{Setting B}}\\[2pt]
T-GAM & 0.1310 (0.0006) & \textbf{0.760 (0.0008)} & \textbf{0.784 (0.0009)} & \textbf{0.1808 (0.0004)} & \textbf{0.736 (0.0007)} & \textbf{0.775 (0.0008)} & \textbf{0.1982 (0.0004)} & \textbf{0.715 (0.0007)} & \textbf{0.763 (0.0009)}\\
Source GAM & 0.1428 (0.0007) & 0.728 (0.0011) & 0.749 (0.0012) & 0.1977 (0.0005) & 0.701 (0.0008) & 0.733 (0.0010) & 0.2170 (0.0004) & 0.681 (0.0007) & 0.719 (0.0009)\\
Pooled GAM & 0.1360 (0.0006) & 0.743 (0.0010) & 0.766 (0.0010) & 0.1891 (0.0004) & 0.716 (0.0007) & 0.751 (0.0008) & 0.2076 (0.0003) & 0.696 (0.0006) & 0.738 (0.0008)\\
Stratified GAM & 0.1354 (0.0006) & 0.743 (0.0010) & 0.766 (0.0010) & 0.1891 (0.0004) & 0.716 (0.0007) & 0.751 (0.0008) & 0.2079 (0.0003) & 0.696 (0.0006) & 0.738 (0.0008)\\
SPT & \textbf{0.1310 (0.0006)} & 0.760 (0.0008) & 0.784 (0.0009) & 0.1811 (0.0004) & 0.735 (0.0008) & 0.773 (0.0009) & 0.1988 (0.0004) & 0.714 (0.0007) & 0.761 (0.0009)\\
Target Cox & 0.1407 (0.0006) & 0.718 (0.0008) & 0.738 (0.0009) & 0.1973 (0.0003) & 0.691 (0.0007) & 0.721 (0.0008) & 0.2174 (0.0003) & 0.672 (0.0006) & 0.708 (0.0008)\\
CoxTL & 0.1407 (0.0006) & 0.718 (0.0008) & 0.738 (0.0009) & 0.1972 (0.0003) & 0.691 (0.0007) & 0.721 (0.0008) & 0.2174 (0.0003) & 0.672 (0.0006) & 0.707 (0.0008)\\
POTL-Cox & 0.1407 (0.0006) & 0.718 (0.0008) & 0.738 (0.0009) & 0.1973 (0.0003) & 0.691 (0.0007) & 0.721 (0.0008) & 0.2174 (0.0003) & 0.672 (0.0006) & 0.707 (0.0008)\\
CoxKL & 0.1408 (0.0006) & 0.718 (0.0008) & 0.738 (0.0009) & 0.1973 (0.0003) & 0.691 (0.0007) & 0.721 (0.0008) & 0.2175 (0.0003) & 0.672 (0.0006) & 0.707 (0.0008)\\
TransCox & 0.1407 (0.0006) & 0.718 (0.0008) & 0.738 (0.0009) & 0.1973 (0.0003) & 0.691 (0.0007) & 0.722 (0.0008) & 0.2174 (0.0003) & 0.672 (0.0006) & 0.708 (0.0008)\\
DiscreteKL & 0.1403 (0.0006) & 0.717 (0.0009) & 0.737 (0.0009) & 0.1975 (0.0003) & 0.690 (0.0007) & 0.720 (0.0008) & 0.2192 (0.0003) & 0.671 (0.0006) & 0.705 (0.0008)\\
Trans-AFT & 0.1467 (0.0007) & 0.697 (0.0022) & 0.715 (0.0024) & 0.2068 (0.0006) & 0.672 (0.0019) & 0.699 (0.0023) & 0.2257 (0.0006) & 0.654 (0.0018) & 0.685 (0.0022)\\
RSF & 0.1365 (0.0006) & 0.741 (0.0008) & 0.763 (0.0009) & 0.1899 (0.0004) & 0.718 (0.0007) & 0.754 (0.0008) & 0.2072 (0.0004) & 0.700 (0.0007) & 0.745 (0.0010)\\
IPCW-XGBoost & 0.1354 (0.0006) & 0.742 (0.0010) & 0.765 (0.0011) & 0.1880 (0.0004) & 0.720 (0.0008) & 0.756 (0.0009) & 0.2069 (0.0004) & 0.697 (0.0009) & 0.741 (0.0012)\\
\textit{Oracle} & \textit{0.1282 (0.0006)} & \textit{0.770 (0.0008)} & \textit{0.795 (0.0008)} & \textit{0.1766 (0.0004)} & \textit{0.745 (0.0006)} & \textit{0.785 (0.0007)} & \textit{0.1922 (0.0003)} & \textit{0.726 (0.0005)} & \textit{0.777 (0.0007)}\\
\addlinespace[4pt]
\multicolumn{10}{l}{\textbf{Setting D}}\\[2pt]
T-GAM & 0.1371 (0.0006) & 0.729 (0.0009) & 0.751 (0.0009) & 0.1911 (0.0004) & 0.711 (0.0008) & 0.748 (0.0009) & 0.2049 (0.0003) & 0.695 (0.0006) & 0.745 (0.0008)\\
Source GAM & 0.1479 (0.0007) & 0.712 (0.0009) & 0.732 (0.0010) & 0.2036 (0.0005) & 0.692 (0.0007) & 0.726 (0.0008) & 0.2166 (0.0004) & 0.679 (0.0006) & 0.725 (0.0008)\\
Pooled GAM & 0.1411 (0.0006) & 0.721 (0.0009) & 0.741 (0.0010) & 0.1963 (0.0004) & 0.701 (0.0007) & 0.737 (0.0008) & 0.2097 (0.0003) & 0.687 (0.0006) & 0.736 (0.0008)\\
Stratified GAM & 0.1406 (0.0006) & 0.721 (0.0009) & 0.741 (0.0010) & 0.1965 (0.0004) & 0.701 (0.0007) & 0.737 (0.0008) & 0.2098 (0.0003) & 0.687 (0.0006) & 0.736 (0.0008)\\
SPT & \textbf{0.1362 (0.0006)} & \textbf{0.734 (0.0008)} & \textbf{0.756 (0.0009)} & \textbf{0.1895 (0.0004)} & \textbf{0.715 (0.0007)} & \textbf{0.752 (0.0008)} & \textbf{0.2024 (0.0003)} & \textbf{0.701 (0.0005)} & \textbf{0.752 (0.0007)}\\
Target Cox & 0.1411 (0.0006) & 0.712 (0.0008) & 0.732 (0.0009) & 0.1975 (0.0004) & 0.693 (0.0007) & 0.728 (0.0008) & 0.2109 (0.0003) & 0.681 (0.0005) & 0.729 (0.0007)\\
CoxTL & 0.1409 (0.0006) & 0.713 (0.0008) & 0.733 (0.0009) & 0.1969 (0.0004) & 0.695 (0.0007) & 0.730 (0.0008) & 0.2102 (0.0003) & 0.682 (0.0005) & 0.731 (0.0007)\\
POTL-Cox & 0.1408 (0.0006) & 0.714 (0.0008) & 0.734 (0.0008) & 0.1967 (0.0004) & 0.695 (0.0006) & 0.731 (0.0008) & 0.2098 (0.0002) & 0.682 (0.0005) & 0.731 (0.0007)\\
CoxKL & 0.1409 (0.0006) & 0.714 (0.0008) & 0.734 (0.0008) & 0.1968 (0.0004) & 0.695 (0.0006) & 0.731 (0.0008) & 0.2100 (0.0003) & 0.683 (0.0005) & 0.731 (0.0007)\\
TransCox & 0.1411 (0.0006) & 0.712 (0.0008) & 0.732 (0.0009) & 0.1974 (0.0004) & 0.693 (0.0007) & 0.728 (0.0008) & 0.2108 (0.0003) & 0.681 (0.0005) & 0.729 (0.0007)\\
DiscreteKL & 0.1405 (0.0006) & 0.714 (0.0008) & 0.734 (0.0008) & 0.1966 (0.0004) & 0.695 (0.0006) & 0.731 (0.0008) & 0.2103 (0.0003) & 0.682 (0.0005) & 0.731 (0.0007)\\
Trans-AFT & 0.1452 (0.0008) & 0.693 (0.0027) & 0.711 (0.0030) & 0.2045 (0.0008) & 0.676 (0.0025) & 0.708 (0.0030) & 0.2188 (0.0007) & 0.664 (0.0023) & 0.708 (0.0030)\\
RSF & 0.1418 (0.0006) & 0.709 (0.0008) & 0.729 (0.0009) & 0.1990 (0.0003) & 0.693 (0.0007) & 0.727 (0.0009) & 0.2133 (0.0003) & 0.679 (0.0006) & 0.726 (0.0009)\\
IPCW-XGBoost & 0.1393 (0.0006) & 0.719 (0.0009) & 0.740 (0.0010) & 0.1938 (0.0004) & 0.703 (0.0008) & 0.739 (0.0009) & 0.2093 (0.0003) & 0.685 (0.0007) & 0.733 (0.0009)\\
\textit{Oracle} & \textit{0.1345 (0.0006)} & \textit{0.739 (0.0008)} & \textit{0.762 (0.0008)} & \textit{0.1872 (0.0004)} & \textit{0.720 (0.0006)} & \textit{0.758 (0.0007)} & \textit{0.1998 (0.0003)} & \textit{0.705 (0.0005)} & \textit{0.757 (0.0007)}\\
\addlinespace[4pt]
\bottomrule\end{tabular}
\par\smallskip\begin{minipage}{0.97\linewidth}\footnotesize
Entries are mean (MCSE), using 100 replications for every method and horizon. Bold identifies the best fitted-method mean within the displayed set; Oracle is excluded. Q25/Q50/Q75 are target-training observed-follow-up quantiles. SPT uses calibrated soft labels and the one-standard-error rule with reference $\tau=0.5$.
All three metrics use censored test outcomes. Target-training/test censoring rates are 50\%/30\%.
\end{minipage}\end{sidewaystable}

%% file: table/supp_table_11.tex
\begin{sidewaystable}[p]
\centering
\caption{Weibull: $n_T=1000$, $p=7$, all reported comparators. Settings B/D.}\label{tab:allG31000weibull0}
\scriptsize\setlength{\tabcolsep}{1.3pt}\renewcommand{\arraystretch}{1.08}
\begin{tabular}{l*{9}{r}}\toprule
 & \multicolumn{3}{c}{Q25} & \multicolumn{3}{c}{Q50} & \multicolumn{3}{c}{Q75}\\
Method & Brier & $C_U(h)$ & AUC & Brier & $C_U(h)$ & AUC & Brier & $C_U(h)$ & AUC\\\midrule
\multicolumn{10}{l}{\textbf{Setting B}}\\[2pt]
T-GAM & 0.1403 (0.0005) & 0.718 (0.0009) & 0.742 (0.0010) & \textbf{0.1847 (0.0003)} & \textbf{0.714 (0.0007)} & \textbf{0.759 (0.0008)} & \textbf{0.1943 (0.0003)} & \textbf{0.707 (0.0006)} & \textbf{0.771 (0.0008)}\\
Source GAM & 0.1495 (0.0006) & 0.694 (0.0011) & 0.715 (0.0012) & 0.1997 (0.0005) & 0.685 (0.0008) & 0.724 (0.0009) & 0.2124 (0.0004) & 0.677 (0.0007) & 0.730 (0.0009)\\
Pooled GAM & 0.1437 (0.0006) & 0.707 (0.0009) & 0.730 (0.0011) & 0.1917 (0.0004) & 0.698 (0.0007) & 0.739 (0.0008) & 0.2033 (0.0003) & 0.690 (0.0006) & 0.748 (0.0008)\\
Stratified GAM & 0.1432 (0.0006) & 0.707 (0.0009) & 0.730 (0.0011) & 0.1915 (0.0004) & 0.698 (0.0007) & 0.739 (0.0008) & 0.2035 (0.0003) & 0.690 (0.0006) & 0.748 (0.0008)\\
SPT & \textbf{0.1401 (0.0005)} & \textbf{0.719 (0.0009)} & \textbf{0.743 (0.0010)} & 0.1850 (0.0004) & 0.713 (0.0007) & 0.757 (0.0008) & 0.1948 (0.0003) & 0.705 (0.0006) & 0.769 (0.0009)\\
Target Cox & 0.1465 (0.0006) & 0.686 (0.0008) & 0.706 (0.0009) & 0.1987 (0.0003) & 0.677 (0.0007) & 0.713 (0.0008) & 0.2137 (0.0003) & 0.667 (0.0006) & 0.717 (0.0008)\\
CoxTL & 0.1464 (0.0006) & 0.686 (0.0008) & 0.707 (0.0009) & 0.1987 (0.0003) & 0.677 (0.0006) & 0.713 (0.0008) & 0.2137 (0.0003) & 0.667 (0.0005) & 0.717 (0.0008)\\
POTL-Cox & 0.1464 (0.0006) & 0.686 (0.0008) & 0.707 (0.0009) & 0.1986 (0.0003) & 0.677 (0.0007) & 0.713 (0.0008) & 0.2137 (0.0003) & 0.667 (0.0005) & 0.717 (0.0008)\\
CoxKL & 0.1464 (0.0006) & 0.686 (0.0008) & 0.707 (0.0009) & 0.1986 (0.0003) & 0.677 (0.0007) & 0.713 (0.0008) & 0.2137 (0.0003) & 0.667 (0.0006) & 0.716 (0.0008)\\
TransCox & 0.1465 (0.0006) & 0.686 (0.0008) & 0.707 (0.0009) & 0.1987 (0.0003) & 0.677 (0.0007) & 0.713 (0.0008) & 0.2137 (0.0003) & 0.667 (0.0005) & 0.717 (0.0008)\\
DiscreteKL & 0.1465 (0.0006) & 0.685 (0.0008) & 0.705 (0.0009) & 0.1992 (0.0003) & 0.675 (0.0007) & 0.711 (0.0008) & 0.2153 (0.0003) & 0.666 (0.0005) & 0.714 (0.0008)\\
Trans-AFT & 0.1530 (0.0007) & 0.661 (0.0023) & 0.679 (0.0026) & 0.2099 (0.0006) & 0.652 (0.0022) & 0.684 (0.0027) & 0.2249 (0.0007) & 0.643 (0.0021) & 0.685 (0.0029)\\
RSF & 0.1444 (0.0005) & 0.703 (0.0008) & 0.725 (0.0009) & 0.1930 (0.0003) & 0.699 (0.0007) & 0.740 (0.0009) & 0.2040 (0.0004) & 0.693 (0.0007) & 0.752 (0.0010)\\
IPCW-XGBoost & 0.1439 (0.0006) & 0.702 (0.0010) & 0.725 (0.0011) & 0.1914 (0.0004) & 0.700 (0.0008) & 0.742 (0.0010) & 0.2035 (0.0004) & 0.690 (0.0008) & 0.749 (0.0011)\\
\textit{Oracle} & \textit{0.1372 (0.0005)} & \textit{0.730 (0.0008)} & \textit{0.755 (0.0009)} & \textit{0.1801 (0.0003)} & \textit{0.723 (0.0006)} & \textit{0.769 (0.0007)} & \textit{0.1882 (0.0003)} & \textit{0.716 (0.0005)} & \textit{0.784 (0.0006)}\\
\addlinespace[4pt]
\multicolumn{10}{l}{\textbf{Setting D}}\\[2pt]
T-GAM & 0.1456 (0.0006) & 0.688 (0.0010) & 0.709 (0.0012) & 0.1952 (0.0004) & 0.689 (0.0008) & 0.731 (0.0009) & 0.2033 (0.0003) & 0.685 (0.0005) & 0.750 (0.0008)\\
Source GAM & 0.1549 (0.0007) & 0.676 (0.0009) & 0.696 (0.0010) & 0.2084 (0.0005) & 0.672 (0.0007) & 0.710 (0.0008) & 0.2164 (0.0004) & 0.668 (0.0006) & 0.727 (0.0009)\\
Pooled GAM & 0.1487 (0.0006) & 0.684 (0.0009) & 0.704 (0.0010) & 0.2004 (0.0004) & 0.680 (0.0007) & 0.720 (0.0008) & 0.2086 (0.0004) & 0.677 (0.0006) & 0.738 (0.0008)\\
Stratified GAM & 0.1481 (0.0006) & 0.684 (0.0009) & 0.704 (0.0010) & 0.2006 (0.0004) & 0.680 (0.0007) & 0.720 (0.0008) & 0.2088 (0.0004) & 0.677 (0.0006) & 0.738 (0.0008)\\
SPT & \textbf{0.1444 (0.0006)} & \textbf{0.695 (0.0008)} & \textbf{0.717 (0.0009)} & \textbf{0.1938 (0.0004)} & \textbf{0.693 (0.0007)} & \textbf{0.735 (0.0008)} & \textbf{0.2012 (0.0003)} & \textbf{0.690 (0.0005)} & \textbf{0.756 (0.0007)}\\
Target Cox & 0.1477 (0.0006) & 0.678 (0.0008) & 0.698 (0.0009) & 0.2003 (0.0004) & 0.675 (0.0007) & 0.714 (0.0008) & 0.2092 (0.0003) & 0.672 (0.0005) & 0.733 (0.0008)\\
CoxTL & 0.1475 (0.0006) & 0.679 (0.0008) & 0.699 (0.0009) & 0.2000 (0.0004) & 0.676 (0.0007) & 0.715 (0.0008) & 0.2088 (0.0003) & 0.673 (0.0005) & 0.734 (0.0008)\\
POTL-Cox & 0.1475 (0.0006) & 0.680 (0.0008) & 0.700 (0.0009) & 0.1997 (0.0003) & 0.677 (0.0007) & 0.716 (0.0008) & 0.2083 (0.0003) & 0.673 (0.0005) & 0.735 (0.0008)\\
CoxKL & 0.1474 (0.0006) & 0.680 (0.0008) & 0.700 (0.0009) & 0.1997 (0.0004) & 0.677 (0.0007) & 0.716 (0.0008) & 0.2084 (0.0003) & 0.673 (0.0005) & 0.735 (0.0008)\\
TransCox & 0.1477 (0.0006) & 0.678 (0.0008) & 0.698 (0.0009) & 0.2003 (0.0004) & 0.675 (0.0007) & 0.714 (0.0008) & 0.2091 (0.0003) & 0.672 (0.0005) & 0.733 (0.0008)\\
DiscreteKL & 0.1477 (0.0006) & 0.679 (0.0008) & 0.699 (0.0009) & 0.2001 (0.0003) & 0.676 (0.0006) & 0.716 (0.0008) & 0.2087 (0.0003) & 0.673 (0.0005) & 0.734 (0.0007)\\
Trans-AFT & 0.1520 (0.0007) & 0.656 (0.0021) & 0.673 (0.0023) & 0.2091 (0.0006) & 0.653 (0.0020) & 0.687 (0.0024) & 0.2204 (0.0007) & 0.650 (0.0019) & 0.703 (0.0025)\\
RSF & 0.1489 (0.0006) & 0.672 (0.0009) & 0.692 (0.0010) & 0.2022 (0.0003) & 0.674 (0.0007) & 0.712 (0.0009) & 0.2121 (0.0003) & 0.671 (0.0006) & 0.731 (0.0008)\\
IPCW-XGBoost & 0.1479 (0.0006) & 0.677 (0.0011) & 0.697 (0.0013) & 0.1981 (0.0004) & 0.681 (0.0008) & 0.721 (0.0010) & 0.2080 (0.0003) & 0.676 (0.0007) & 0.737 (0.0009)\\
\textit{Oracle} & \textit{0.1428 (0.0006)} & \textit{0.701 (0.0008)} & \textit{0.723 (0.0009)} & \textit{0.1910 (0.0003)} & \textit{0.698 (0.0006)} & \textit{0.741 (0.0007)} & \textit{0.1978 (0.0003)} & \textit{0.695 (0.0005)} & \textit{0.762 (0.0007)}\\
\addlinespace[4pt]
\bottomrule\end{tabular}
\par\smallskip\begin{minipage}{0.97\linewidth}\footnotesize
Entries are mean (MCSE), using 100 replications for every method and horizon. Bold identifies the best fitted-method mean within the displayed set; Oracle is excluded. Q25/Q50/Q75 are target-training observed-follow-up quantiles. SPT uses calibrated soft labels and the one-standard-error rule with reference $\tau=0.5$.
All three metrics use censored test outcomes. Target-training/test censoring rates are 50\%/30\%.
\end{minipage}\end{sidewaystable}

%% file: table/supp_table_12.tex
\begin{sidewaystable}[p]
\centering
\caption{Loglogistic: $n_T=500$, $p=20$, all reported comparators. Settings B/D.}\label{tab:allG5500loglogistic0}
\scriptsize\setlength{\tabcolsep}{1.3pt}\renewcommand{\arraystretch}{1.08}
\begin{tabular}{l*{9}{r}}\toprule
 & \multicolumn{3}{c}{Q25} & \multicolumn{3}{c}{Q50} & \multicolumn{3}{c}{Q75}\\
Method & Brier & $C_U(h)$ & AUC & Brier & $C_U(h)$ & AUC & Brier & $C_U(h)$ & AUC\\\midrule
\multicolumn{10}{l}{\textbf{Setting B}}\\[2pt]
T-GAM & 0.1347 (0.0008) & 0.743 (0.0014) & 0.766 (0.0015) & 0.1863 (0.0006) & 0.723 (0.0011) & 0.759 (0.0014) & 0.2068 (0.0006) & 0.698 (0.0013) & 0.742 (0.0015)\\
Source GAM & 0.1421 (0.0008) & 0.727 (0.0011) & 0.748 (0.0012) & 0.1971 (0.0005) & 0.701 (0.0008) & 0.733 (0.0009) & 0.2170 (0.0004) & 0.681 (0.0008) & 0.719 (0.0010)\\
Pooled GAM & 0.1379 (0.0008) & 0.736 (0.0011) & 0.758 (0.0011) & 0.1918 (0.0005) & 0.710 (0.0008) & 0.744 (0.0009) & 0.2113 (0.0004) & 0.689 (0.0007) & 0.730 (0.0009)\\
Stratified GAM & 0.1369 (0.0008) & 0.736 (0.0011) & 0.758 (0.0011) & 0.1917 (0.0005) & 0.710 (0.0008) & 0.744 (0.0009) & 0.2119 (0.0004) & 0.689 (0.0007) & 0.730 (0.0009)\\
SPT & \textbf{0.1329 (0.0008)} & \textbf{0.750 (0.0012)} & \textbf{0.773 (0.0013)} & \textbf{0.1839 (0.0005)} & \textbf{0.728 (0.0009)} & \textbf{0.765 (0.0011)} & \textbf{0.2036 (0.0005)} & \textbf{0.705 (0.0010)} & \textbf{0.750 (0.0012)}\\
Target Cox & 0.1430 (0.0008) & 0.701 (0.0012) & 0.720 (0.0013) & 0.2032 (0.0005) & 0.676 (0.0010) & 0.704 (0.0011) & 0.2254 (0.0005) & 0.658 (0.0009) & 0.691 (0.0011)\\
CoxTL & 0.1424 (0.0008) & 0.703 (0.0011) & 0.722 (0.0011) & 0.2022 (0.0005) & 0.677 (0.0009) & 0.705 (0.0010) & 0.2244 (0.0004) & 0.659 (0.0008) & 0.691 (0.0011)\\
POTL-Cox & 0.1420 (0.0008) & 0.706 (0.0011) & 0.724 (0.0011) & 0.2013 (0.0004) & 0.679 (0.0008) & 0.707 (0.0010) & 0.2233 (0.0004) & 0.660 (0.0008) & 0.692 (0.0010)\\
CoxKL & 0.1422 (0.0008) & 0.705 (0.0011) & 0.724 (0.0011) & 0.2015 (0.0005) & 0.678 (0.0009) & 0.706 (0.0010) & 0.2234 (0.0004) & 0.659 (0.0008) & 0.691 (0.0011)\\
TransCox & 0.1428 (0.0008) & 0.702 (0.0012) & 0.720 (0.0013) & 0.2029 (0.0005) & 0.677 (0.0010) & 0.705 (0.0011) & 0.2250 (0.0005) & 0.659 (0.0009) & 0.692 (0.0011)\\
DiscreteKL & 0.1417 (0.0008) & 0.705 (0.0010) & 0.723 (0.0011) & 0.2021 (0.0004) & 0.678 (0.0008) & 0.705 (0.0010) & 0.2261 (0.0004) & 0.658 (0.0008) & 0.690 (0.0011)\\
Trans-AFT & 0.1480 (0.0009) & 0.679 (0.0022) & 0.695 (0.0024) & 0.2106 (0.0006) & 0.655 (0.0020) & 0.679 (0.0023) & 0.2303 (0.0006) & 0.638 (0.0018) & 0.667 (0.0023)\\
RSF & 0.1445 (0.0008) & 0.705 (0.0017) & 0.724 (0.0018) & 0.2042 (0.0006) & 0.687 (0.0013) & 0.717 (0.0015) & 0.2229 (0.0006) & 0.671 (0.0013) & 0.707 (0.0017)\\
IPCW-XGBoost & 0.1423 (0.0009) & 0.708 (0.0019) & 0.727 (0.0020) & 0.1992 (0.0007) & 0.689 (0.0016) & 0.720 (0.0019) & 0.2248 (0.0007) & 0.653 (0.0020) & 0.687 (0.0024)\\
\textit{Oracle} & \textit{0.1275 (0.0007)} & \textit{0.770 (0.0008)} & \textit{0.795 (0.0008)} & \textit{0.1761 (0.0004)} & \textit{0.745 (0.0007)} & \textit{0.785 (0.0008)} & \textit{0.1921 (0.0003)} & \textit{0.727 (0.0005)} & \textit{0.777 (0.0007)}\\
\addlinespace[4pt]
\multicolumn{10}{l}{\textbf{Setting D}}\\[2pt]
T-GAM & 0.1411 (0.0008) & 0.712 (0.0015) & 0.733 (0.0016) & 0.1968 (0.0005) & 0.696 (0.0013) & 0.730 (0.0015) & 0.2132 (0.0005) & 0.677 (0.0013) & 0.724 (0.0016)\\
Source GAM & 0.1481 (0.0008) & 0.711 (0.0010) & 0.731 (0.0011) & 0.2035 (0.0006) & 0.692 (0.0007) & 0.726 (0.0008) & 0.2170 (0.0004) & 0.678 (0.0006) & 0.724 (0.0008)\\
Pooled GAM & 0.1435 (0.0007) & 0.716 (0.0009) & 0.737 (0.0010) & 0.1988 (0.0005) & 0.697 (0.0007) & 0.731 (0.0008) & 0.2128 (0.0004) & 0.683 (0.0007) & 0.730 (0.0009)\\
Stratified GAM & 0.1427 (0.0007) & 0.716 (0.0009) & 0.737 (0.0010) & 0.1993 (0.0005) & 0.697 (0.0007) & 0.731 (0.0008) & 0.2130 (0.0004) & 0.683 (0.0007) & 0.730 (0.0009)\\
SPT & \textbf{0.1378 (0.0007)} & \textbf{0.729 (0.0009)} & \textbf{0.750 (0.0009)} & \textbf{0.1913 (0.0004)} & \textbf{0.711 (0.0007)} & \textbf{0.747 (0.0008)} & \textbf{0.2046 (0.0003)} & \textbf{0.697 (0.0006)} & \textbf{0.747 (0.0008)}\\
Target Cox & 0.1440 (0.0008) & 0.696 (0.0011) & 0.715 (0.0012) & 0.2032 (0.0005) & 0.679 (0.0010) & 0.711 (0.0011) & 0.2183 (0.0004) & 0.667 (0.0008) & 0.712 (0.0011)\\
CoxTL & 0.1415 (0.0008) & 0.711 (0.0009) & 0.731 (0.0009) & 0.1976 (0.0005) & 0.693 (0.0008) & 0.727 (0.0009) & 0.2114 (0.0003) & 0.680 (0.0007) & 0.728 (0.0009)\\
POTL-Cox & 0.1412 (0.0008) & 0.711 (0.0008) & 0.732 (0.0008) & 0.1971 (0.0004) & 0.693 (0.0007) & 0.728 (0.0008) & 0.2108 (0.0003) & 0.680 (0.0006) & 0.728 (0.0008)\\
CoxKL & 0.1415 (0.0008) & 0.712 (0.0008) & 0.732 (0.0009) & 0.1974 (0.0004) & 0.693 (0.0007) & 0.728 (0.0008) & 0.2110 (0.0003) & 0.680 (0.0006) & 0.729 (0.0008)\\
TransCox & 0.1438 (0.0008) & 0.697 (0.0011) & 0.717 (0.0012) & 0.2027 (0.0005) & 0.680 (0.0009) & 0.712 (0.0011) & 0.2177 (0.0004) & 0.668 (0.0008) & 0.713 (0.0010)\\
DiscreteKL & 0.1409 (0.0007) & 0.711 (0.0008) & 0.732 (0.0008) & 0.1972 (0.0004) & 0.693 (0.0007) & 0.728 (0.0008) & 0.2115 (0.0003) & 0.680 (0.0006) & 0.728 (0.0008)\\
Trans-AFT & 0.1463 (0.0008) & 0.682 (0.0024) & 0.700 (0.0026) & 0.2063 (0.0007) & 0.666 (0.0022) & 0.696 (0.0026) & 0.2215 (0.0007) & 0.654 (0.0020) & 0.696 (0.0026)\\
RSF & 0.1494 (0.0008) & 0.668 (0.0016) & 0.684 (0.0017) & 0.2123 (0.0004) & 0.658 (0.0013) & 0.686 (0.0015) & 0.2281 (0.0004) & 0.651 (0.0011) & 0.690 (0.0014)\\
IPCW-XGBoost & 0.1468 (0.0009) & 0.681 (0.0017) & 0.699 (0.0018) & 0.2060 (0.0005) & 0.670 (0.0012) & 0.701 (0.0014) & 0.2231 (0.0004) & 0.653 (0.0013) & 0.693 (0.0016)\\
\textit{Oracle} & \textit{0.1344 (0.0007)} & \textit{0.740 (0.0008)} & \textit{0.762 (0.0009)} & \textit{0.1868 (0.0004)} & \textit{0.720 (0.0006)} & \textit{0.758 (0.0007)} & \textit{0.1997 (0.0003)} & \textit{0.705 (0.0005)} & \textit{0.758 (0.0007)}\\
\addlinespace[4pt]
\bottomrule\end{tabular}
\par\smallskip\begin{minipage}{0.97\linewidth}\footnotesize
Entries are mean (MCSE), using 100 replications for every method and horizon. Bold identifies the best fitted-method mean within the displayed set; Oracle is excluded. Q25/Q50/Q75 are target-training observed-follow-up quantiles. SPT uses calibrated soft labels and the one-standard-error rule with reference $\tau=0.5$.
All three metrics use censored test outcomes. Target-training/test censoring rates are 50\%/30\%.
\end{minipage}\end{sidewaystable}

%% file: table/supp_table_13.tex
\begin{sidewaystable}[p]
\centering
\caption{Weibull: $n_T=500$, $p=20$, all reported comparators. Settings B/D.}\label{tab:allG5500weibull0}
\scriptsize\setlength{\tabcolsep}{1.3pt}\renewcommand{\arraystretch}{1.08}
\begin{tabular}{l*{9}{r}}\toprule
 & \multicolumn{3}{c}{Q25} & \multicolumn{3}{c}{Q50} & \multicolumn{3}{c}{Q75}\\
Method & Brier & $C_U(h)$ & AUC & Brier & $C_U(h)$ & AUC & Brier & $C_U(h)$ & AUC\\\midrule
\multicolumn{10}{l}{\textbf{Setting B}}\\[2pt]
T-GAM & 0.1433 (0.0008) & 0.702 (0.0014) & 0.724 (0.0015) & 0.1901 (0.0005) & 0.701 (0.0011) & 0.743 (0.0014) & 0.2033 (0.0006) & 0.691 (0.0013) & 0.750 (0.0016)\\
Source GAM & 0.1486 (0.0008) & 0.693 (0.0011) & 0.714 (0.0012) & 0.1993 (0.0005) & 0.685 (0.0008) & 0.723 (0.0009) & 0.2125 (0.0004) & 0.677 (0.0007) & 0.730 (0.0009)\\
Pooled GAM & 0.1450 (0.0008) & 0.701 (0.0011) & 0.723 (0.0012) & 0.1944 (0.0004) & 0.693 (0.0008) & 0.732 (0.0009) & 0.2070 (0.0004) & 0.684 (0.0006) & 0.740 (0.0008)\\
Stratified GAM & 0.1441 (0.0007) & 0.701 (0.0011) & 0.723 (0.0012) & 0.1941 (0.0004) & 0.693 (0.0008) & 0.732 (0.0009) & 0.2075 (0.0004) & 0.684 (0.0006) & 0.740 (0.0008)\\
SPT & \textbf{0.1414 (0.0008)} & \textbf{0.710 (0.0012)} & \textbf{0.733 (0.0013)} & \textbf{0.1876 (0.0005)} & \textbf{0.707 (0.0009)} & \textbf{0.749 (0.0011)} & \textbf{0.1998 (0.0005)} & \textbf{0.697 (0.0009)} & \textbf{0.758 (0.0012)}\\
Target Cox & 0.1484 (0.0008) & 0.673 (0.0011) & 0.692 (0.0012) & 0.2043 (0.0005) & 0.664 (0.0009) & 0.698 (0.0010) & 0.2211 (0.0004) & 0.655 (0.0008) & 0.701 (0.0011)\\
CoxTL & 0.1478 (0.0008) & 0.674 (0.0011) & 0.693 (0.0012) & 0.2034 (0.0004) & 0.665 (0.0008) & 0.698 (0.0009) & 0.2205 (0.0004) & 0.655 (0.0007) & 0.701 (0.0010)\\
POTL-Cox & 0.1473 (0.0008) & 0.676 (0.0011) & 0.695 (0.0012) & 0.2025 (0.0004) & 0.666 (0.0008) & 0.700 (0.0009) & 0.2193 (0.0004) & 0.657 (0.0007) & 0.703 (0.0010)\\
CoxKL & 0.1474 (0.0008) & 0.676 (0.0010) & 0.695 (0.0011) & 0.2025 (0.0004) & 0.666 (0.0008) & 0.699 (0.0009) & 0.2195 (0.0003) & 0.656 (0.0007) & 0.702 (0.0009)\\
TransCox & 0.1482 (0.0008) & 0.674 (0.0011) & 0.693 (0.0012) & 0.2039 (0.0005) & 0.665 (0.0009) & 0.698 (0.0010) & 0.2207 (0.0004) & 0.656 (0.0008) & 0.702 (0.0011)\\
DiscreteKL & 0.1475 (0.0008) & 0.675 (0.0010) & 0.694 (0.0011) & 0.2034 (0.0004) & 0.665 (0.0007) & 0.698 (0.0009) & 0.2218 (0.0004) & 0.655 (0.0007) & 0.700 (0.0010)\\
Trans-AFT & 0.1534 (0.0009) & 0.646 (0.0021) & 0.662 (0.0023) & 0.2131 (0.0006) & 0.637 (0.0020) & 0.665 (0.0024) & 0.2298 (0.0007) & 0.628 (0.0019) & 0.666 (0.0026)\\
RSF & 0.1509 (0.0008) & 0.672 (0.0016) & 0.690 (0.0017) & 0.2060 (0.0005) & 0.672 (0.0012) & 0.707 (0.0014) & 0.2197 (0.0006) & 0.668 (0.0011) & 0.718 (0.0015)\\
IPCW-XGBoost & 0.1507 (0.0009) & 0.663 (0.0018) & 0.681 (0.0020) & 0.2031 (0.0007) & 0.669 (0.0015) & 0.704 (0.0018) & 0.2200 (0.0007) & 0.655 (0.0018) & 0.703 (0.0024)\\
\textit{Oracle} & \textit{0.1362 (0.0007)} & \textit{0.730 (0.0009)} & \textit{0.755 (0.0009)} & \textit{0.1797 (0.0004)} & \textit{0.723 (0.0006)} & \textit{0.769 (0.0008)} & \textit{0.1881 (0.0003)} & \textit{0.716 (0.0005)} & \textit{0.784 (0.0006)}\\
\addlinespace[4pt]
\multicolumn{10}{l}{\textbf{Setting D}}\\[2pt]
T-GAM & 0.1495 (0.0008) & 0.669 (0.0016) & 0.687 (0.0018) & 0.2012 (0.0005) & 0.674 (0.0012) & 0.712 (0.0014) & 0.2111 (0.0006) & 0.671 (0.0012) & 0.730 (0.0016)\\
Source GAM & 0.1547 (0.0008) & 0.676 (0.0010) & 0.695 (0.0010) & 0.2084 (0.0005) & 0.671 (0.0007) & 0.709 (0.0008) & 0.2166 (0.0004) & 0.668 (0.0006) & 0.726 (0.0008)\\
Pooled GAM & 0.1506 (0.0008) & 0.680 (0.0009) & 0.700 (0.0010) & 0.2032 (0.0005) & 0.676 (0.0007) & 0.714 (0.0008) & 0.2120 (0.0004) & 0.672 (0.0006) & 0.732 (0.0008)\\
Stratified GAM & 0.1496 (0.0008) & 0.680 (0.0009) & 0.700 (0.0010) & 0.2035 (0.0005) & 0.676 (0.0007) & 0.714 (0.0008) & 0.2124 (0.0004) & 0.672 (0.0006) & 0.732 (0.0008)\\
SPT & \textbf{0.1457 (0.0008)} & \textbf{0.690 (0.0009)} & \textbf{0.711 (0.0010)} & \textbf{0.1956 (0.0004)} & \textbf{0.689 (0.0007)} & \textbf{0.730 (0.0008)} & \textbf{0.2033 (0.0003)} & \textbf{0.687 (0.0006)} & \textbf{0.751 (0.0008)}\\
Target Cox & 0.1502 (0.0008) & 0.666 (0.0010) & 0.684 (0.0011) & 0.2058 (0.0005) & 0.662 (0.0009) & 0.698 (0.0011) & 0.2165 (0.0004) & 0.659 (0.0008) & 0.716 (0.0011)\\
CoxTL & 0.1478 (0.0008) & 0.677 (0.0009) & 0.697 (0.0010) & 0.2009 (0.0004) & 0.674 (0.0008) & 0.712 (0.0009) & 0.2103 (0.0003) & 0.671 (0.0007) & 0.731 (0.0009)\\
POTL-Cox & 0.1476 (0.0008) & 0.678 (0.0008) & 0.698 (0.0009) & 0.2003 (0.0004) & 0.674 (0.0007) & 0.713 (0.0008) & 0.2094 (0.0003) & 0.671 (0.0006) & 0.732 (0.0008)\\
CoxKL & 0.1476 (0.0008) & 0.678 (0.0008) & 0.698 (0.0009) & 0.2003 (0.0004) & 0.675 (0.0007) & 0.713 (0.0008) & 0.2095 (0.0003) & 0.672 (0.0005) & 0.732 (0.0007)\\
TransCox & 0.1500 (0.0008) & 0.667 (0.0010) & 0.685 (0.0011) & 0.2054 (0.0004) & 0.663 (0.0009) & 0.699 (0.0010) & 0.2160 (0.0004) & 0.660 (0.0008) & 0.717 (0.0011)\\
DiscreteKL & 0.1480 (0.0008) & 0.678 (0.0008) & 0.697 (0.0009) & 0.2009 (0.0004) & 0.674 (0.0007) & 0.712 (0.0008) & 0.2101 (0.0003) & 0.671 (0.0006) & 0.731 (0.0008)\\
Trans-AFT & 0.1525 (0.0009) & 0.647 (0.0021) & 0.664 (0.0023) & 0.2106 (0.0007) & 0.644 (0.0020) & 0.675 (0.0024) & 0.2232 (0.0007) & 0.641 (0.0020) & 0.690 (0.0027)\\
RSF & 0.1546 (0.0008) & 0.637 (0.0014) & 0.652 (0.0016) & 0.2142 (0.0004) & 0.642 (0.0012) & 0.673 (0.0014) & 0.2272 (0.0004) & 0.645 (0.0011) & 0.696 (0.0014)\\
IPCW-XGBoost & 0.1532 (0.0009) & 0.642 (0.0019) & 0.658 (0.0021) & 0.2097 (0.0005) & 0.649 (0.0013) & 0.682 (0.0015) & 0.2219 (0.0005) & 0.648 (0.0012) & 0.699 (0.0016)\\
\textit{Oracle} & \textit{0.1425 (0.0008)} & \textit{0.701 (0.0008)} & \textit{0.723 (0.0009)} & \textit{0.1907 (0.0004)} & \textit{0.698 (0.0006)} & \textit{0.740 (0.0007)} & \textit{0.1978 (0.0003)} & \textit{0.695 (0.0005)} & \textit{0.762 (0.0007)}\\
\addlinespace[4pt]
\bottomrule\end{tabular}
\par\smallskip\begin{minipage}{0.97\linewidth}\footnotesize
Entries are mean (MCSE), using 100 replications for every method and horizon. Bold identifies the best fitted-method mean within the displayed set; Oracle is excluded. Q25/Q50/Q75 are target-training observed-follow-up quantiles. SPT uses calibrated soft labels and the one-standard-error rule with reference $\tau=0.5$.
All three metrics use censored test outcomes. Target-training/test censoring rates are 50\%/30\%.
\end{minipage}\end{sidewaystable}

%% file: table/supp_table_14.tex
\begin{table}[p]\centering
\caption{Observed-outcome and DGP-risk evaluation: a compact comparison of T-GAM, SPT and probability-only regularization.}\label{tab:dgp_check}
\footnotesize\setlength{\tabcolsep}{3pt}\renewcommand{\arraystretch}{1.1}
\begin{tabular}{l*{8}{r}}\toprule
 & \multicolumn{4}{c}{Log-logistic} & \multicolumn{4}{c}{Weibull}\\
\cmidrule(lr){2-5}\cmidrule(lr){6-9}
 & \multicolumn{2}{c}{Brier} & \multicolumn{2}{c}{AUC} & \multicolumn{2}{c}{Brier} & \multicolumn{2}{c}{AUC}\\
Method & IPCW & DGP & IPCW & DGP & IPCW & DGP & IPCW & DGP\\\midrule
\multicolumn{9}{l}{\textbf{Main design: $n_T=500$, $p=7$, settings A--D}}\\
T-GAM & 0.17293 & 0.17259 & 0.7635 & 0.7643 & 0.17604 & 0.17567 & 0.7471 & 0.7480\\
SPT & \textbf{0.16876} & \textbf{0.16829} & \textbf{0.7759} & \textbf{0.7771} & \textbf{0.17191} & \textbf{0.17142} & \textbf{0.7596} & \textbf{0.7609}\\
PR-GAM & 0.17146 & 0.17110 & 0.7673 & 0.7682 & 0.17473 & 0.17435 & 0.7504 & 0.7514\\
\textit{Oracle} & \textit{0.16443} & \textit{0.16397} & \textit{0.7868} & \textit{0.7880} & \textit{0.16723} & \textit{0.16674} & \textit{0.7709} & \textit{0.7722}\\
\addlinespace[3pt]
\multicolumn{9}{l}{\textbf{Smaller target sample: $n_T=200$, $p=7$, settings B/D}}\\
T-GAM & 0.18830 & 0.18799 & 0.7194 & 0.7199 & 0.19149 & 0.19106 & 0.7022 & 0.7030\\
SPT & \textbf{0.17890} & \textbf{0.17845} & \textbf{0.7491} & \textbf{0.7502} & \textbf{0.18216} & \textbf{0.18164} & \textbf{0.7328} & \textbf{0.7340}\\
PR-GAM & 0.18489 & 0.18459 & 0.7274 & 0.7279 & 0.18817 & 0.18775 & 0.7106 & 0.7113\\
\textit{Oracle} & \textit{0.16983} & \textit{0.16940} & \textit{0.7724} & \textit{0.7735} & \textit{0.17297} & \textit{0.17248} & \textit{0.7558} & \textit{0.7571}\\
\addlinespace[3pt]
\multicolumn{9}{l}{\textbf{Larger target sample: $n_T=1000$, $p=7$, settings B/D}}\\
T-GAM & 0.17382 & 0.17340 & 0.7608 & 0.7621 & 0.17722 & 0.17678 & 0.7437 & 0.7449\\
SPT & \textbf{0.17316} & \textbf{0.17267} & \textbf{0.7630} & \textbf{0.7645} & \textbf{0.17654} & \textbf{0.17605} & \textbf{0.7461} & \textbf{0.7475}\\
PR-GAM & 0.17376 & 0.17332 & 0.7609 & 0.7622 & 0.17713 & 0.17668 & 0.7439 & 0.7451\\
\textit{Oracle} & \textit{0.16975} & \textit{0.16929} & \textit{0.7722} & \textit{0.7735} & \textit{0.17285} & \textit{0.17235} & \textit{0.7557} & \textit{0.7571}\\
\addlinespace[3pt]
\multicolumn{9}{l}{\textbf{Irrelevant covariates: $n_T=500$, $p=20$, settings B/D}}\\
T-GAM & 0.17984 & 0.17952 & 0.7422 & 0.7429 & 0.18309 & 0.18273 & 0.7244 & 0.7253\\
SPT & \textbf{0.17567} & \textbf{0.17523} & \textbf{0.7554} & \textbf{0.7566} & \textbf{0.17888} & \textbf{0.17842} & \textbf{0.7386} & \textbf{0.7399}\\
PR-GAM & 0.17789 & 0.17754 & 0.7476 & 0.7485 & 0.18122 & 0.18084 & 0.7299 & 0.7309\\
\textit{Oracle} & \textit{0.16943} & \textit{0.16899} & \textit{0.7724} & \textit{0.7736} & \textit{0.17250} & \textit{0.17202} & \textit{0.7555} & \textit{0.7569}\\
\addlinespace[3pt]
\bottomrule\end{tabular}
\par\smallskip\begin{minipage}{0.98\linewidth}\footnotesize The main design uses GCV-selected structures; the extensions use prespecified working structures. Entries average the means from $R=100$ replications equally over settings and Q25/Q50/Q75 within each distribution: 12 tasks for the main design and six for each extension. Task-specific means and MCSEs are reported in the full result tables. IPCW uses censored test outcomes; DGP integrates over true horizon-specific event probabilities. Bold identifies the best unrounded average among the three fitted methods in each panel and column. Oracle is an italicized reference, excluded from ranking.\end{minipage}
\end{table}

%% file: table/supp_table_15.tex
\begin{table}[p]\centering
\caption{Training-selected target interactions in the main simulation.}\label{tab:structure_selection}
\small\setlength{\tabcolsep}{8pt}
\begin{tabular}{llrrr}\toprule
Setting & Distribution & Q25 & Q50 & Q75\\\midrule
A & Log-logistic & 99\% & 100\% & 100\%\\
A & Weibull & 100\% & 100\% & 100\%\\
B & Log-logistic & 100\% & 100\% & 100\%\\
B & Weibull & 100\% & 100\% & 100\%\\
C & Log-logistic & 94\% & 96\% & 98\%\\
C & Weibull & 94\% & 97\% & 99\%\\
D & Log-logistic & 42\% & 40\% & 40\%\\
D & Weibull & 47\% & 38\% & 50\%\\
\bottomrule\end{tabular}
\par\smallskip\begin{minipage}{0.95\linewidth}\footnotesize
Entries are percentages of 100 full-training target fits selecting the interaction by GCV before REML estimation. Target selection is also repeated independently within every inner training fold. The source selects the interaction in all 2400 fits (four settings, two distributions, three horizons and 100 replications). Both candidates include all main effects.
\end{minipage}\end{table}

%% file: table/supp_table_16.tex
\begin{table}[p]\centering
\caption{Projection versus direct-score borrowing under structural mismatch: paired SPT-minus-DST differences in setting D.}\label{tab:structural_contrasts}
\small\setlength{\tabcolsep}{4pt}\renewcommand{\arraystretch}{1.15}
\begin{tabular}{llrrr}\toprule
Distribution & Horizon & $\Delta$Brier & $\Delta C_U(h)$ (pp) & $\Delta$AUC (pp)\\\midrule
Log-logistic & Q25 & -0.00095 (0.00010) & +0.400 (0.045) & +0.435 (0.049)\\
Log-logistic & Q50 & -0.00210 (0.00020) & +0.505 (0.050) & +0.589 (0.058)\\
Log-logistic & Q75 & -0.00303 (0.00028) & +0.598 (0.056) & +0.729 (0.068)\\
Weibull & Q25 & -0.00100 (0.00012) & +0.471 (0.055) & +0.526 (0.061)\\
Weibull & Q50 & -0.00194 (0.00019) & +0.486 (0.048) & +0.589 (0.058)\\
Weibull & Q75 & -0.00215 (0.00025) & +0.428 (0.049) & +0.564 (0.065)\\
\bottomrule\end{tabular}
\par\smallskip\begin{minipage}{0.95\linewidth}\footnotesize
Entries are paired mean differences (MCSE) across 100 replications. Negative Brier and positive concordance/AUC differences favor SPT. Both methods use the same training-selected target representation and source fit, with calibrated soft labels and the one-standard-error rule with reference $\tau=0.5$.
\end{minipage}\end{table}